\documentclass[showpacs,amssymb,floatfix]{revtex4}
\usepackage{graphicx}
\usepackage{dcolumn}
\usepackage{bm}
\begin{document}
\title{Light-induced breaking of symmetry in
photonic crystal waveguides  with nonlinear defects as a key for
all-optical switching circuits}
\author{Evgeny Bulgakov, Almas Sadreev, and
Konstantin N. Pichugin}
\address{Kirensky Institute of Physics, 660036, Krasnoyarsk, Russia}
\date{\today}
\begin{abstract}
We consider light transmission in 2D photonic crystal waveguide
coupled with two identical nonlinear defects positioned
symmetrically aside the waveguide. We show that with growth of
injected light power there is a breaking of symmetry by two ways.
In the first way the symmetry is broken because of different light
intensities at the defects. In the second way the intensities at
the defects are equaled but phases of complex amplitudes are
different. That results in a vortical power flow between the
defects similar to the DC Josephson effect if the input power over
the waveguide is applied and the defects are coupled. As
application of these phenomena we consider the symmetry breaking
for the light transmission in a T-shaped photonic waveguide with
two nonlinear defects. We demonstrate as this phenomenon can be
explored for all-optical switching of light transmission from the
left output waveguide to the right one by application of input
pulses. Finally we consider the symmetry breaking in the waveguide
coupled with single defect presented however by two dipole modes.
\end{abstract}
\maketitle
\section{Introduction}
Symmetry breaking in a nonlinear quantum system is a fundamental
effect caused by the interplay of nonlinearity with linear
potential which defines the symmetry. It is commonly known that
the ground state in one-dimensional linear quantum mechanics  is
nodeless and follows the symmetry of the potential. However the
self-attractive nonlinearity in the nonlinear Schr\"odinger
equation  breaks the symmetry of the ground state, replacing it by
a new asymmetric state minimizing the system's energy. For
example, the nonlinear Schr\"odinger equation in  double-well
potential reveals anti-symmetric ground state with variation of
normalization of the state \cite{kivshar}. The phenomenon of the
spontaneous symmetry breaking in analog with the double-well
potential are realized in a nonlinear dual-core directional fiber
\cite{akhmediev,tasgal,gubeskys}. Spontaneous symmetry breaking
was demonstrated recently by Brazhnyi and Malomed in a linear
discrete chain (Schr\"odinger lattice) with two nonlinear sites
\cite{brazhnyi}. They have shown as analytically as well as
numerically the existence of symmetric, anti-symmetric, and
non-symmetric eigen-modes with eigen-frequencies below the
propagation band of the chain, and that a variation of the
population of modes can give rise to a bifurcation form one to
another mode. The system has direct relation to photonic crystal
(PhC) waveguides with two in-channel nonlinear cavities where the
population of the cavities might be governed by external source of
the light.

Indeed, the phenomenon of symmetry breaking is studied in the
nonlinear optics with injection of input power with the
establishment of one or more asymmetric states which no longer
preserve  the symmetry properties of the original state
\cite{Haelterman,Peschel,Babushkin,Torres,Longchambon}. In
particular Maes {\it et al.} \cite{maes1,maes2} considered the
symmetry breaking for the nonlinear cavities aligned along the
waveguide, that is a Fabry-P\'{e}rot architecture close to the
system considered in Ref. \cite{Babushkin}. That system is
symmetric relative to the inversion of the transport axis if equal
power is injected on both sides of the coupled cavities. The
symmetry breaking was found also for the case of many coupled
nonlinear optical cavities in ring-like architecture
\cite{Otsuka,Huybrechts}. In the section II we write the equations
of motion for the nonlinear optical cavities coupled with PhC
waveguides by using an analogy of the two-dimensional PhC with
quantum mechanics \cite{joanbook}. As for an application we
consider three simple PhC systems which undergo the symmetry
breaking phenomena for variation of the light frequency or the
input power. The first simplest system is two identical nonlinear
defects positioned symmetrically aside the straight forward linear
waveguide (section III). Each defect is presented by single
monopole mode. We show two types of the symmetry breaking
\cite{BPS1,BPS2}. In the first type the symmetry is broken because
of different light intensities at the defects. In the second type
of the symmetry breaking  the intensities at the cavities are
equaled but phases of complex amplitudes are different. That
results in a vortical power flow between the defects similar to
the DC Josephson effect if the input power over the waveguide is
applied, and the defects are coupled.

In section IV  we consider as the phenomenon of the symmetry
breaking can be explored for so called all-optical switching
\cite{jensen,mayer,friberg,gibbs} by use of the T-shaped photonic
waveguide with two identical nonlinear cavities positioned
symmetrically. That system combines two systems. The first one is
the Fabry-P\'{e}rot interferometer (FPI) consisting of two
nonlinear off-channel cavities aligned along the straightforward
waveguide considered in Refs. \cite{maes1,maes2,FPR}. As was shown
in Ref. \cite{FPR} there is a discrete set of the a self-induced
bound (localized) states in continuum (BSC) which are the standing
waves between off-channel cavities. In the second system two
nonlinear cavities are aligned perpendicular to the input
waveguide. As was said above there is the anti-bonding bound state
in continuum (BSC). Here we show that both types of the bound
states might be important for the breaking of symmetry. All these
phenomena agree well with computations based on an expansion  of
the electromagnetic field into optimally adapted photonic Wannier
functions in two-dimensional PhC \cite{marzari,busch}.

Finally, in section V we consider the single nonlinear defect with
two dipole eigen-modes which belong the propagation band of the
PhC straightforward waveguide (section III). We demonstrate the
symmetry breaking provided that the system is excited with equal
powers from both sides similar to that Maes {\it et al.} has shown
in the system of two coupled nonlinear cavities
\cite{maes1,maes2}.
\section{basic equations}
The light propagation in linear PhC is described by the Maxwell
equations
\begin{eqnarray}
\label{maxwell}
\nabla\times\vec{E}=-\frac{\partial\vec{H}}{\partial t}\nonumber\\
\nabla\times\vec{H}=\frac{\partial\vec{D}}{\partial t},\\
\vec{D}(\vec{r},t)=\epsilon_0(\vec{r})\vec{E}(\vec{r},t).\nonumber
 \end{eqnarray}
We take the light velocity to be equal to unit. However if there
are defects with instantaneous Kerr nonlinearity, the displacement
electric vector interior to the defects has a nonlinear
contribution $\vec{D}(\vec{r},t)=\epsilon_0(\vec{r})
\vec{E}(\vec{r},t)+\chi^{(3)}[\vec{E}(\vec{r},t)]^2\vec{E}(\vec{r},t)$
\cite{LLel,abad}. A substitution of the electric field in the form
$[\vec{E}(\vec{r},t)=\frac{1}{2}[\vec{E}(\vec{r})e^{i\omega
t}+\vec{E}^{*}(\vec{r})e^{-i\omega t}]$ into Eq. (\ref{maxwell})
and neglect by highly oscillating terms such as $e^{2i\omega}$
allows us to write the Maxwell equations in the same form as Eq.
(\ref{maxwell}) with
\begin{equation}\label{eps}
  \epsilon(\vec{r})=
  \epsilon_0(\vec{r})+\frac{1}{4}\chi^{(3)}(\omega)|\vec{E}(\vec{r})|^2\vec{E}(\vec{r})+
\frac{1}{2}\chi^{(3)}(\omega)\vec{E}^2(\vec{r})\vec{E}^{*}(\vec{r}).
\end{equation}
In what follows we consider the 2D PhC with arrays of infinitely
long dielectric rods as shown in Fig. \ref{fig1}(a) in which the
electric field is directed along the rods while the magnetic field
is directed perpendicular to the rods [in the plane of Fig.
\ref{fig1}(a)]. Then Eq. (\ref{eps}) simplifies as follows
\cite{LLel}
\begin{equation}\label{eps1}
\epsilon(\vec{r})=
\epsilon_0(\vec{r})+\frac{3}{4}\chi^{(3)}(\omega)|\vec{E}(\vec{r})|^2\vec{E}(\vec{r}).
\end{equation}

There is a remarkable analogy of electrodynamics in dielectric
media with quantum mechanics \cite{joanbook,skor}. In particular,
if the nonlinear contribution to the dielectric constant is small
we can use the well-known methods of quantum mechanical
perturbation theory.  Let $|\psi\rangle=\left(\begin{array}{l}
\overrightarrow{E}\cr \overrightarrow{H}\cr \end{array}\right)$ be
the electromagnetic state in the PhC. Then the Maxwell equations
(\ref{maxwell}) can be written as the Schr\"odinger equation
$i\dot{|\psi\rangle}=\widehat{H}|\psi\rangle$ indeed with the
Hamiltonian \cite{joanbook,skor,Winn}
\begin{equation}\label{hamilt}
\widehat{H}=\left(\begin{array}{cc} 0 &
\frac{i}{\epsilon(\vec{r})}\nabla\times  \cr -i\nabla\times  & 0
\end{array}\right).
\end{equation}
Because of the perturbation of the dielectric constant
(\ref{eps1}) the Hamiltonian can be presented as
$\widehat{H}=\widehat{H}_0+\widehat{V}$ where
\begin{equation}\label{perturb}
\widehat{H}_0=\left(\begin{array}{cc} 0 &
\frac{i}{\epsilon_0(\vec{r})}\nabla\times  \cr -i\nabla\times & 0
\end{array}\right),
~\widehat{V}=\left(\begin{array}{cc} 0 &
i\delta\left(\frac{1}{\epsilon(\vec{r})}\right)\nabla\times \cr 0
& 0
\end{array}\right),
\end{equation}
and
\begin{equation}\label{deltaeps}
\delta\left(\frac{1}{\epsilon(\vec{r})}\right)=\frac{1}{\epsilon(\vec{r})}-\frac{1}{\epsilon_0(\vec{r})}.
\end{equation}

Let us introduce (following, for example, Refs. \cite{Winn,cowan})
the following inner product for the unperturbed system:
\begin{equation}
\langle\psi|\psi'\rangle=\frac{1}{2}\int[\epsilon_0(\vec{r})\vec{E}^{*}\vec{E}'+
\vec{H}^{*}\vec{H}']d^3\vec{r}. \label{IP}
\end{equation}
which obeys the following normalization and orthogonality
conditions for the bound eigen-states of the unperturbed
Hamiltonian $\widehat{H}_0|\psi_m\rangle=\omega_m|\psi_m\rangle$
\begin{equation}
\langle\psi_n|\psi_{n'}\rangle=\frac{1}{2}\int[\epsilon_0(\vec{r})\vec{E}_n^{*}\vec{E}_{n'}+
\vec{H}_n^{*}\vec{H}_{n'}]d^3\vec{r}=\int\epsilon_0(\vec{r})\vec{E}_n^{*}\vec{E}_{n'}d^3\vec{r}
=\delta_{nn'}. \label{norma}
\end{equation}
Then the matrix elements for the perturbation calculated by use of
these eigen-states are
\begin{equation}\label{Vmn}
\langle m|V|n\rangle=\frac{\omega_n}{2}\int
d^3\vec{r}\epsilon_0^2(\vec{r})\delta\left(\frac{1}{\epsilon(\vec{r})}\right)\vec{E}_m^{*}(\vec{r})
\vec{E}_n(\vec{r}).
\end{equation}
One can see that the matrix (\ref{Vmn}) is not Hermitian as was
noted in Ref. \cite{Winn}. The origin is that the unperturbed
states obey the inner product (\ref{IP})  with the dielectric
constant $\epsilon_0(\vec{r})$ while the eigen-states of the full
Hamiltonian $\widehat{H}_0+\widehat{V}$ obey the inner product
with a different dielectric constant $\epsilon(\vec{r})$.
Respectively, the Hamiltonian $\widehat{H}$ is non-Hermitian with
the inner product (\ref{norma}).

In order to avoid this problem we must use the inner product which
is not tied to a specific choice of the dielectric constant. One
way, given in Ref. \cite{joanbook}, is by using only the magnetic
field for the state. Another way is to absorb the dielectric
constant in the scalar product by a new function as
$\vec{F}=\sqrt{\epsilon(\vec{r})}\vec{E}$. Then the inner product
becomes
\begin{equation}
\langle\psi|\psi'\rangle=\frac{1}{2}\int[\vec{F}^{*} \vec{F}^{'}+
\vec{H}^{*}\vec{H}^{'}]d^3\vec{r}. \label{inner}
\end{equation}
The value
$\langle\psi|\psi\rangle=\frac{1}{2}\int[\epsilon(\vec{r})|\vec{E}|^2
+ |\vec{H}|^2]d^3\vec{r}$ is proportional to the energy of EM
field which is important for the derivation of the forthcoming
coupled mode theory (CMT) equations. That technique changes the
Maxwell equations as follows:
\begin{eqnarray}
\nabla\times\frac{\vec{F}}{\sqrt{\epsilon(\vec{r})}}=-\dot{\vec{H}}\nonumber\\
\frac{1}{\sqrt{\epsilon(\vec{r})}}\nabla\times\vec{H}=\dot{\vec{F}}.\\
\nonumber
 \label{maxwellnew}
\end{eqnarray}
The Hamiltonian takes the following form
\begin{equation}\label{perturbnew}
\widehat{H}_0=\left(\begin{array}{cc} 0 &
\frac{i}{\sqrt{\epsilon_0(\vec{r})}}\nabla\times  \cr
-i\nabla\times \frac{1}{\sqrt{\epsilon_0(\vec{r})}}& 0
\end{array}\right),
~\widehat{V}=\left(\begin{array}{cc} 0 &
i\delta\left(\frac{1}{\sqrt{\epsilon(\vec{r})}}\right)\nabla\times
\cr
-i\nabla\times\delta\left(\frac{1}{\sqrt{\epsilon(\vec{r})}}\right)
& 0
\end{array}\right).
\end{equation}

Now the eigen-states of the full Hamiltonian can be expanded over
the eigen-states $|m\rangle=\left(\begin{array}{c} \vec{F}_m \cr
\vec{H}_m
\end{array}\right)$ of the unperturbed Hamiltonian $\widehat{H}_0$ where
\begin{eqnarray}
\nabla\times\frac{\vec{F}_m}{\sqrt{\epsilon_0(\vec{r})}}=i\omega_m\vec{H}_m\nonumber\\
\nabla\times\vec{H}_m=-i\omega_m\sqrt{\epsilon_0(\vec{r})}\vec{F}_m.\\
\nonumber
 \label{maxwellstat}
\end{eqnarray}
Then we obtain from (\ref{Vmn})
\begin{equation}\label{Vmn1}
\langle m|V|n\rangle=\frac{(\omega_m+\omega_n)}{2}\int
d^3\vec{r}\epsilon_0^{3/2}(\vec{r})\delta\left(\frac{1}{\sqrt{\epsilon(\vec{r})}}\right)
E_m^{*}(\vec{r}) E_n(\vec{r}).
\end{equation}
One can see that the full Hamiltonian is Hermitian now.

If the nonlinear defect rods are thin enough, the dielectric
constant (\ref{eps1}) can be rewritten as follows
\begin{equation}\label{Kerr}
  \epsilon_j({\bf x})=(\epsilon_0+\frac{3}{4}\chi^{(3)}(\omega)|E({\bf x})|^2)
  \sum_j\theta({\bf x}-{\bf x}_j).
\end{equation}
Here $j$ enumerates the defects, $\theta=1$ inside the defect rod
and $\theta=0$ outside. As was shown for the simple square lattice
2D PhC from thin GaAs dielectric rods \cite{busch} the resonance
spectra in the PhC waveguide  are located in a rather narrow
frequency domain. Therefore, we neglect the frequency dependence
in the nonlinear susceptibility $\chi^{(3)}(\omega)$ in the
following. Assuming that the nonlinear contribution in Eq.
(\ref{Kerr}) is small compared to $\epsilon_0$ we obtain for the
matrix elements (\ref{Vmn1}) per unit length of the defect rods
\begin{equation}\label{MEV}
  \langle m|V|n\rangle\approx
 - \frac{3}{16\epsilon_0^{3/2}}\chi^{(3)}(\omega_m+\omega_n)\sum_j\int_{\sigma_j}
 d^2{\bf x}|E({\bf   x})|^2E_m({\bf x})^{*}E_n({\bf x}).
\end{equation}

In order to find electric fields at the defects we must constitute
a way to excite the defect modes. Here we consider that the EM
field propagates from the left along the straight forward
waveguide, interacts with the nonlinear defects, reflects back and
transmits to the right. Then the transmission process can be
described by the CMT stationary equations
\cite{haus,manol,fan-suh,suh}
\begin{equation}\label{Am}
[\omega-\sum_n(\omega_m\delta_{mn}+V_{mn}+i\Gamma_n]A_m=i\sqrt{\Gamma_m}E_{in}.
\end{equation}
These CMT equations, in fact, are the Lippmann-Schwinger equation
\cite{fanPRB,photonic}
\begin{equation}\label{LS}
(\omega-\widehat{H}_{eff})\Psi=i\widehat{W}E_{in}.
\end{equation}
where the complex matrix $\widehat{H}_{eff}$  equals
\begin{equation}\label{Heff}
\widehat{H}_{eff}=\widehat{H}_0+\widehat{V}-i\widehat{W}\widehat{W}^{+},
\end{equation}
the columns of the matrix $\widehat{W}$ consists of coupling
constants of the m-th eigen-mode with the p-th injecting wave
$\sqrt{\Gamma_{mp}}$, and the column $\Psi$ consists of the mode
amplitudes $A_m$. The solution $\Psi$ is given by inverse of the
matrix $\omega-\widehat{H}_{eff}$ where the matrix elements of the
effective Hamiltonian $\widehat{H}_{eff}$ in turn depend on the
mode amplitudes $A_m$. In order to write the equations of
self-consistency for the amplitudes at the defects we expand the
electric field at the j-th defect over eigen-modes $E({\bf
x}_j)=\sum_m A_m\psi_m({\bf x}_j)$. That defines the equations of
self-consistency after substitution into Eq. (\ref{Am}).

Finally, we present the transmission amplitude in the framework of
the CMT \cite{manol,fan-suh}
\begin{equation}\label{t}
t=E_{in}-\widehat{W}^{+}\Psi.
\end{equation}
\section{Linear optical waveguide coupled with two nonlinear off-channel
cavities aligned symmetrically}
\begin{figure}[ht]
\includegraphics[scale=0.3]{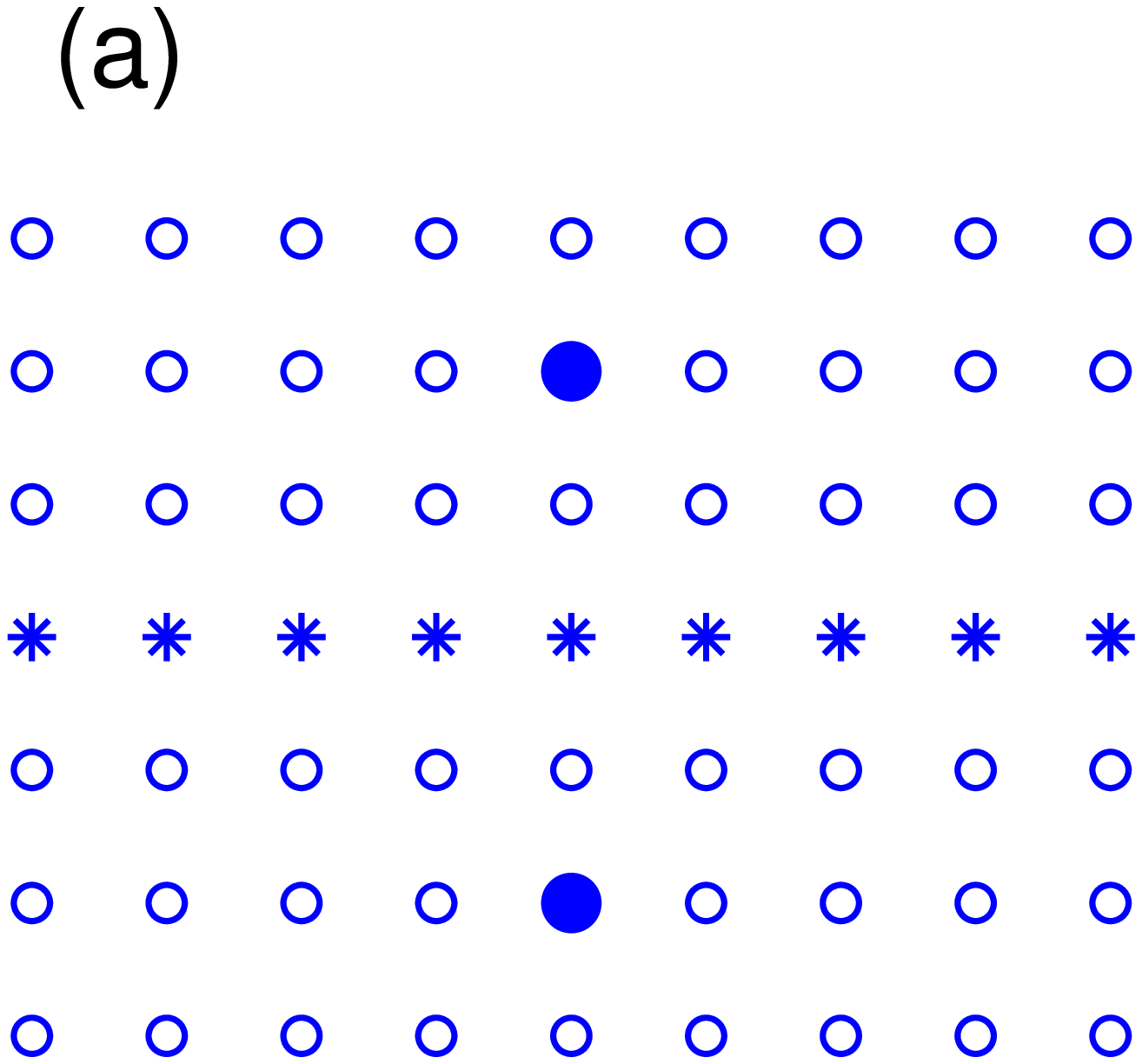}
\includegraphics[scale=0.3]{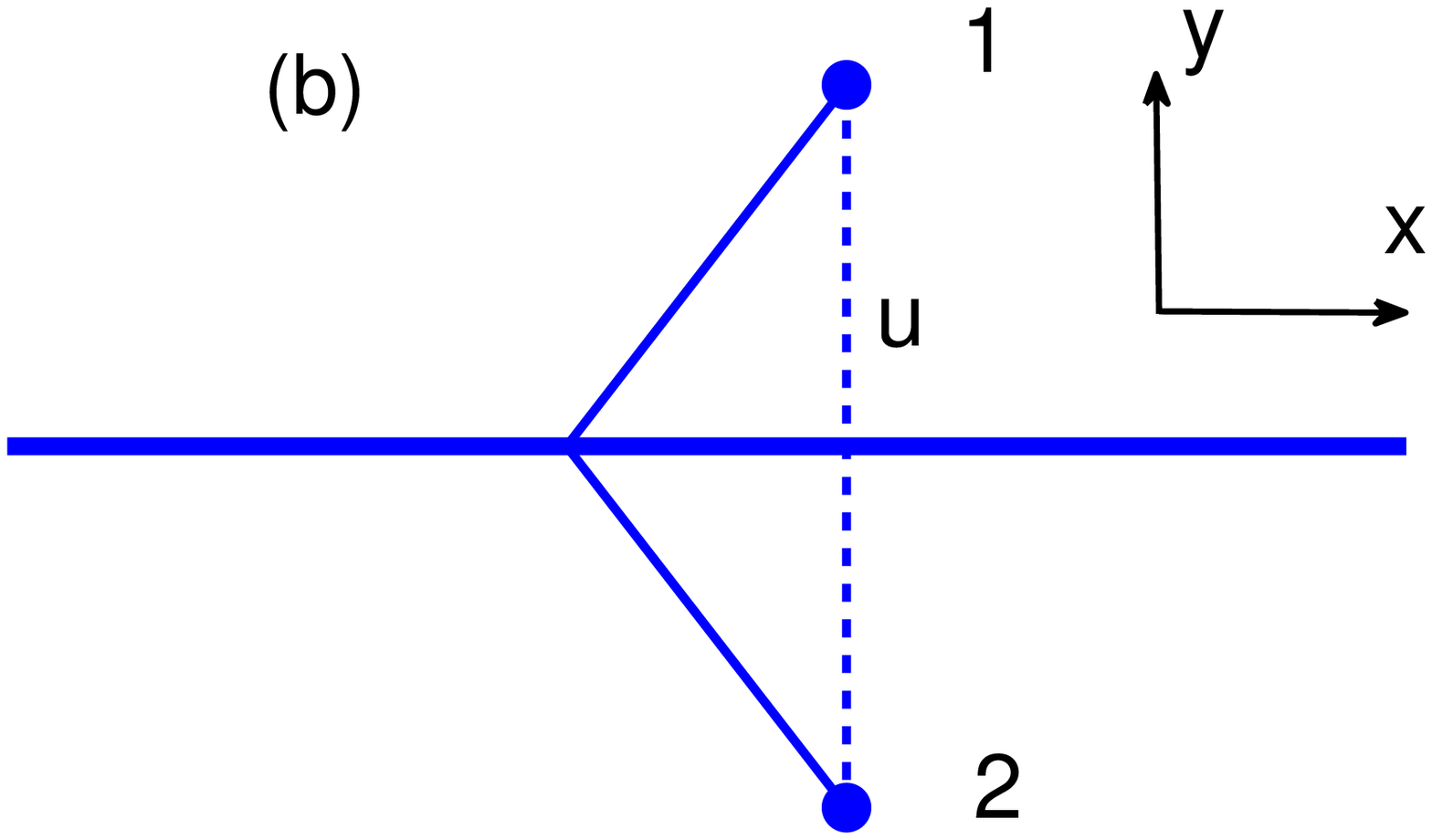}
\caption{(a) Two defect rods made from a Kerr media marked by
filled circles are inserted into the square lattice PhC of
dielectric rods with the  lattice constant $a = 0.5\mu m$, the
cylindrical dielectric rods have radius $0.18a$ and dielectric
constant $\epsilon_0= 11.56$. The 1D waveguide is formed by
substitution of linear chain of rods by the rods with dielectric
constant $\epsilon_W+\epsilon_0$ marked by stars. (b) Schematic
system consisting of a waveguide aside coupled to two single-mode
cavities. The cavities are coupled each other via $u$.}
\label{fig1}
\end{figure}

Two identical nonlinear defects positioned symmetrically relative
to the single straight forward waveguide is one of the simplest
systems in which the breaking of symmetry occurs \cite{BPS1,BPS2}.
The system can easily be realized in 2D PhC as shown in Fig.
\ref{fig1} (a).
The system is symmetric relative to the inversion of the y axis,
as shown in Fig. \ref{fig1}(b), and thereby supplements the system
in which two nonlinear cavities are aligned along the waveguide
considered by Maes {\it et al} \cite{maes1,maes2}. That system is
symmetric relative to the inversion of the x axis if equal power
is injected on both sides of the waveguide.

\begin{figure}[ht]
\includegraphics[scale=0.36]{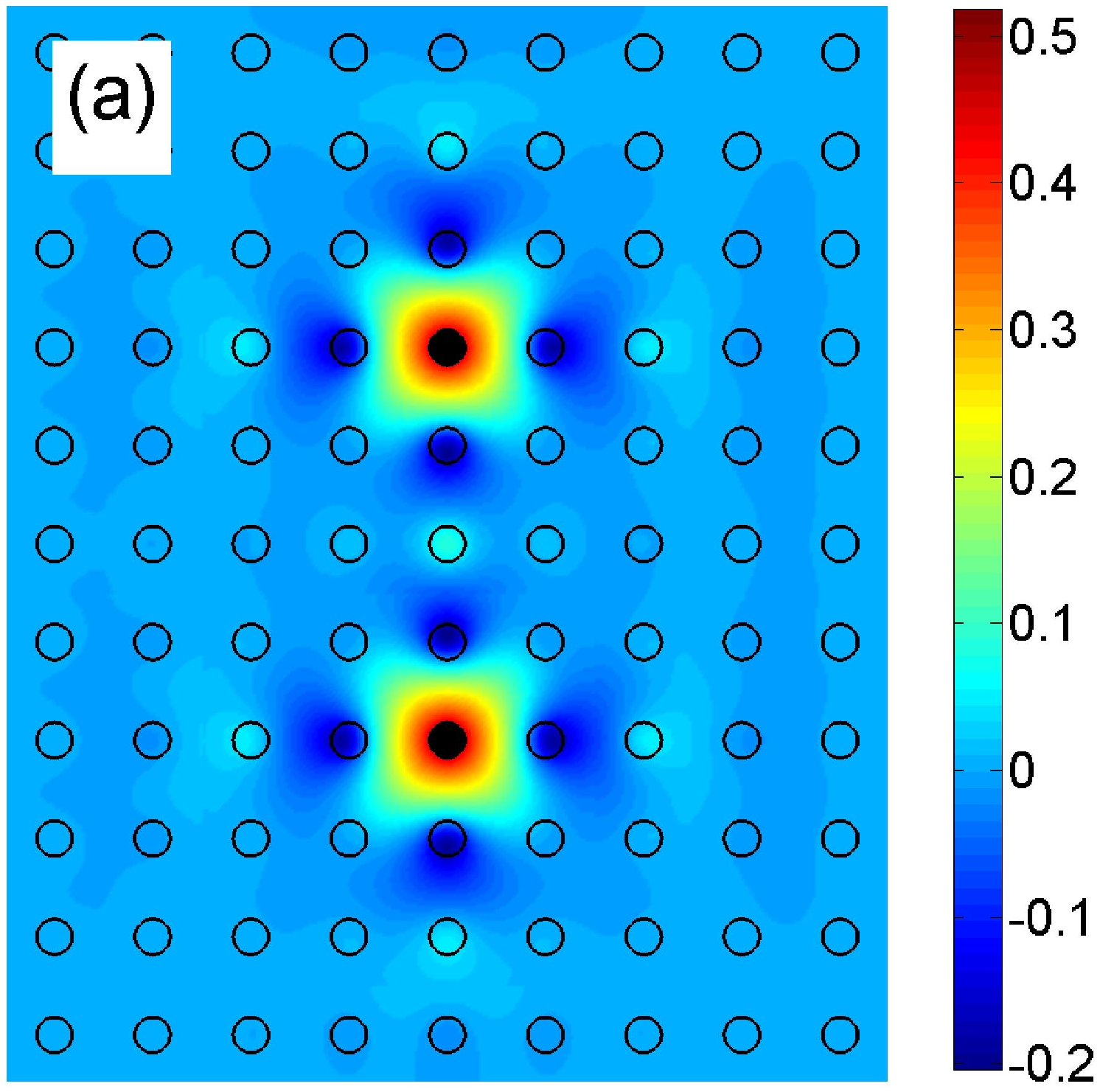}
\includegraphics[scale=0.36]{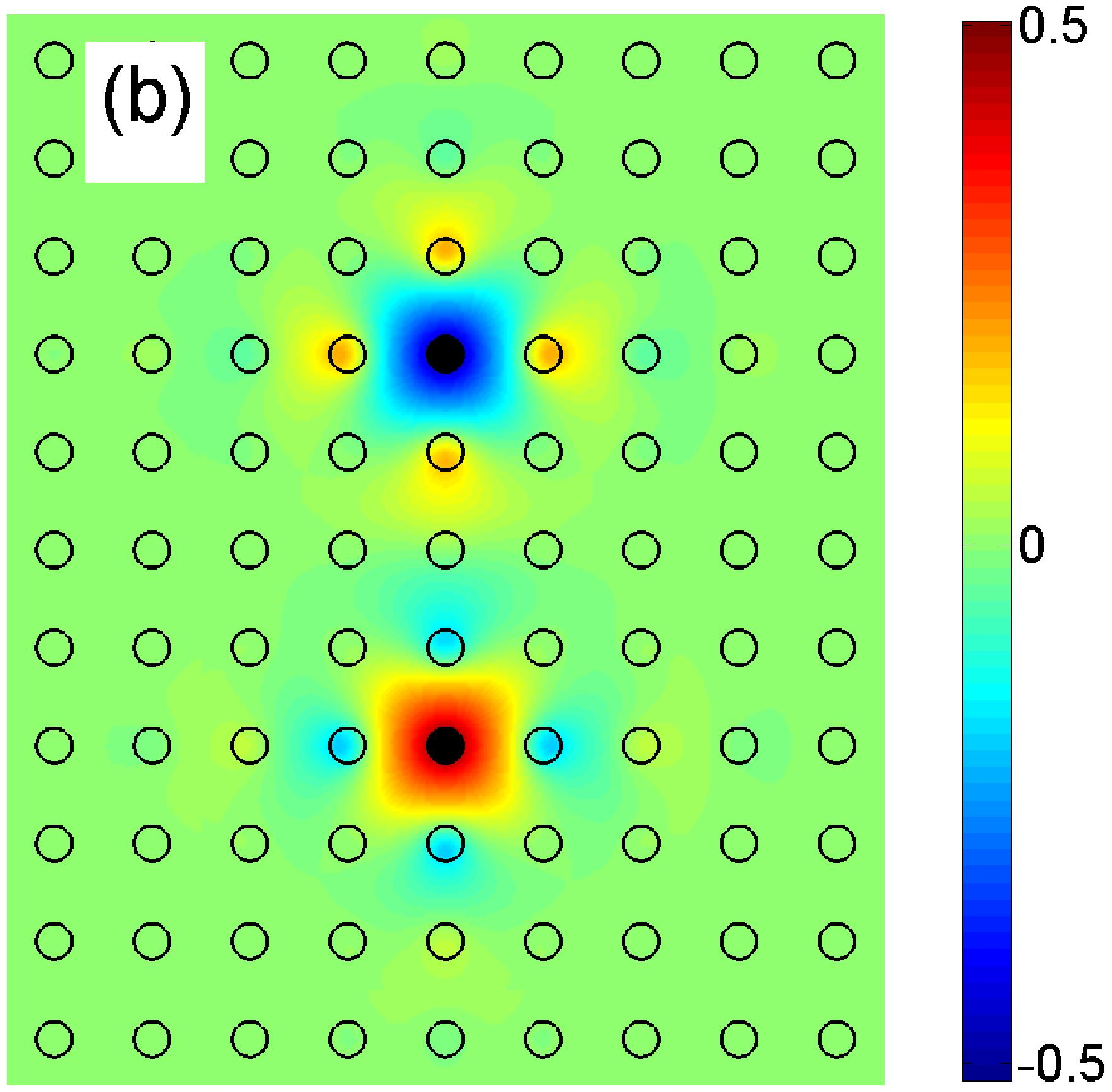}
\caption{(a) Bonding (even) mode and (b) anti-bonding (odd) mode
of two overlapped linear defects in the 2D PhC.  The defects have
the same radius as the radius of rest rods but different
dielectric constant $\epsilon_0=3$. The frequency of the isolated
defect equals $0.3593$ in terms of a value $2\pi c/a$. For the
case of two overlapped defects shown here the frequency is split
to be equal $0.3603$ (bonding) and $0.3584$ (anti-bonding).}
\label{modes}
\end{figure}
Let each defect supports a localized non degenerate monopole
solution for the TM mode only, which has the electric field
component parallel to the infinitely long rods
\cite{joanbook,busch}. Other solutions, (dipole, quadrupole, etc.)
are assumed to be extended in the photonic crystal for the
appropriate cavity radius and the dielectric constant
\cite{busch,ville} and are thereby excluded from the
consideration. Therefore, we have a two-level description for
$\hat{H}_0$ with the eigen-frequencies
\begin{equation}\label{wsa}
    \omega_{s,a}=\omega_0\pm u
\end{equation}
where $u$ is the coupling constant $u$. We denote the
corresponding even (bonding) and odd (anti-bonding) eigen-modes as
$\psi_{s,a}({\bf x})$. Both modes for specific PhC are shown in
Fig. \ref{modes}. We pay attention that the frequency of the
bonding (nodeless) mode is higher than the frequency of the
anti-bonding mode with one nodal line.

Next, we assume that the EM wave which propagates along the
waveguide obeys the symmetry of the total system. Therefore the
wave might be only symmetrical relative to $y\rightarrow -y$ or
anti-symmetrical. Respectively, the symmetric wave could excite
only the bonding mode while the anti-bonding mode would remain as
a hidden mode. Other words, the anti-bonding mode is the bound
state in symmetrical continuum \cite{photonic}. The same refers to
the case of the anti-symmetric wave and the defect bonding mode.
Therefore, for the linear case one can see the only resonance dip
at $\omega_s$, if the symmetric wave propagates along the
waveguide. However due to the nonlinearity the light transmission
acquires much more rich behavior because of spontaneous breaking
of symmetry.

Substituting two eigen-functions $\psi_{s,a}({\bf x})$ into Eq.
(\ref{MEV}) and considering a radius of the defect rods are very
thin compared to the characteristic scale of wave function we
obtain
\begin{equation}\label{V2}
  \langle m|V|n\rangle\approx
 - \frac{3}{16}\sigma\chi^{(3)}(\omega_m+\omega_n)\sum_{j=1,2}|E({\bf
  x}_j)|^2\psi_m({\bf x}_j)^{*}\psi_n({\bf x}_j),
\end{equation}
where $\sigma$ is the cross-section of the defects. Finally, we
obtain from Eq. (\ref{V2})
\begin{equation}\label{Vmn2}
\widehat{V}=\lambda\left(\begin{array}{cc}
\omega_s\phi_s^2(I_1+I_2) & \omega_0\phi_s\phi_a(I_1-I_2)\cr
\omega_0\phi_s\phi_a(I_1-I_2) & \omega_a\phi_a^2(I_1+I_2)
\end{array}\right)
\end{equation}
where $\phi_s=\psi_s({\bf x}_1)\sqrt{\sigma}=\psi_s({\bf
x}_2)\sqrt{\sigma}, \phi_a=\psi_a({\bf
x}_1)\sqrt{\sigma}=-\psi_a({\bf x}_2)\sqrt{\sigma}$, ${\bf x}_1$
and ${\bf x_2}$ are the positions of the defects in the
two-dimensional PhC, and $I_j=|E({\bf x}_j)|^2, j=1,2$ are the
intensities of the electric field at the nonlinear defects,
$\lambda=-\frac{3}{4}\chi^{(3)}$.

In order to find electric fields at the defects we must constitute
a way to excite the defect modes. Here we consider that the EM
field propagates from the left along the waveguide, interacts with
the nonlinear defects, reflects back, and transmits to the right.
Then the transmission process can be described by the CMT
stationary equations \cite{haus,manol,fan-suh} for the bonding
mode amplitude $A_s$ and the anti-bonding amplitude $A_a$
\begin{eqnarray}\label{A1A2sym}
&[\omega-\omega_s-\lambda\omega_s\phi_s^2(I_1+I_2)+i\Gamma]A_s-\lambda\omega_0\phi_s\phi_a(I_1-I_2)
A_a= i\sqrt{\Gamma}E_{in},&\nonumber\\
&-\lambda\omega_0\phi_s\phi_a(I_1-I_2) A_s+
[\omega-\omega_a-\lambda\omega_a\phi_a^2(I_1+I_2)]A_a=0,&
\end{eqnarray}
where only the bonding mode is coupled with the waveguide because
of the symmetry. The equivalent model is shown in Fig. \ref{fig1}
(b).

The amplitudes $A_s$ and $A_a$ are given by inverse of the matrix
given in the right hand of Eq. (\ref{A1A2sym})  whose matrix
elements in turn depend on the intensities $I_1, I_2$. In order to
write the equations of self-consistency for the intensities at the
defects $I_j, j=1,2$ we expand the electric field $E({\bf x})$ at
the thin j-th defect over eigen-modes $\phi_s({\bf x})$ as $E({\bf
x}_j)=\sum_{m=s,a} A_m\phi_m({\bf x}_j)$. The expansion can be
specified as follows
\begin{equation}\label{expansion}
E_1=E({\bf x}_1)=\phi_sA_s+\phi_aA_a, ~~E_2=E({\bf
x}_2)=\phi_sA_s-\phi_aA_a
\end{equation}
where symmetry properties of the eigen modes $\phi_m({\bf x})$
were taken into account. Respectively,
\begin{equation}\label{expansionI}
I_1=|\phi_sA_s+\phi_aA_a|^2, I_2=|\phi_sA_s-\phi_aA_a|^2
\end{equation}
which defines the equations of self-consistency after substitution
into Eq. (\ref{A1A2sym}). In general they are rather cumbersome.
Let us, first, consider the more simple case of the isolated
defects so that the overlapping $u$ can be neglected. Then the
values of the eigen-functions at the defects are equal
$\phi_s=\phi_a$. Even in that simplified case the solution of Eqs.
(\ref{A1A2sym}) has cardinal features different from the case of
the single nonlinear defect considered in Refs.
\cite{mcgurn,flach,miros,ming2,longhi,miros1,miros2}. These
features are the result of the mutual interference of wave flows
reflected by the nonlinear defects. If
$det(\widehat{H}_{eff}-\omega)\neq 0$ the amplitudes of the mode
excitement for the transmission can be easily found from Eq.
(\ref{A1A2sym}) as follows
\begin{eqnarray}\label{AsAa}
A_s=\frac{i\sqrt{\Gamma}E_{in}[\omega-\omega_0(1+2\lambda
I)]}{(\omega-\omega_0(1+2\lambda I))^2
-\omega_0^2\Delta^2+i\Gamma(\omega-\omega_0(1+2\lambda
I))},\nonumber\\
A_a=\frac{i\sqrt{\Gamma}E_{in}\omega_0\Delta}{(\omega-\omega_0(1+2\lambda
I))^2 -\omega_0^2\Delta^2+i\Gamma(\omega-\omega_0(1+2\lambda I))},
\end{eqnarray}
where the values $I=(I_1+I_2)/2, \Delta=\lambda(I_1-I_2)$ in turn
depend on the mode amplitudes according to (\ref{expansionI}).
Substituting these solutions into Eq. (\ref{expansionI}) we obtain
the following nonlinear equations of self-consistency
\begin{eqnarray}\label{I1I2}
I_1=\frac{\Gamma E_{in}^2[\omega-\omega_0(1+2\lambda
I_2)]^2}{[\omega-\omega_0(1+2\lambda
I_1)]^2[\omega-\omega_0(1+2\lambda
I_2)]^2+\Gamma^2[\omega-\omega_0(1+2\lambda I)]^2} ,\nonumber\\
I_2=\frac{\Gamma E_{in}^2[\omega-\omega_0(1+2\lambda
I_1)]^2}{[\omega-\omega_0(1+2\lambda
I_1)]^2[\omega-\omega_0(1+2\lambda
I_2)]^2+\Gamma^2[\omega-\omega_0(1+2\lambda I)]^2}.
\end{eqnarray}
The solution of these equations gives the steady state for the
transmission in the waveguide coupled with two nonlinear defects.
Finally, we write from Eq. (\ref{t}) equation for the transmission
amplitude:
\begin{equation}\label{ts}
t=E_{in}-\sqrt{\Gamma}A_s.
\end{equation}
The odd amplitude $A_a$ does not contribute into the transmission
amplitude because of the symmetry.

In the forthcoming CMT calculations we fix the parameters of the
CMT model as follows: $ \omega_0=1, \Gamma=0.01, \lambda=-0.01$.
We consider the case of isolated defects $u=0, \phi_s=\phi_a=1$
and the case of coupled defects with $u=0.01, \phi_s=1,
\phi_a=1.1$. Rigorously speaking these values $u$ and $\phi_s,
\phi_a$ correlate with each other. However, in our model case, we
disregard this correlation.
\subsection{Symmetry preserving solution}
\begin{figure}
\includegraphics[scale=.35]{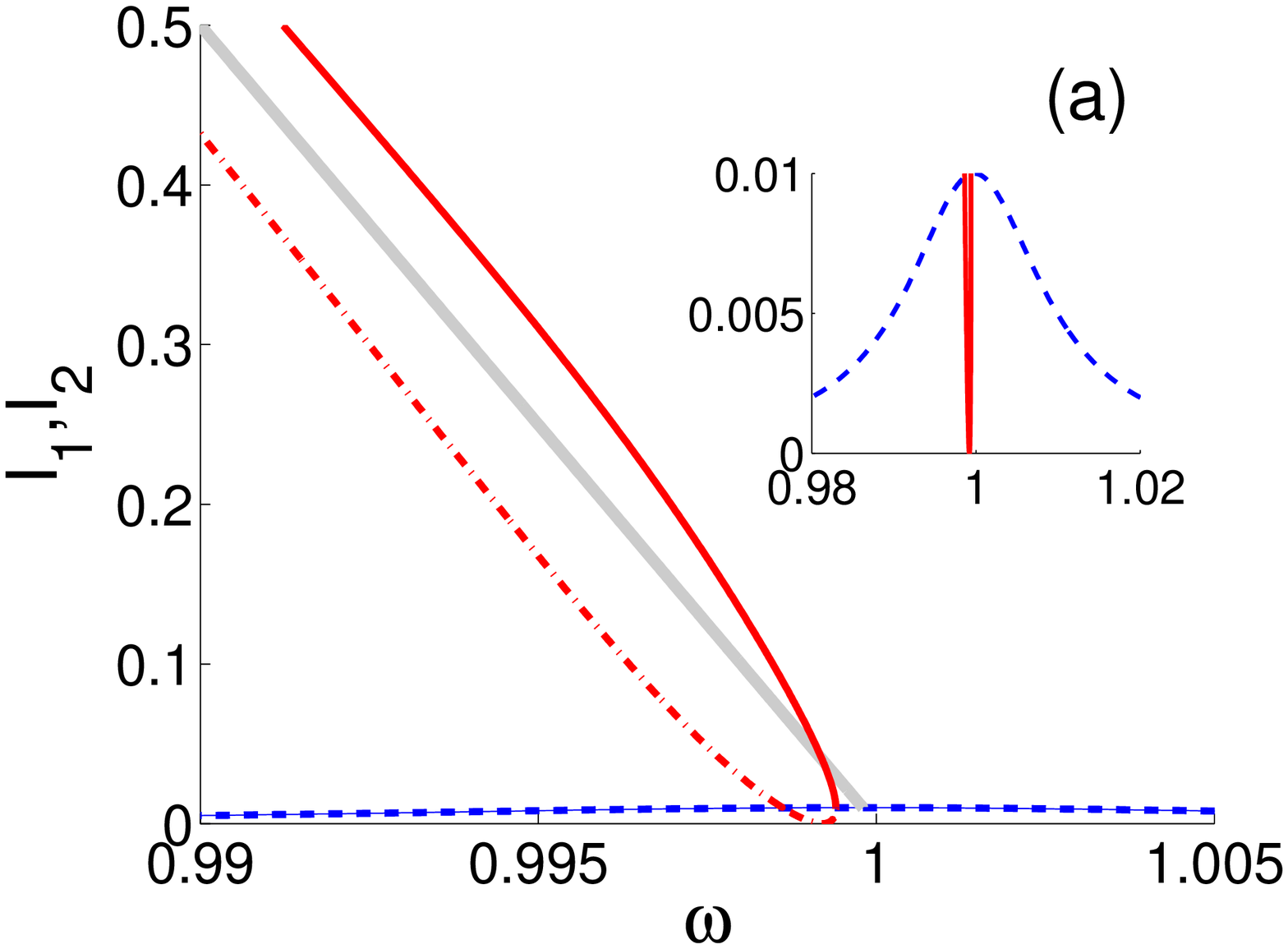}
\includegraphics[scale=.35]{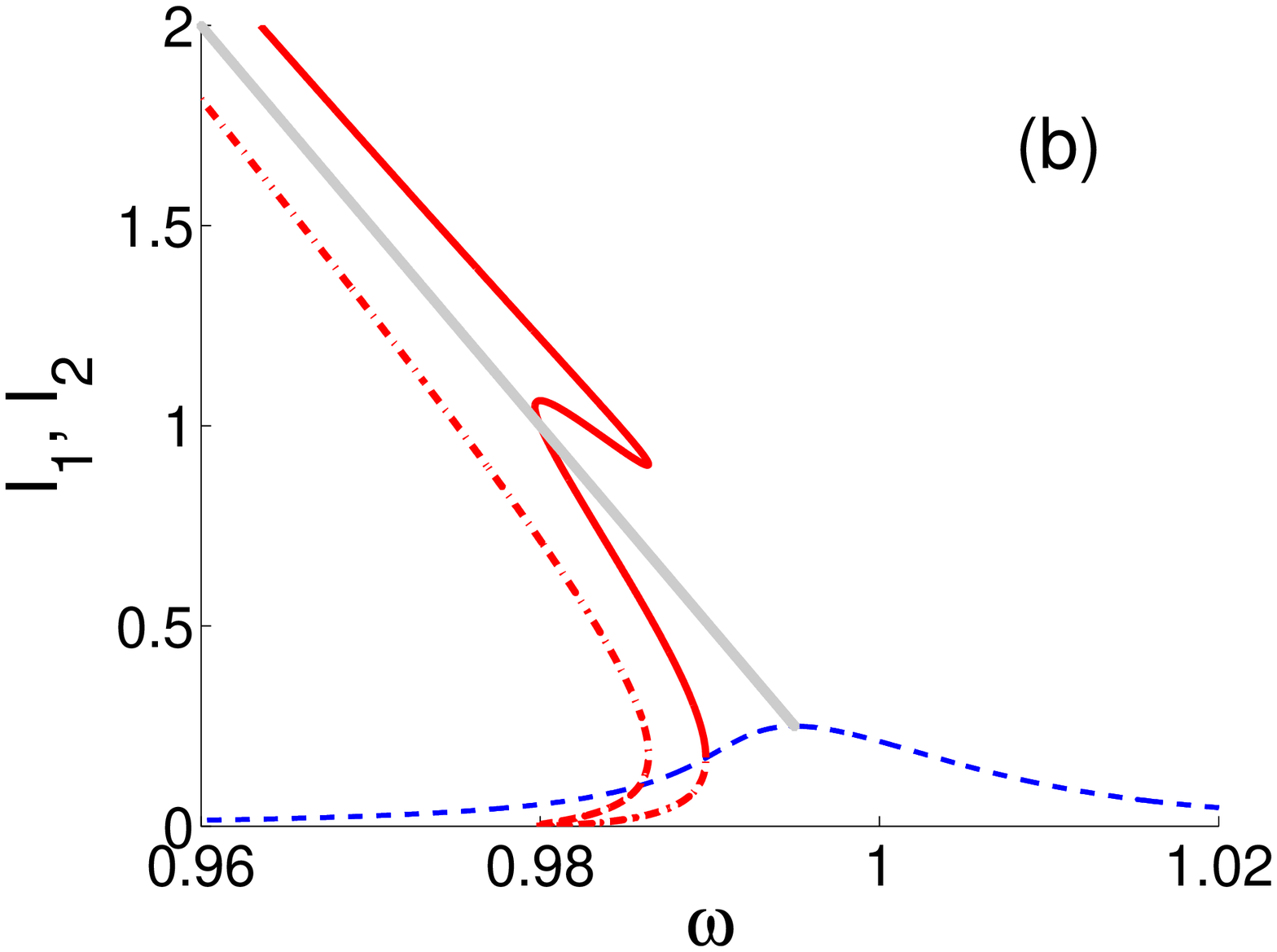}
\caption{Frequency behavior of the intensities at the isolated
defects $u=0$. (a) $E_{in}=0.01$, (b) $E_{in}=0.05$. Here and in
the forthcoming figures dashed blue line shows the symmetry
preserving solution. Solid and dash-dotted red lines show the
symmetry breaking solutions which has different intensities at the
defects $I_1$ and $I_2$. Gray thick solid line shows a new phase
parity breaking solution at which
$det(\omega-\widehat{H}_{eff})=0$.} \label{intu0}
\end{figure}

We start with the solution $E_1=E_2$ that preserves the symmetry.
In this case the incident wave excites only the symmetric even
mode $A_s$
\begin{equation}\label{symbr}
 A_s=\frac{i\sqrt{\Gamma}E_{in}}{\omega-\omega_0(1+2\lambda I)+i\Gamma}
\end{equation}
as follows from Eq. (\ref{AsAa}) with the only resonance frequency
$\omega_0(1+2\lambda I)$ and the width $2\Gamma$. The
self-consistency equation for the symmetry preserving solution
$I=I_1=I_2$ simplifies
\begin{equation}\label{triv}
I[(\omega-\omega_0(1+2\lambda I))^2+\Gamma^2]=\Gamma E_{in}^2.
\end{equation}
That coincides with the equation of self-consistency for the
single off-channel nonlinear defect obtained in Ref. \cite{flach}.
The solution of this cubic nonlinear equation is shown in Fig.
\ref{intu0} by dashed blue lines. The frequency behavior of the
intensities inherits the linear case, as shown in the inset.  With
growth of the input power the resonance frequency shifts to the
left because of the nonlinear contribution $2\lambda I$ as seen
from Eq. (\ref{symbr}).
\begin{figure}
\includegraphics[scale=0.35]{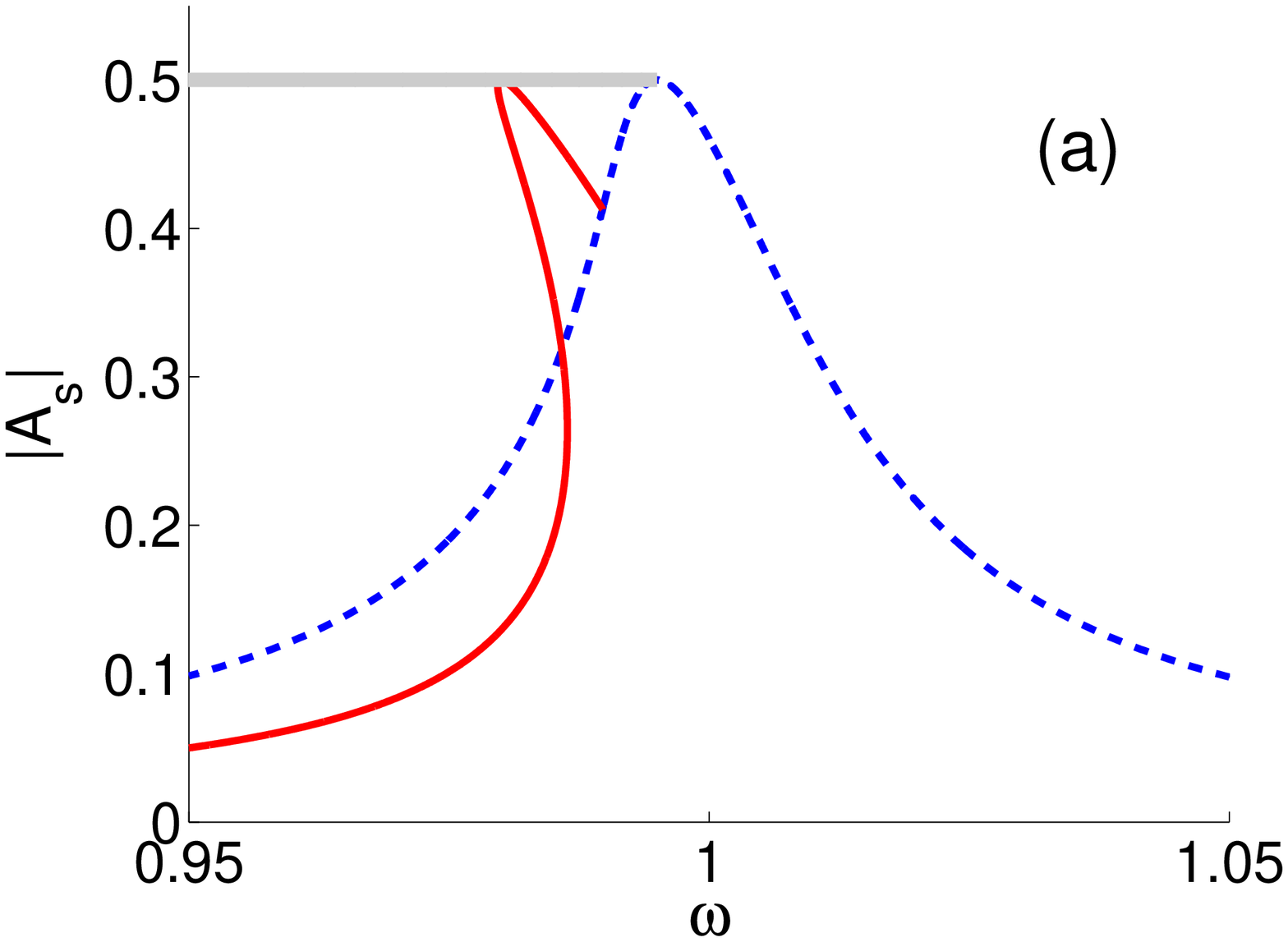}
\includegraphics[scale=.35]{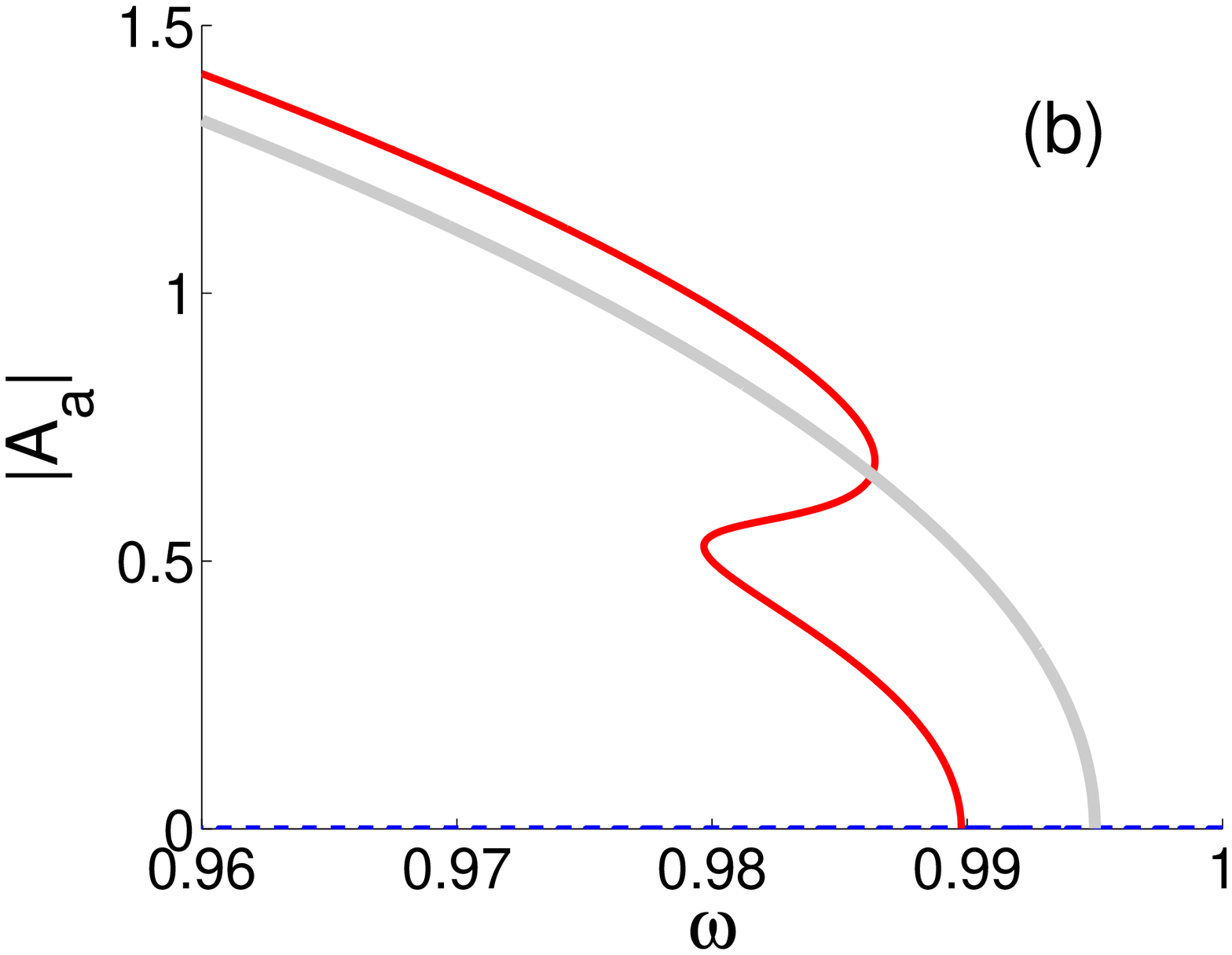}
\caption{Frequency behavior of (a) the bonding amplitude $|A_s| $
and (b) the anti-bonding amplitude $|A_a|$ of the model shown in
Fig. \ref{fig1} (b) for $u=0, E_{in}=0.05$.} \label{AAu0}
\end{figure}
\begin{figure}
\includegraphics[scale=0.35]{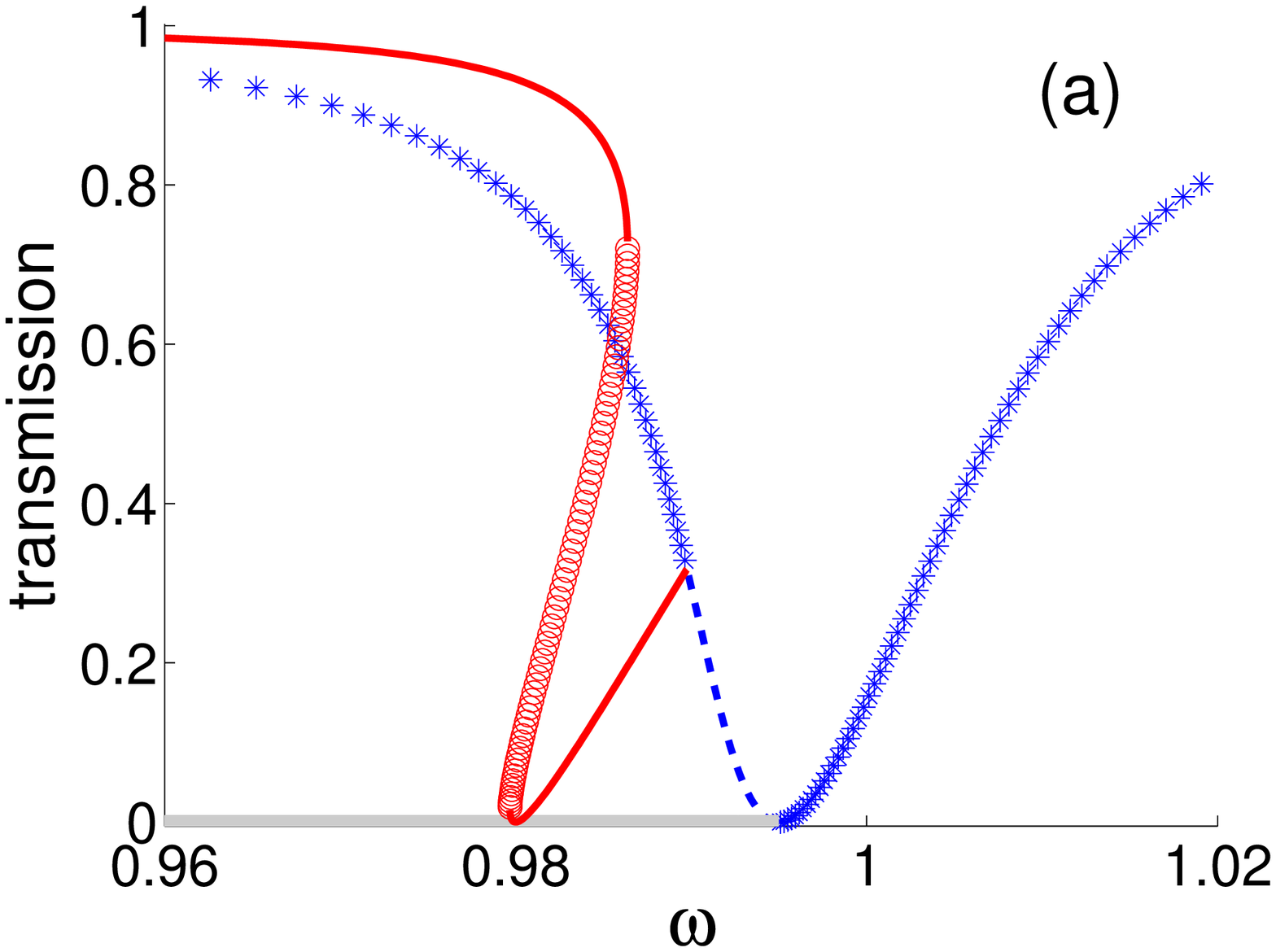}
\includegraphics[scale=0.35]{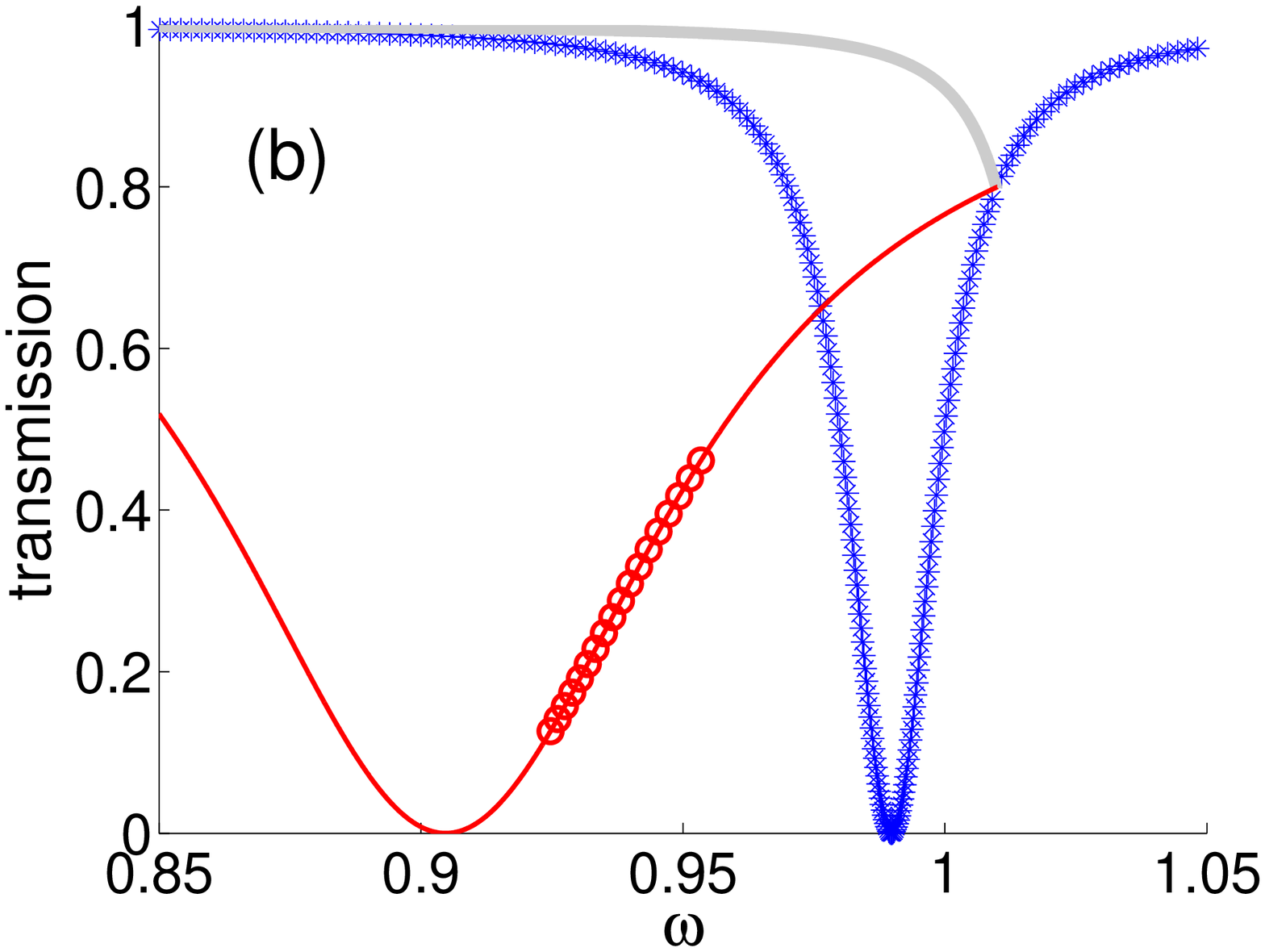}
\caption{Frequency behavior of the transmission for the isolated
defects for (a) $E_{in}=0.05$ and (b) for the coupled defects for
$E_{in}=0.01$. Stars and open circles show stable domains of the
solutions.} \label{tr}
\end{figure}
\begin{figure}
\includegraphics[scale=.35]{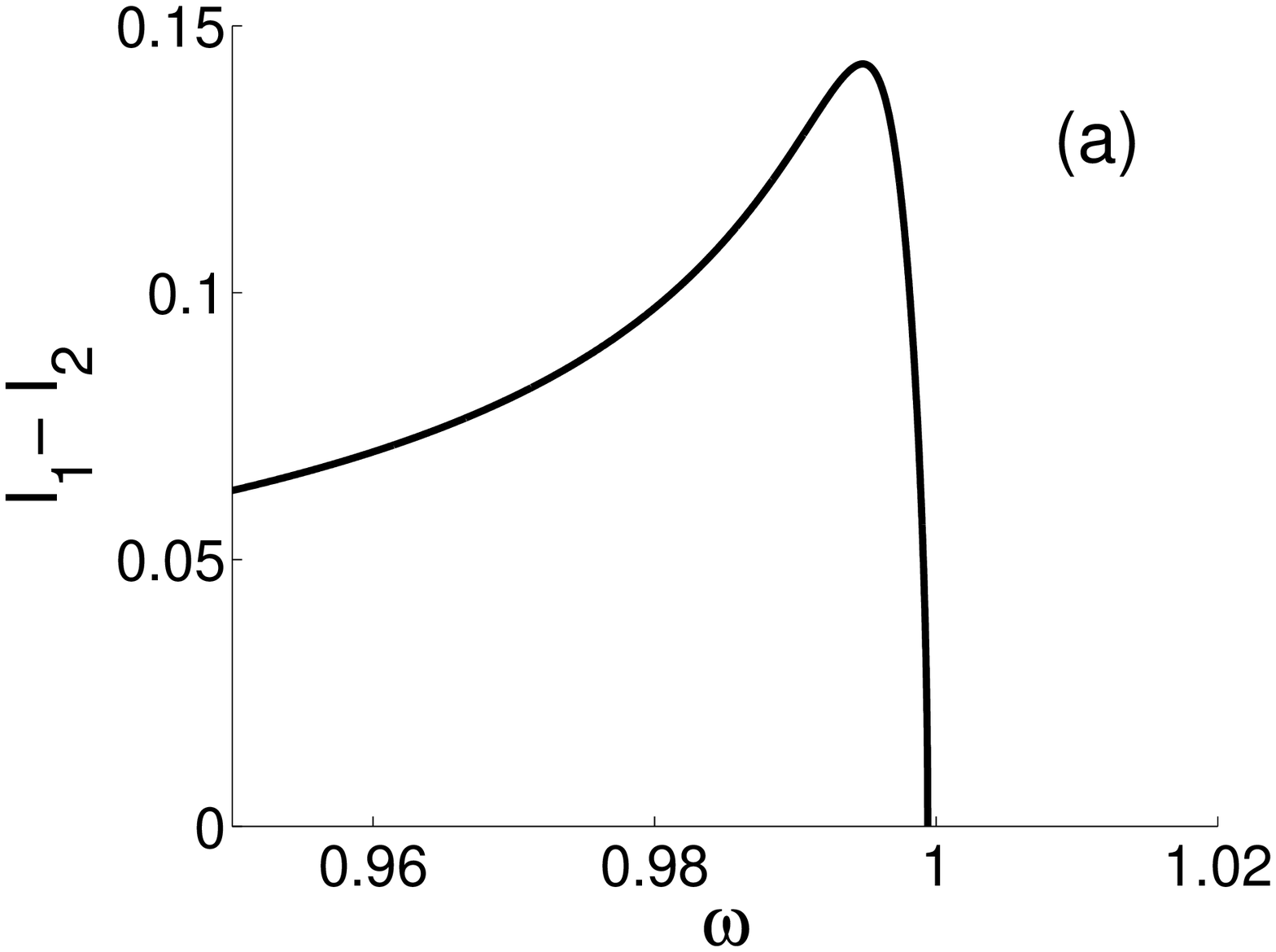}
\includegraphics[scale=.35]{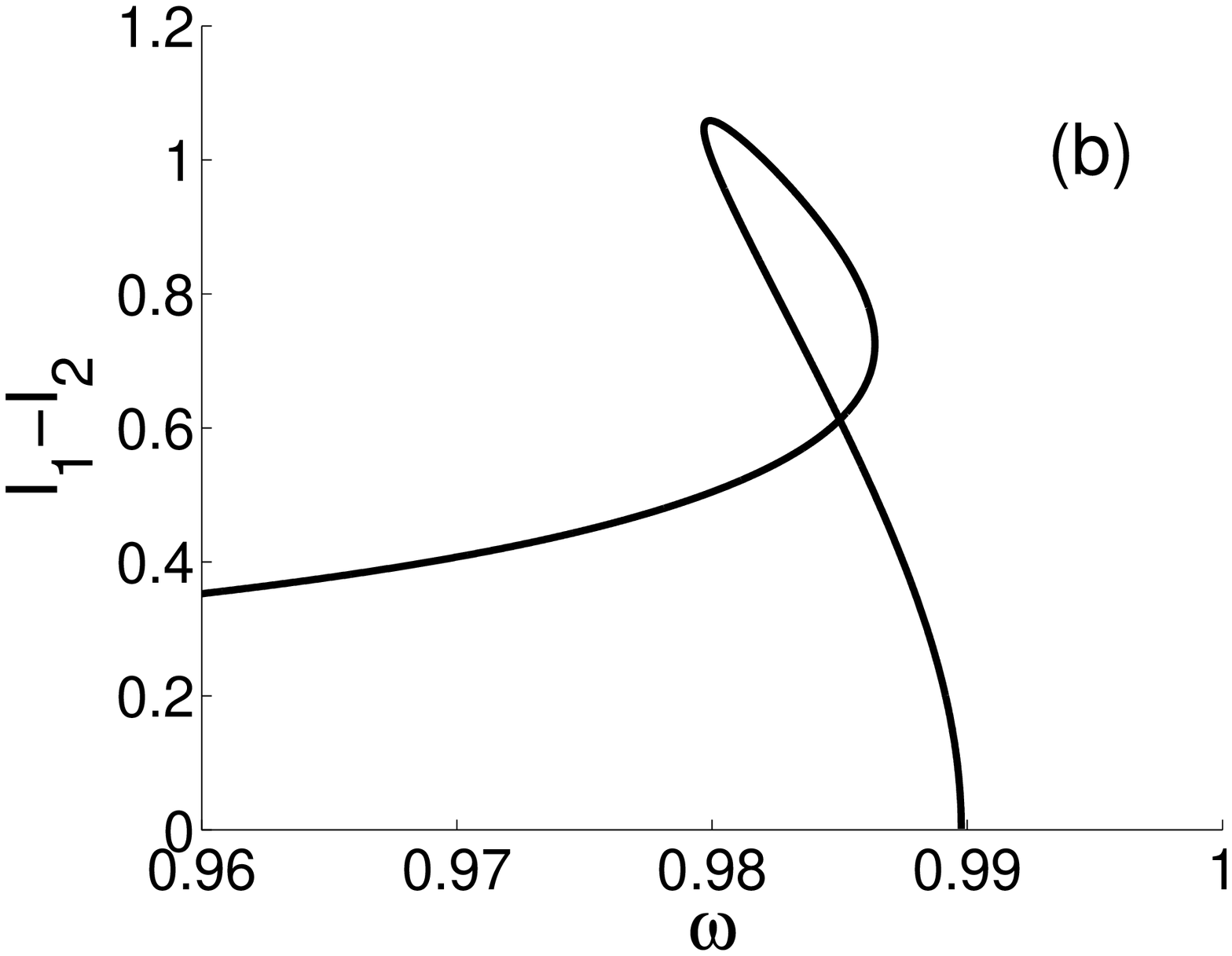}
\caption{The difference between the intensities at the defects for
$u=0$: (a) $E_{in}=0.01$ and (b) $E_{in}=0.05$. Only the symmetry
breaking solution is shown.} \label{Deltau0}
\end{figure}

 The frequency behavior of mode excitations $|A_s|, ~|A_a|$ is
 shown in Fig. \ref{AAu0} by blue dashed
lines. As seen from Fig. \ref{AAu0} (a) $A_s$ has a resonance
peak. Respectively, the transmission $T=|t|^2/E_{in}^2$ has a
resonance dip at the frequency $\omega_0(1+2\lambda
I)=\omega_0(1+2\lambda E_{in}^2/\Gamma)$ as shown in Fig.
\ref{tr}(a) by the dashed line. The last equality follows from Eq.
(\ref{triv}).
\begin{figure}
\includegraphics[scale=0.35]{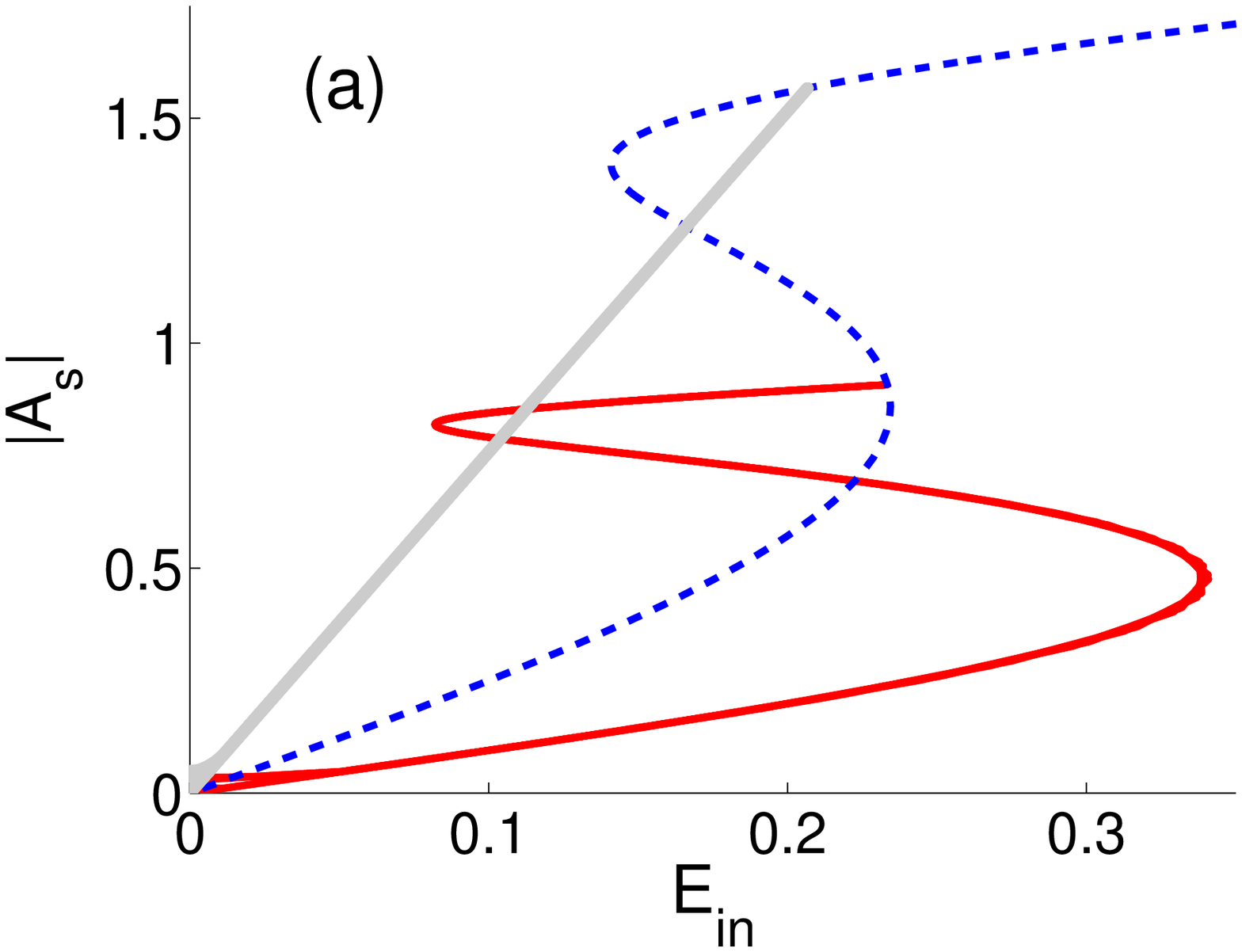}
\includegraphics[scale=0.35]{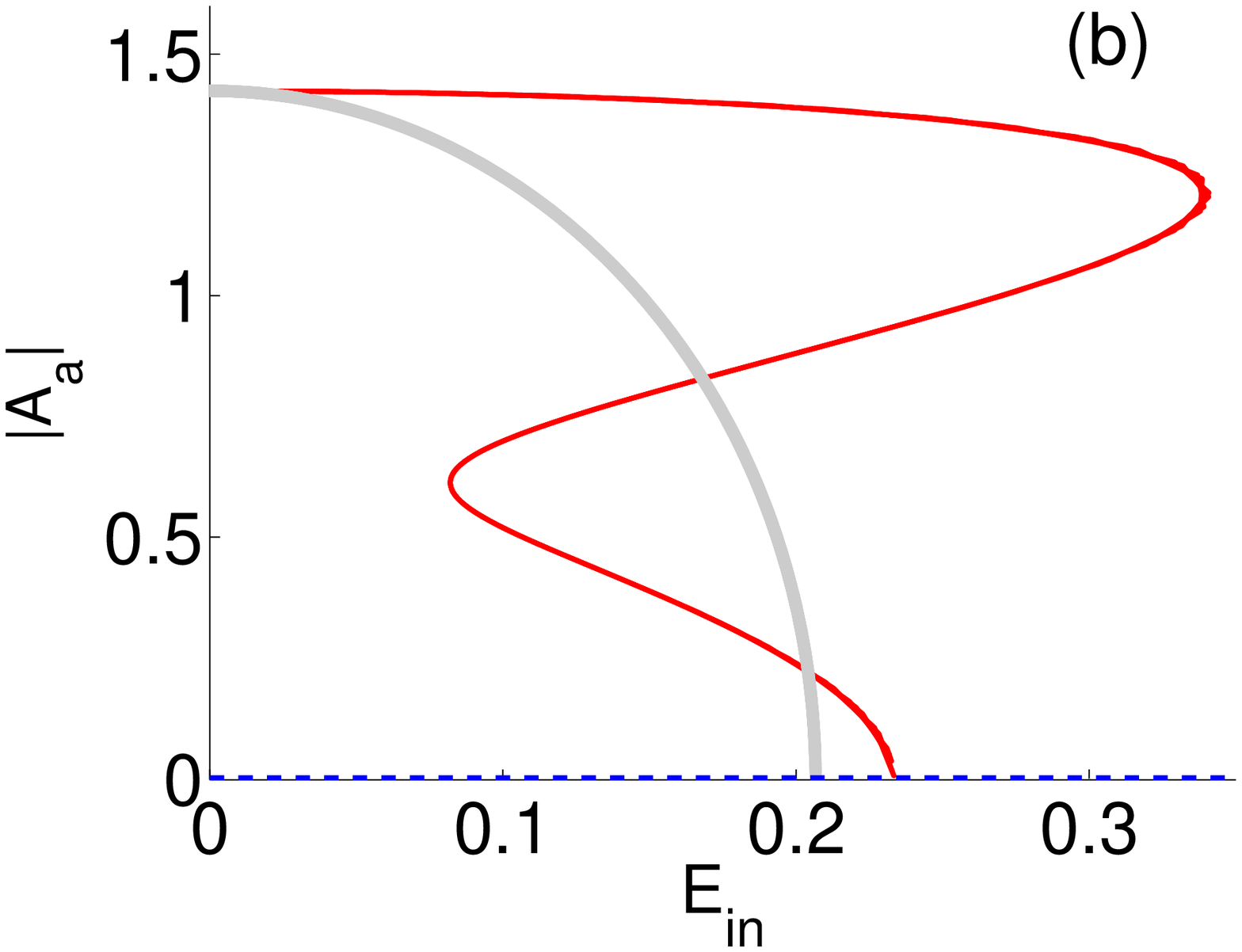}
\caption{Amplitudes (a) $|A_s|$ and (b) $|A_a|$ as a function of
the incident wave amplitude $E_{in}$ for the coupled defects with
the parameters $\omega=0.95, u=0.01, \phi_s=1, \phi_a=1.1$.}
\label{AAEin}
\end{figure}
\subsection{Symmetry breaking solution}
For the transmission through the nonlinear symmetric media the
symmetry might be broken
\cite{Haelterman,Peschel,Babushkin,Torres,Longchambon,maes1,maes2,Otsuka,Huybrechts}.
Numerical solution of Eq. (\ref{A1A2sym}), indeed, reveals the
solution with $I_1>I_2$, i.e., the nonlinearity gives rise to a
breaking of the symmetry below (above) the critical frequency
$\omega_c$ for $\lambda<0$ ($\lambda>0)$. The symmetry breaking
solution is shown in Fig. \ref{intu0} by solid lines for $I_1$ and
dash-dotted lines for $I_2$. There is also the solution that
differs from the former in that $E_1\leftrightarrow E_2$. If the
solutions are stable, a choice of the solution happens
incidentally, as it does for a phase transition  of the second
order in cooperative systems. As shown in Fig. \ref{Deltau0}, a
value $I_1-I_2$ or the odd mode amplitude $A_a$, indeed, might
serve as the order parameter that characterizes the symmetry
breaking.

It is surprising that there is the frequency at which the
intensity at one of the nonlinear defects turns to zero as shown
in Fig. \ref{intu0}. According to Eqs. (\ref{I1I2}) that occurs at
the frequency
\begin{equation}\label{zerot}
  \omega_{dip}=\omega_0(1+8\lambda E_{in}^2/\Gamma).
\end{equation}
At this frequency we have $A_s=A_a=E_1/2=E_{in}/\sqrt{\Gamma}$ in
accordance with Eqs. (\ref{AsAa}) and (\ref{expansion}). By
substituting this equality into Eq. (\ref{t}) we immediately
obtain that the frequency (\ref{zerot}) defines the position of
resonance dip for the symmetry breaking solution. As will be
shown, that result of full extinction of one of the nonlinear
defects is observed in the PhC system as well [Fig.
\ref{wavebroken}(b)].

In Figs. \ref{AAu0}(a) and \ref{AAu0} (b) we show the frequency
dependence of the even and odd mode amplitudes $|A_s|$ and $|A_a|$
respectively for $E_{in}=0.05$. One can see that, first, the
incident wave begins to excite the odd mode below $\omega_c$ for
$\lambda<0$, and, second, $|A_s|$ and $|A_a|$ show the
bistability. The even mode $A_s$ displays a resonance peak (solid
line) with the resonance width twice less than the resonance width
of the peak for the symmetry preserving solution (dashed line).
Correspondingly, the transmission in Fig. \ref{tr} demonstrates a
narrow dip for the symmetry breaking solution. In order to
understand that phenomenon let us consider the resonance poles of
the even and odd amplitudes given by zeros of the denominators in
Eq. (\ref{AsAa})
\begin{equation}\label{poles}
  z_{1,2}=\omega_0(1+2\lambda I)-\frac{i\Gamma}{2}\pm\sqrt{\omega_0^2\Delta^2-\frac{\Gamma^2}{4}}.
\end{equation}
For the solution with $\Delta=0$ we had the only resonance pole
with the resonance half width $\Gamma$. As Fig. \ref{Deltau0}
shows there is the frequency domain roughly between 0.98 and 0.99
where $\omega_0\Delta>\Gamma/2$ and where the resonance half-width
is twice less than $\Gamma$ according to formula (\ref{poles}).
Therefore, in this frequency domain we can expect the resonance
dip to be twice narrower compared to the symmetry preserving
solution with $\Delta=0$.

The lesser the width of resonance, the more unstable the resonance
\cite{joanbook}. One can thereby see that the bistability of the
symmetry breaking solution is more profound in comparison to the
symmetry preserving solution. The resonance peak in $|A_s|$ for
the symmetry breaking solution terminates at that frequency where
the odd mode amplitude $|A_a|$ arises as seen from Fig.
\ref{AAu0}(b). Close to this frequency the amplitude $A_a$ has a
square root behavior typical for the order parameter in phase
transition of the second order. The dependence of $A_a$ on the
amplitude of the incident wave demonstrates the same behavior [see
below Fig. \ref{AAEin} (b)].
\subsection{Phase parity breaking solution}
At last, there is the solution that has equal intensities at the
defects but nevertheless a symmetry is broken because of phases of
the complex amplitudes $E_1$ and $E_2$. This solution refers to
the special case of Eq. (\ref{A1A2sym}) when the determinant of
the matrix $\omega-\widehat{H}_{eff}$ equals zero, (i.e., the
inverse of matrix does not exist). It occurs at
\begin{equation}\label{det0}
 I_1=I_2=I,~~ \omega=\omega_a(1+2\lambda \phi_a^2 I).
\end{equation}
Then the solution of Eq. (\ref{A1A2sym}) for the even mode
amplitude $A_s$ is
\begin{equation}\label{Asu0}
  A_s=\frac{E_{in}}{\sqrt{\Gamma}},
\end{equation}
while $A_a$ is undetermined yet.

Let us take for a while, the defects to be linear. Then the second
equation in (\ref{det0}) shrinks to the isolated point
$\omega=\omega_a$. As given by the CMT equations (\ref{A1A2sym})
and as seen from Fig. \ref{fig1} this odd mode has zero
overlapping with the waveguide and Eq. (\ref{det0}) thereby
defines the bound state in continuum (BSC)
\cite{photonic,neumann,ostrovsky,friedrich,ring,sadreev_review}.
The solution of the Eq. (\ref{A1A2sym}) $\left(\begin{array}{c}
A_s\cr A_a
\end{array}\right)$ with $A_s$ given by Eq. (\ref{Asu0}) and
arbitrary $A_a$ is therefore a superposition of the transport
solution and the BSC.

For the nonlinear defects the situation changes dramatically.
First, there is the whole frequency region $\omega\geq \omega_a$
for $\lambda>0$ or $\omega\leq \omega_a$ for $\lambda<0$ where
$det(\omega-\widehat{H}_{eff})=0$ as seen from Eq. (\ref{det0}).
Equation (\ref{det0}) thereby defines the BSC with eigen frequency
in whole region as dependent on the BSC intensity. Second, the BSC
can not be independently superposed to the transport solution for
the nonlinear case. The BSC begins to couple with the incident
wave and can not be defined as the bound state if $E_{in}\neq 0$.

\begin{figure}
\includegraphics[scale=.4]{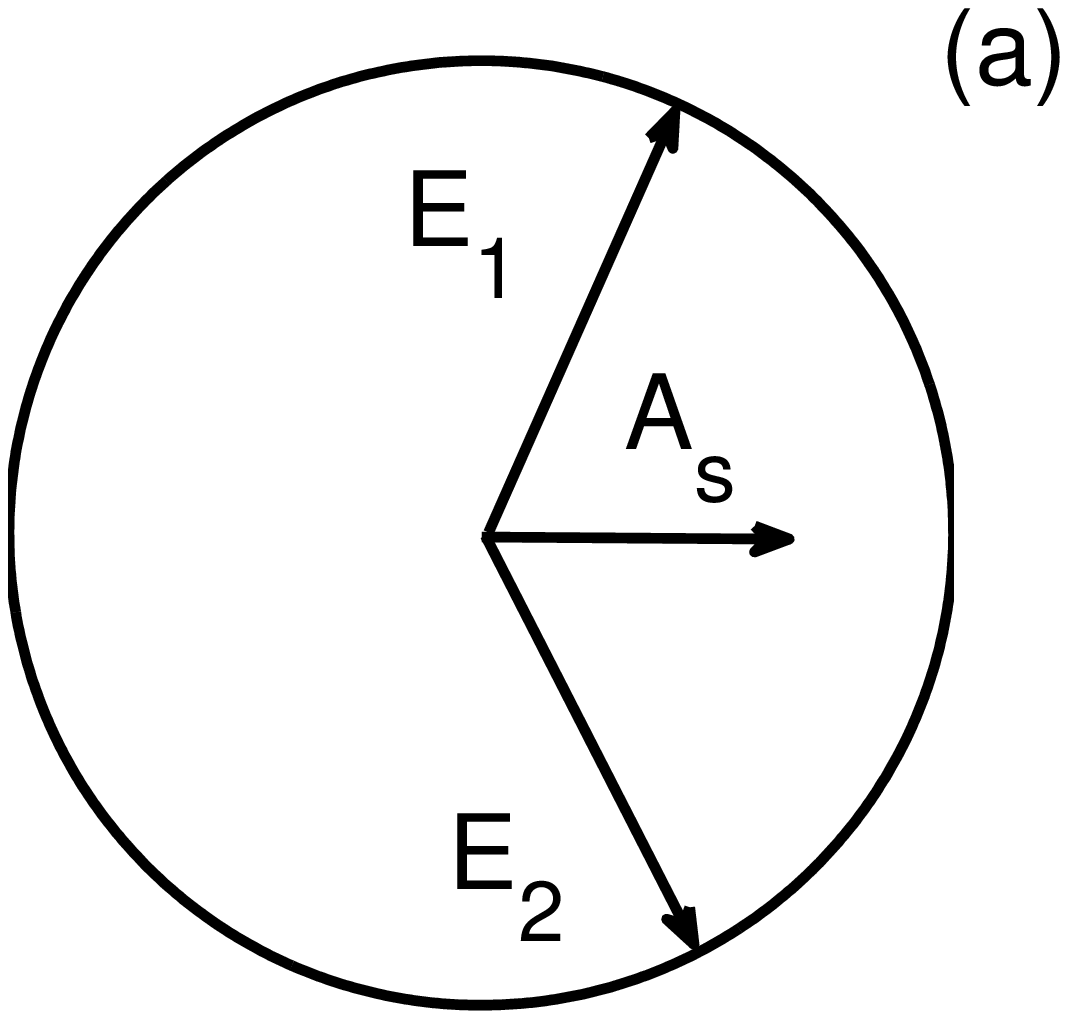}
\includegraphics[scale=.4]{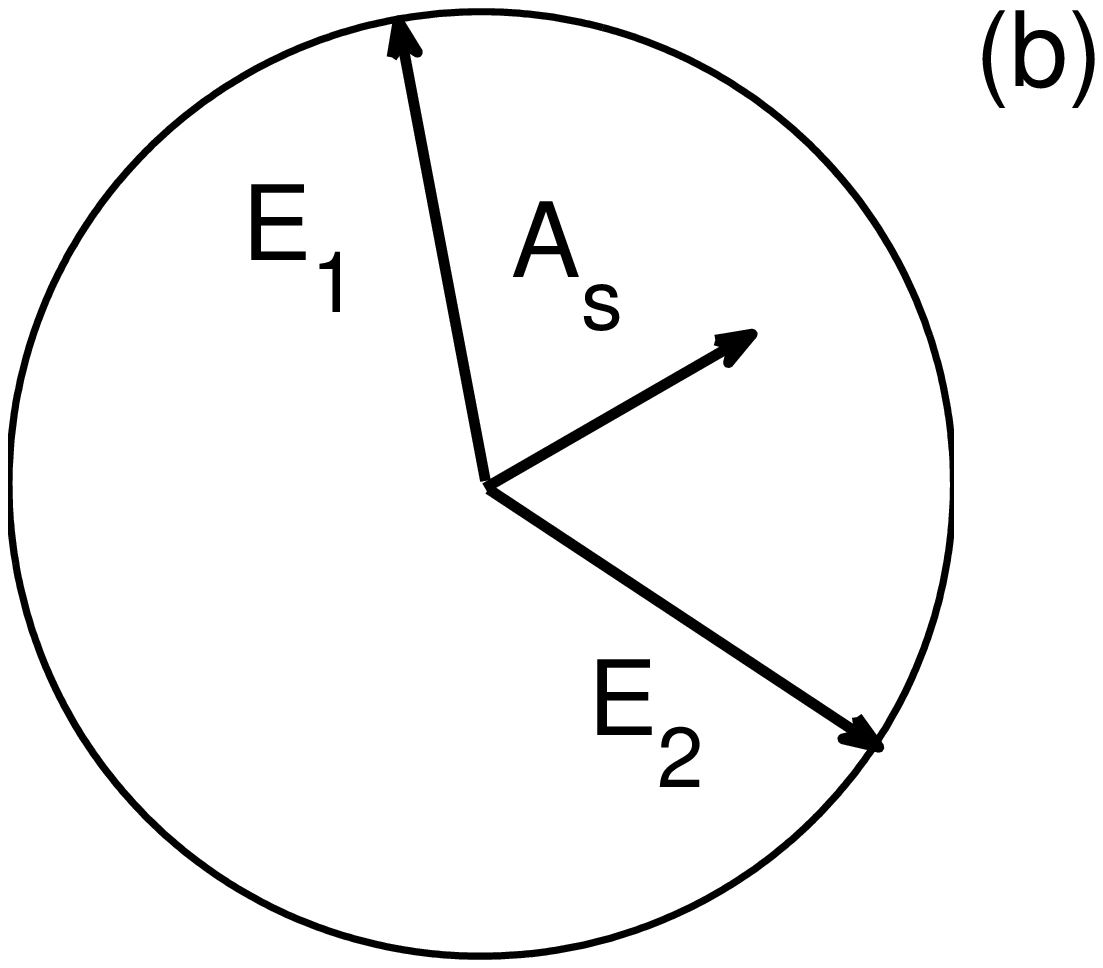}
\caption{Graphic solutions of (a) Eqs. (\ref{As}) (a) and
(\ref{Asuu}) (b), respectively. Radius of circle is $\sqrt{I}$.}
\label{triangle}
\end{figure}

\begin{figure}
\includegraphics[scale=.35]{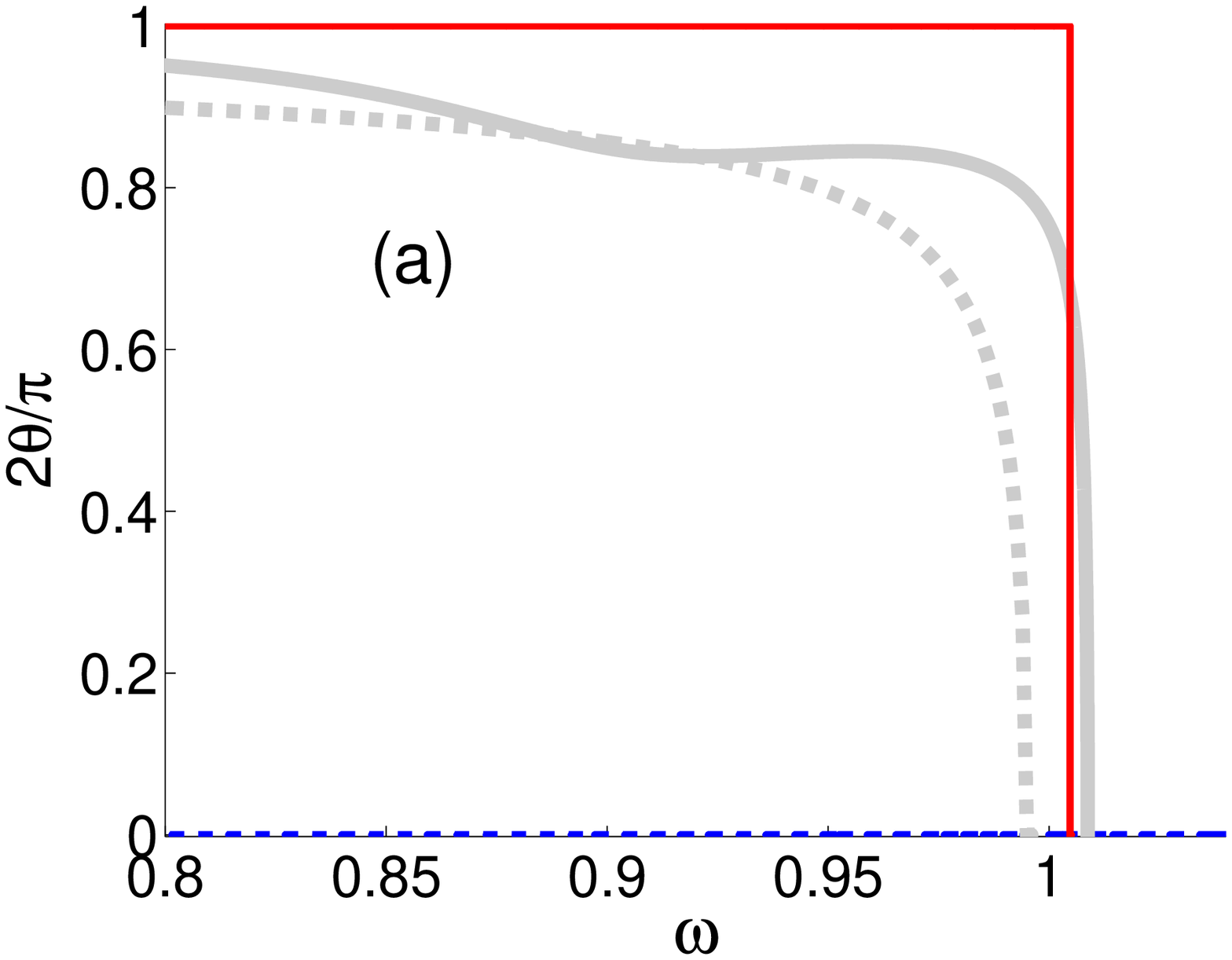}
\includegraphics[scale=.35]{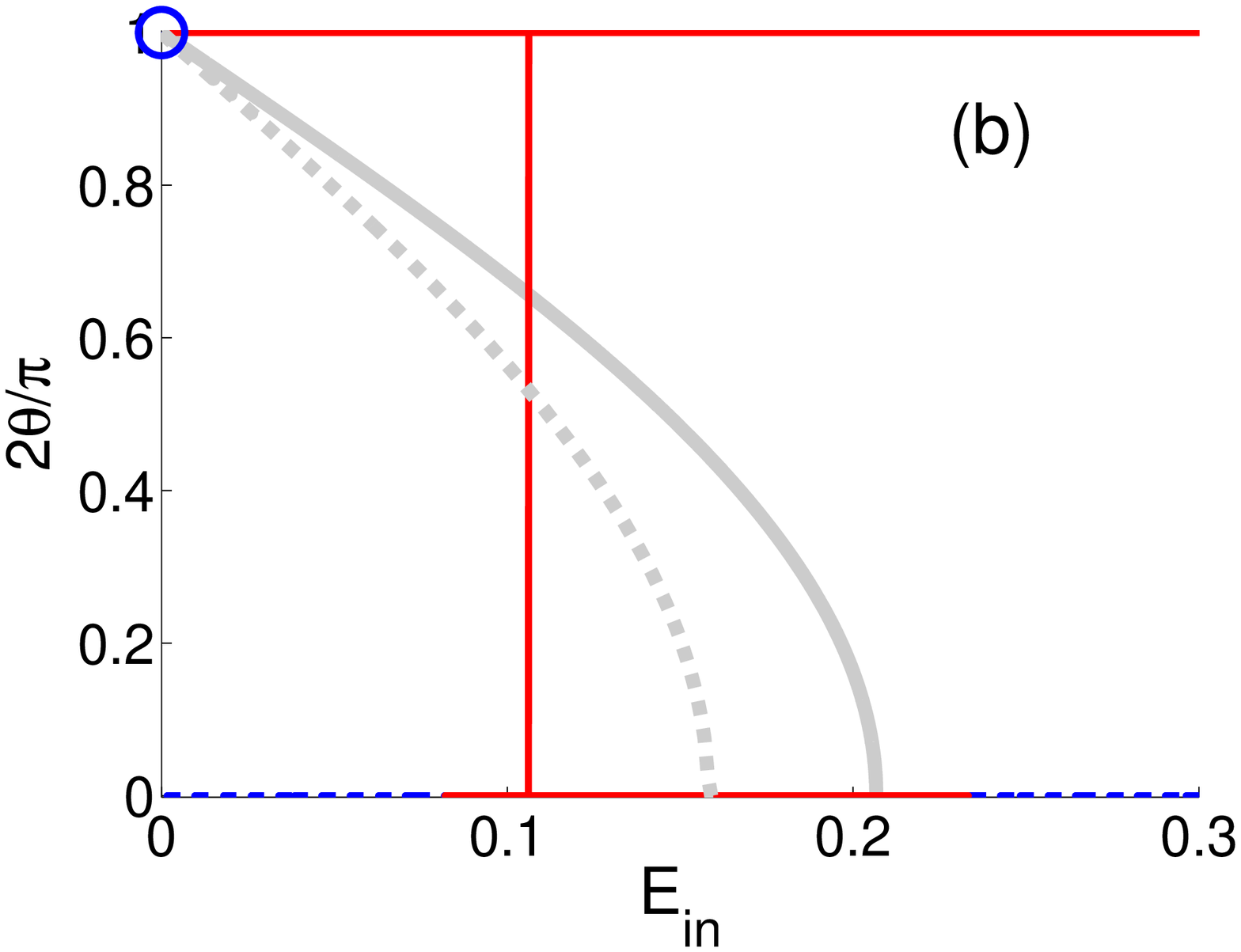}
\caption{Difference between phases of the amplitudes $E_1$ and
$E_2$ for $u=0.01$ as a function of (a) the frequency for
$E_{in}=0.05$ and (b) the amplitude of the incident wave for
$\omega=0.95$. Dashed blue line shows the symmetry preserving
solution, solid red line shows the symmetry breaking solution, and
gray lines show the phase parity breaking solution, $u=0$ dashed
and $u=0.01$ solid. The BSC point is shown by open bold circle.}
\label{phase}
\end{figure}

(i) Let the defects be isolated; (i.e. $u=0, \phi_s=\phi_a=1$. On
the one hand, we obtain from Eq. (\ref{Asu0})
\begin{equation}\label{As}
 A_s=\frac{E_1+E_2}{2}=E_{in}/\sqrt{\Gamma},
\end{equation}
according to Eq. (\ref{expansion}). That is the bonding mode
amplitude is constant over the frequency as shown in Fig.
\ref{AAu0} (a) by the gray thick solid line. On the other hand,
Eq. (\ref{det0}) directly shows that the intensities at the
defects do not depend on $E_{in}$,
\begin{equation}\label{Iph}
  I=\frac{\omega-\omega_0}{2\lambda}.
\end{equation}
Since $|E_1|=|E_2|=\sqrt{I}$ the only way to satisfy Eqs.
(\ref{As}) and (\ref{Iph}) is to consider that the amplitudes at
the defects are $E_1=\sqrt{I}\exp(i\theta),
E_2=\sqrt{I}\exp(-i\theta)$. That is illustrated in Fig.
\ref{triangle}(a).
With the use of Eqs. (\ref{As}) and (\ref{Iph}), we obtain
\begin{equation}\label{theta}
  \cos^2 \theta=\frac{2\lambda E_{in}^2}{\Gamma(\omega-\omega_0)}.
\end{equation}
For $E_{in}\rightarrow 0$ we have the following limits:
$\theta\rightarrow \pi/2, E_1\rightarrow i\sqrt{I}, E_2\rightarrow
-i\sqrt{I}, E_1+E_2\rightarrow 0$ as seen from Eq. (\ref{theta}).
As soon as $E_{in}\neq 0$ the defects amplitudes are seized to
oscillate in fully anti-symmetric way as shown in Fig.
\ref{triangle}(a). We emphasize that phase difference $2\theta$
has nontrivial behavior if the defects are nonlinear ($\lambda\neq
0$) and the incident wave is applied ($E_{in}\neq 0$) as follows
from Eq. (\ref{theta}). For the symmetry preserving solution
$\theta=0$ (dashed line in Fig. \ref{phase}), for the symmetry
breaking solution $2\theta=0$ or $\pi$ (solid line in Fig.
\ref{phase}) while for the present solution the phase difference
$2\theta$ behaves as an order parameter (gray thick dashed line in
Fig. \ref{phase}) similar to $A_a$ shown in Figs. \ref{AAu0}(b) or
\ref{AAEin}(b).

We define the present solution of the CMT equations
(\ref{A1A2sym}) with the zero determinant $det(\omega-H_{eff})=0$
as the phase parity breaking solution . It exists for $\omega\leq
\omega_0+2\lambda E_{in}^2/\Gamma$ for $\lambda<0$. Knowledge of
the phase $\theta$ allows us now to find the anti-bonding
amplitude
\begin{equation}\label{Aa}
  A_a=(E_1-E_2)/2=i\sqrt{I}\sin\theta.
\end{equation}
The frequency behavior of the even and odd amplitudes $|A_s|,
|A_a|$ are shown in Figs. \ref{AAu0}(b) and \ref{AAEin}(b).
Finally, by substituting Eq. (\ref{As}) into Eq. (\ref{ts}) we
obtain $t=0$ for the phase parity breaking solution as shown in
Fig. \ref{tr}(a) by gray thick dashed line.

(ii) Coupled defects. For the PhC structure shown in Fig.
\ref{fig1}(a) the coupling between the defects $u$ is rather small
compared to the coupling between the waveguide and defects
$\sqrt{\Gamma}$. Nevertheless, an account of the coupling between
the defects has a principal importance as will be seen below. As
was given earlier, the parameters of the coupled defects are
specified as follows $u=0.01, ~ \phi_s=1, ~\phi_a=1.1$.

A substitution of Eq. (\ref{det0}) into Eq. (\ref{Asu0}) gives
\begin{equation}\label{Asuu}
A_s=\frac{\omega_0+u}{\omega_0(1-\alpha)+u(1+\alpha)}\cdot
\frac{\sqrt{\Gamma}E_{in}}{\omega-\omega_r+i\Gamma_r}=\frac{A_{s0}}{\omega-\omega_r+i\Gamma_r}
\end{equation}
where
\begin{eqnarray}\label{omr}
  \omega_r=\frac{(1-\alpha)\omega_s\omega_a}{\omega_0(1-\alpha)+u(1+\alpha)},\\
  \Gamma_r=\Gamma\frac{\omega_a}{\omega_0(1-\alpha)+u(1+\alpha)},\\
A_{s0}=\frac{\sqrt{\Gamma}E_{in}\omega_a}{\omega_0(1-\alpha)+u(1+\alpha)}
  \end{eqnarray}
$\alpha=\phi_s^2/\phi_a^2$. Therefore for the coupled defects the
amplitude $A_s$ acquires typical Bright-Wigner resonance behavior
in which the nonlinearity is excluded. Respectively, a
substitution of the solution (\ref{Asuu}) into Eq. (\ref{ts})
immediately results in the transmission having the resonance dip
at the frequency $\omega_r$ with the half width $\Gamma_r$ which
depends on ratio $\alpha$ and $u$. That result is shown in Fig.
\ref{tr}(b) by the gray thick line. If $u\rightarrow 0,
\phi_a\rightarrow\phi_s, \alpha\rightarrow 1$ the frequency of the
resonance dip goes away, and $\Gamma_r\rightarrow\infty$; that is,
the resonance at the phase parity breaking solution disappears,
and the corresponding transmission tends to zero as seen from Fig.
\ref{tr}(a).

Equation (\ref{det0}) fixes intensity at the defects
\begin{equation}\label{Iu}
  I=\frac{\omega-\omega_a}{2\lambda\phi_a^2\omega_a}
\end{equation}
which is similar to the former case given by Eq. (\ref{Iph}). On
the other hand, we have according to Eq. (\ref{expansion})
$E_1+E_2=A_s/2\phi_s$ where $A_s$ is given by Eq. (\ref{Asuu}). A
graphic illustration of the solution of this equation with modules
of $E_j, j=1,2$ fixed by Eq. (\ref{Iu}), is shown in Fig.
\ref{triangle}(b). By presenting
$E_1=\sqrt{I}\exp(i(\beta+\theta))$ and
$E_2=\sqrt{I}\exp(i(\beta-\theta))$ we obtain from Eqs.
(\ref{Asuu})
\begin{eqnarray}\label{phaseu}
&\cos^2\theta=\frac{\lambda\Gamma_r\omega_a\omega_rE_{in}^2}
{2\alpha(1-\alpha)\omega_s(\omega-\omega_a)[(\omega-\omega_r)^2+\Gamma_r^2]},&
\nonumber\\ &\tan\beta=\frac{\omega-\omega_r}{\Gamma_r}.&
\end{eqnarray}
The behavior of the phase difference $2\theta$ on the frequency or
the incident wave amplitude $E_{in}$ for $u=0.01$ is shown in Fig.
\ref{phase}.

However, the most remarkable feature of the phase parity breaking
solution for $u\neq 0$  is related in a current circulated between
the defects. When the phase difference $2\theta$ exists between
two quantum dots (QD) or superconductors, connected by a weak
link, a tunneling or Josephson current $J=J_0\sin2\theta$ will
flow between them. The value of the current $J_0$ is proportional
to the coupling constant between QDs or superconductors
\cite{Tilley}. In order to explicitly write the expression for a
current flowing between defects we use the Green function approach
developed in Refs. \cite{mcgurn,miros,MingaleevGF} for the 2D PhC
of dielectric rods with the dielectric constant $\epsilon_0$. The
PhC holds the 1D cavity (waveguide) and two 0D defects (nonlinear
cavity rods) as shown in Fig. \ref{fig1} (a). Then the dielectric
constant of full system $\epsilon({\bf x})$ is a sum of periodic
perfect PC and cavity-induced terms $\epsilon({\bf x})=
\epsilon_{PC}({\bf x})+\delta\epsilon({\bf x}|E)$, where
$\delta\epsilon({\bf x}|E)=\epsilon_W({\bf x})+\epsilon_d({ \bf
x}|E)$ is contributed by the waveguide and the two nonlinear
defects:
\begin{equation}\label{epsil}
\epsilon_d({\bf
x}|E)=\epsilon_W\sum\limits_{n=-\infty}^{\infty}\theta({\bf
x}-{\bf x}_n)+\sum\limits_{j=1,2}\epsilon_j.
\end{equation}
Here $\theta=1$ inside the cavity rod and $\theta=0$ outside, and
the nonlinear contributions $\epsilon_j$ are given by Eq.
(\ref{Kerr}). Then the TM electric field directed along the rods
of the PhC $E({\bf x},t)=E({\bf x})e^{i\omega t}$ is satisfied the
integral equation
\begin{equation}\label{GE}
  E({\bf x})=\frac{\omega^2}{c^2}\int d^2{\bf y}G({\bf x},{\bf y}|\omega)
  \delta\epsilon({\bf y}|E) E({\bf y})
\end{equation}
where $G({\bf x},{\bf y}|\omega)$ is the Green function of the
ideal 2D PC of the rods which was calculated in Ref.
\cite{MingaleevGF} for the square lattice PhC. If the radius of
the defects rods is sufficiently small in comparison to the
wavelength of the EM wave, we can write Eq. (\ref{GE}) as the
discrete nonlinear equation \cite{miros,MingaleevGF}
\begin{equation}\label{fullTB}
E_{\bf n}=\sum_{\bf m}J_{\bf n-m}(\omega)\delta\epsilon_{\bf
m}E_{\bf m}
\end{equation}
where $J_{\bf n-m}(\omega)=
\sigma\frac{\omega^2}{c^2}G(\bf{x_n},\bf{x_m}|\omega)$, $\sigma$
is the cross-section of the rods, and ${\bf n, m}$ runs over sites
of the defects [marked by stars and filled circles in Fig.
\ref{fig1} (a)].

\begin{figure}[ht]
\includegraphics[scale=0.35]{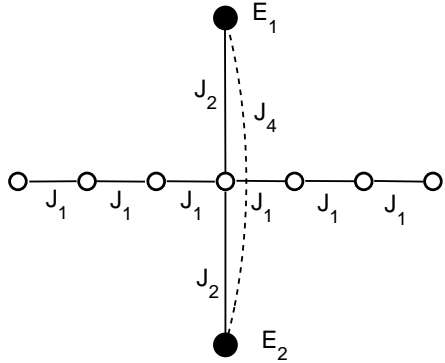}
\caption{Tight-binding version of the PhC system shown in Fig.
\ref{fig1} (a): $J_2$ couples the chain and the defects and $J_4$
connects the defects to each other.} \label{tb}
\end{figure}

We use the nearest-neighbor approximation and write (\ref{fullTB})
as the tight-binding linear chain coupled with two nonlinear
defects
\begin{eqnarray}\label{TB}
&[\frac{1}{\epsilon_W}-J_0(\omega)]E_n=J_1(E_{n+1}+E_{n-1})+
\delta_{n,0}\frac{J_2}{\epsilon_W}(\delta\epsilon_1E_1+\delta\epsilon_2E_2),&\nonumber\\
&[1-\delta\epsilon_1J_0(\omega)]E_1=J_2\epsilon_WE_0+J_4\delta\epsilon_2E_2,&\nonumber\\
 &[1-\delta\epsilon_2J_0(\omega)]E_2=J_2\epsilon_WE_0+J_4\delta\epsilon_1E_1. &
\end{eqnarray}
The model is shown in Fig. \ref{tb}  and consists of a  linear
infinitely long tight-binding chain presented by amplitudes $E_n$
whose spectrum is given by dispersion equation $J_0(\omega)
=\frac{1}{\epsilon_W}-2J_1\cos k$, and two nonlinear defects
presented by amplitudes $\phi_1, ~\phi_2$. The coupling $J_2$
connects the defects and the chain and the coupling $J_4$ connects
the defects.

By multiplying Eq. (\ref{TB}) by $E_0^{*}=t^{*}$ and subtracting
the complex conjugated terms one can obtain the value of the power
flow current flowing between the chain at the "0"-th site and the
defects enumerated as $j=1,2$ as follows
\begin{equation}\label{current}
j_{0\rightarrow 1,2}=\epsilon_WJ_2Im(tE_{1,2}^{*}).
\end{equation}
Similar manipulations with the cavity's amplitudes give the
current between the defects
\begin{equation}\label{current0}
  j_{1\rightarrow 2}=J_4\delta\epsilon Im(E_1E_2^{*})=
J_4\delta\epsilon I\sin(2\theta).
\end{equation}
It follows also that the current from the "-1"-th site to the
"0"-th site of the chain coincide with the current from the "0"-th
site to the "1"-th one. Therefore the currents (\ref{current}) and
(\ref{current0}) coincide also in accordance to the Kirchhoff
rule. Thus, the input power induces vortical current between the
waveguide and defects via the couplings $J_2$ and $J_4$. The
current is excited by the incident wave provided the defects are
nonlinear. Thus, our analysis shows that the symmetry can be
broken not only because of different intensities at the defects
but also by a circulating current between the defects although the
intensities at the defects are equal. This model result of the
Josephson like current between the defects with different phase is
reflected in computations of the Poyinting vector in the PhC
structure as will be shown below.
\subsection{Stability of solutions}
\begin{figure}
\includegraphics[scale=.35]{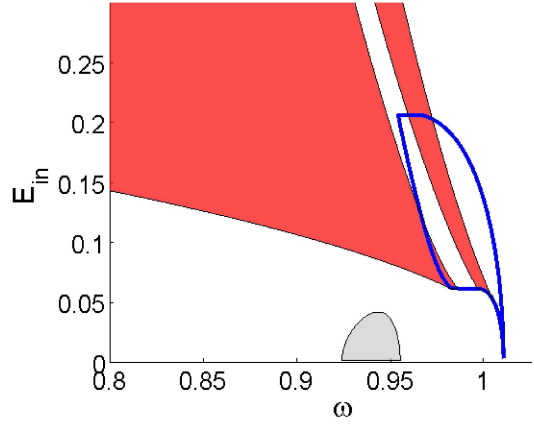}
\caption{Regions of stability of the solution. The symmetry
preserving solution is stable everywhere except interior of the
closed region shown by solid blue line. The stability of the
symmetry breaking solution is shown by red, while the phase parity
breaking solution is shown by gray. The parameters are $u=0.01,
\omega_0=1, \Gamma=0.01, \lambda=-0.01, \phi_s=1, \phi_a=1.1$.}
\label{PD}
\end{figure}

Furthermore, we studied stability of different solutions by
standard methods given for example in Ref. \cite{cowan}. The
stability of the solution can be found from the temporal CMT
equations
\begin{eqnarray}\label{temporalA1A2}
&i\dot{a}_s=[\omega_s+\lambda\omega_s\phi_s^2(I_1+I_2)-i\Gamma]a_s+\lambda\omega_0\phi_s\phi_a(I_1-I_2)
a_a+\sqrt{\Gamma}E_{in}e^{-i\omega t},&\nonumber\\
&i\dot{a}_a=\lambda\omega_0\phi_s\phi_a(I_1-I_2)a_s+
[\omega_a+\lambda\omega_a\phi_a^2(I_1+I_2)]a_a.&
\end{eqnarray}
By presenting $a_s(t)=(A_s+\xi_s(t))e^{-i\omega t}, ~~
a_a(t)=(A_a+\xi_a(t))e^{-i\omega t}$ with $A_s, ~A_a$ as the
steady state obeying the stationary CMT equations (\ref{A1A2sym})
and $|\xi_s(t)|\ll |A_s|, ~|\xi_a(t)|\ll |A_a|$ we obtain the
linearized time-dependent equations for complex $\xi_s, \xi_a$
\begin{equation}\label{stab}
\left(\begin{array}{c} Re(\dot{\xi}_s)\cr Im(\dot{\xi}_s)\cr
Re(\dot{\xi}_a)\cr
Im(\dot{\xi}_a)\end{array}\right)=\widehat{L}\left(\begin{array}{c}
Re(\xi_s)\cr Im(\xi_s)\cr Re(\xi_a)\cr
Im(\xi_a)\end{array}\right).
\end{equation}
Their stability is determined by eigen values of the matrix
$\widehat{L}$ which is time independent. The results of our
calculation of stability are presented in Fig. \ref{tr} which
shows that the stability of the phase parity breaking solution
appears if only the defects are coupled and $\phi_s\neq \phi_a$.
We collected the results of stability of all three solutions in
Fig. \ref{PD} in the form of phase diagrams in plane of the
incident wave amplitude and the frequency.
One can see that the phase parity breaking solution is stable in
some small area of the phase diagram.
\subsection{Numerical calculations in photonic crystal}
We numerically solve the Maxwell equations (\ref{maxwell}) for the
TM mode in the PhC with defect nonlinear rods by expansion of
electromagnetic field over maximally localized photonic Wannier
functions \cite{busch,marzari,photonic}. The square lattice 2D PhC
has the same parameters as given earlier [see Fig. \ref{fig1}(a)].
For the case of isolated linear defects with the same radius as
the radius of host rods and the dielectric constant $\epsilon_d=3$
their eigen frequency $\omega_0=0.3593$ in terms of $2\pi c/a$.
Overlapping of the defect's monopole modes gives rise to splitting
of this frequency $\omega_s=0.3603, \omega_a=0.3584$ as numerical
computation of equations (\ref{maxwell}) gives. Respectively we
obtain that the value of coupling $u=-0.001$. The corresponding
bonding and anti-bonding modes for the nearest distance $a$
between defects were found in Ref. \cite{ming1}. For more distance
$4a$ they are shown in Fig. \ref{modes}. By the normalization
condition (\ref{norma}) the heights of the amplitude modes at the
defects equal $\phi_s=0.5569, ~\phi_a=0.6179$. Let us evaluate the
dimensionless nonlinearity constant $\lambda$. We take in
numerical calculations the incident power per length of order
$100mW/a$ which corresponds to the incident intensity
$I_0=100mW/a^2$. For chosen PhC lattice with period $a=0.5\mu m$
we obtain that the incident intensity equals $0.04GW/cm^2$.   With
the use of $\epsilon=\epsilon_0+2\sqrt{\epsilon_0}n_2I_0$ we
obtain
\begin{equation}\label{lambda}
\lambda=- 2\sqrt{\epsilon_0}n_2I_0.
\end{equation}
We take the linear and nonlinear refractive indexes of the defect
rods to be, respectively, $n_0=\sqrt{\epsilon_0}=\sqrt{3},
n_2=2\times 10^{-12}cm^2/W$. By substituting all of these
estimates into (\ref{lambda}) we obtain $\lambda\sim -0.9\times
10^{-2}$ which is close to that used in the CMT consideration.
Finally, we estimate the coupling of the defect mode with the
propagation mode of the PhC waveguide $\sqrt{\Gamma}$. There are
many ways to calculate $\Gamma$ using for example Refs.
\cite{cowan,michaelis,lecamp}. In the present paper we estimated
$\Gamma$ numerically by using the following approach. We took the
single linear defect aside the PC waveguide as shown in Fig.
\ref{fig1}(a), and calculated the transmission spectra. By the
resonance width of the spectra we evaluated $\Gamma=0.00185$.

\begin{figure}
\includegraphics[scale=0.4]{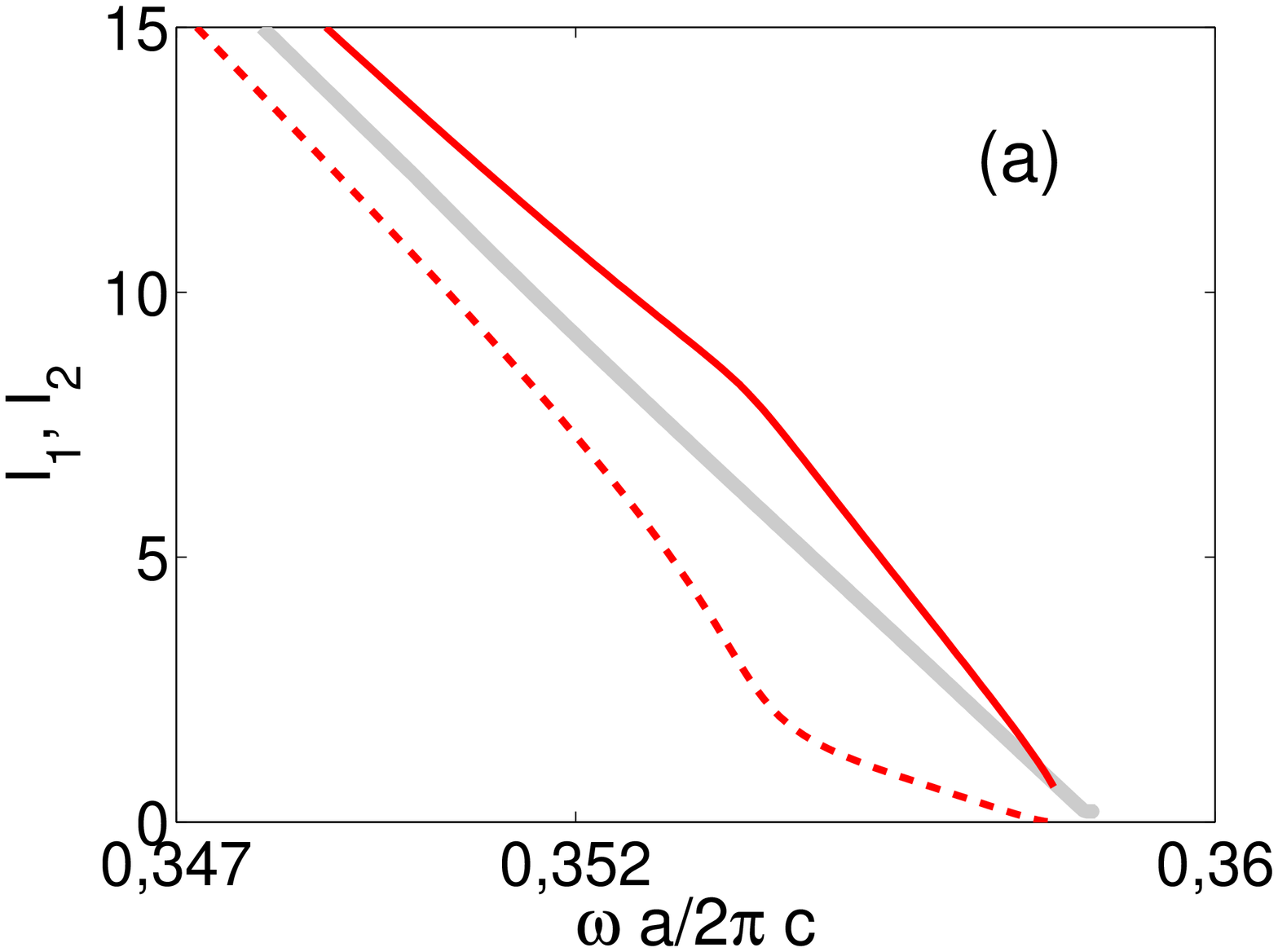}
\includegraphics[scale=0.4]{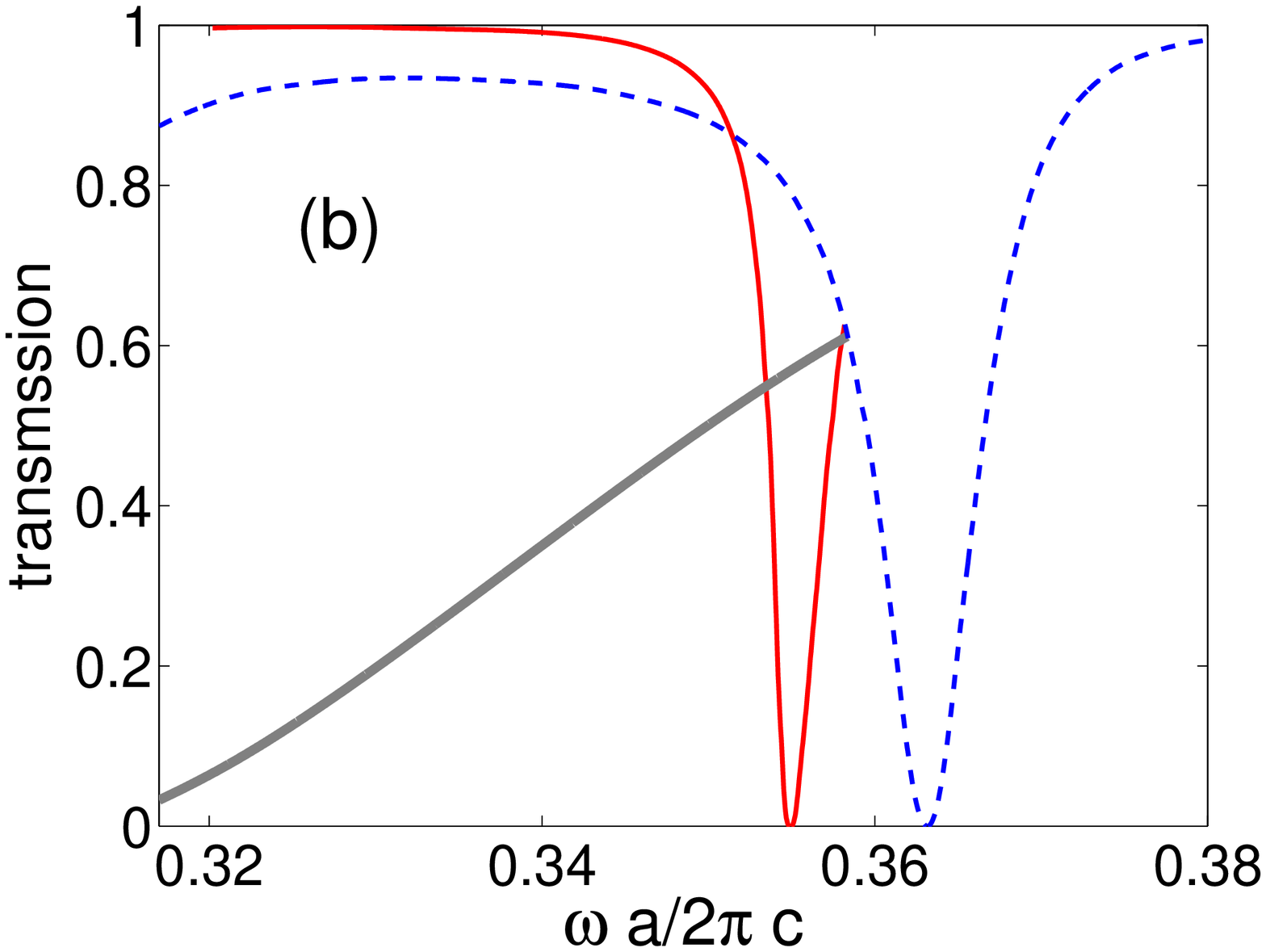}
\caption{Self-consistent solution for (a) the intensities of the
EM field at the nonlinear defects and (b) transmission spectra in
the PhC structure in the PhC structure shown in Fig. \ref{fig1}.
The parameters of the PhC and defects are given in Fig.
\ref{modes}. The input power per length equals $100mW/a$.
$n_2=2\times 10^{-12}cm^2/W, \lambda=-0.009$.} \label{PCint}
\end{figure}
\begin{figure}
\includegraphics[width=.3\textwidth,height=0.21\textheight]{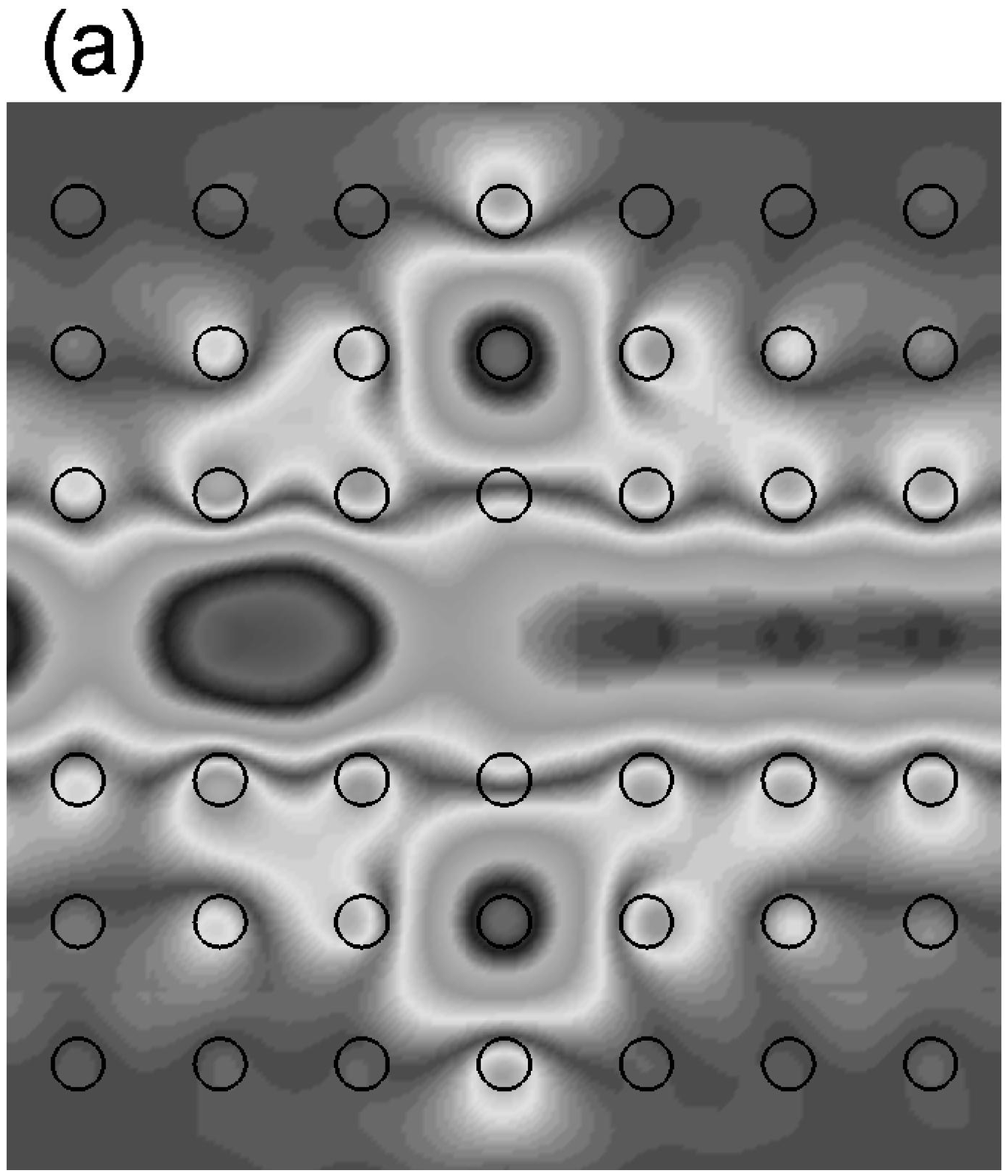}
\includegraphics[width=.3\textwidth,height=0.22\textheight]{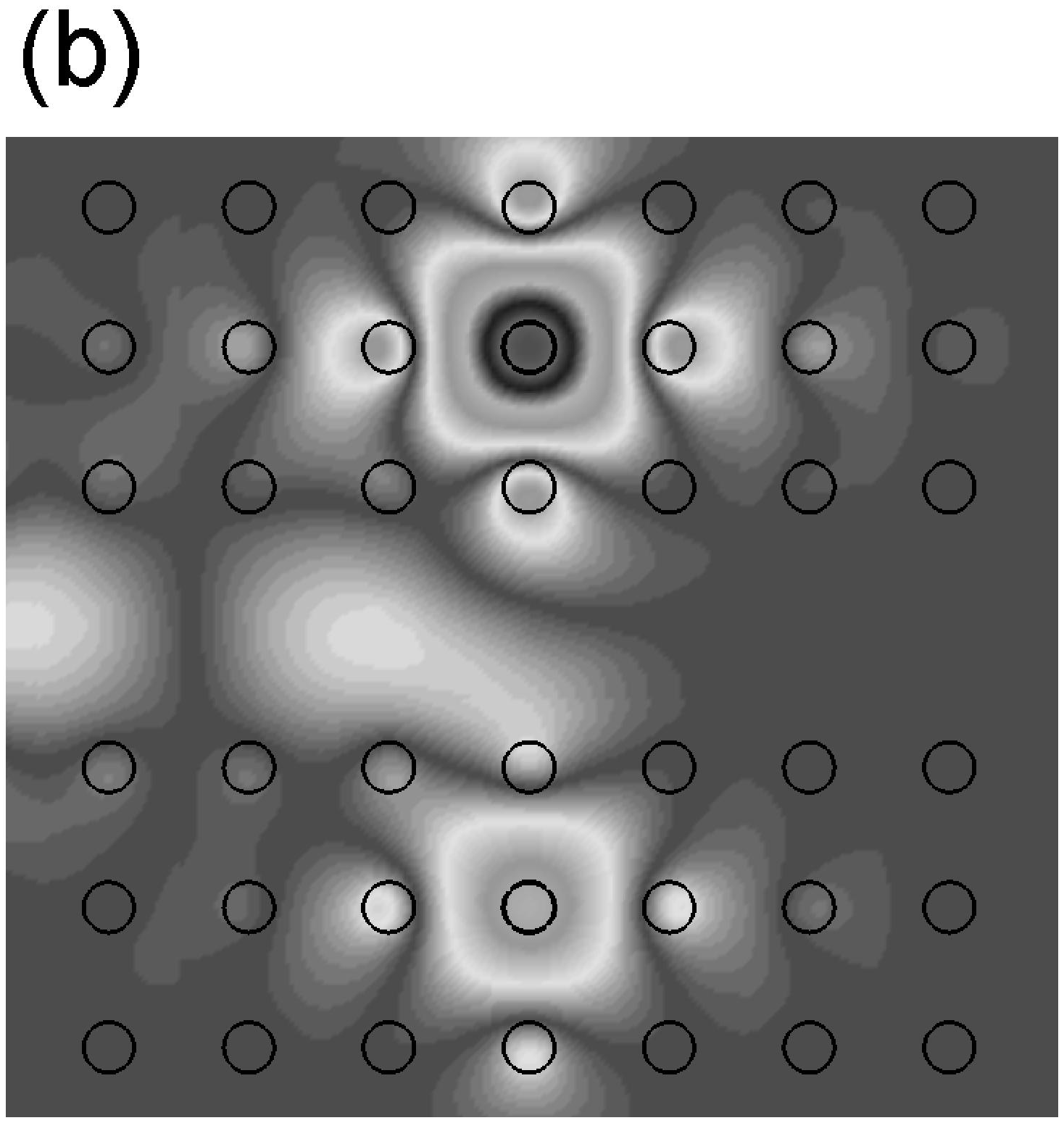}
\includegraphics[width=.3\textwidth,height=0.22\textheight]{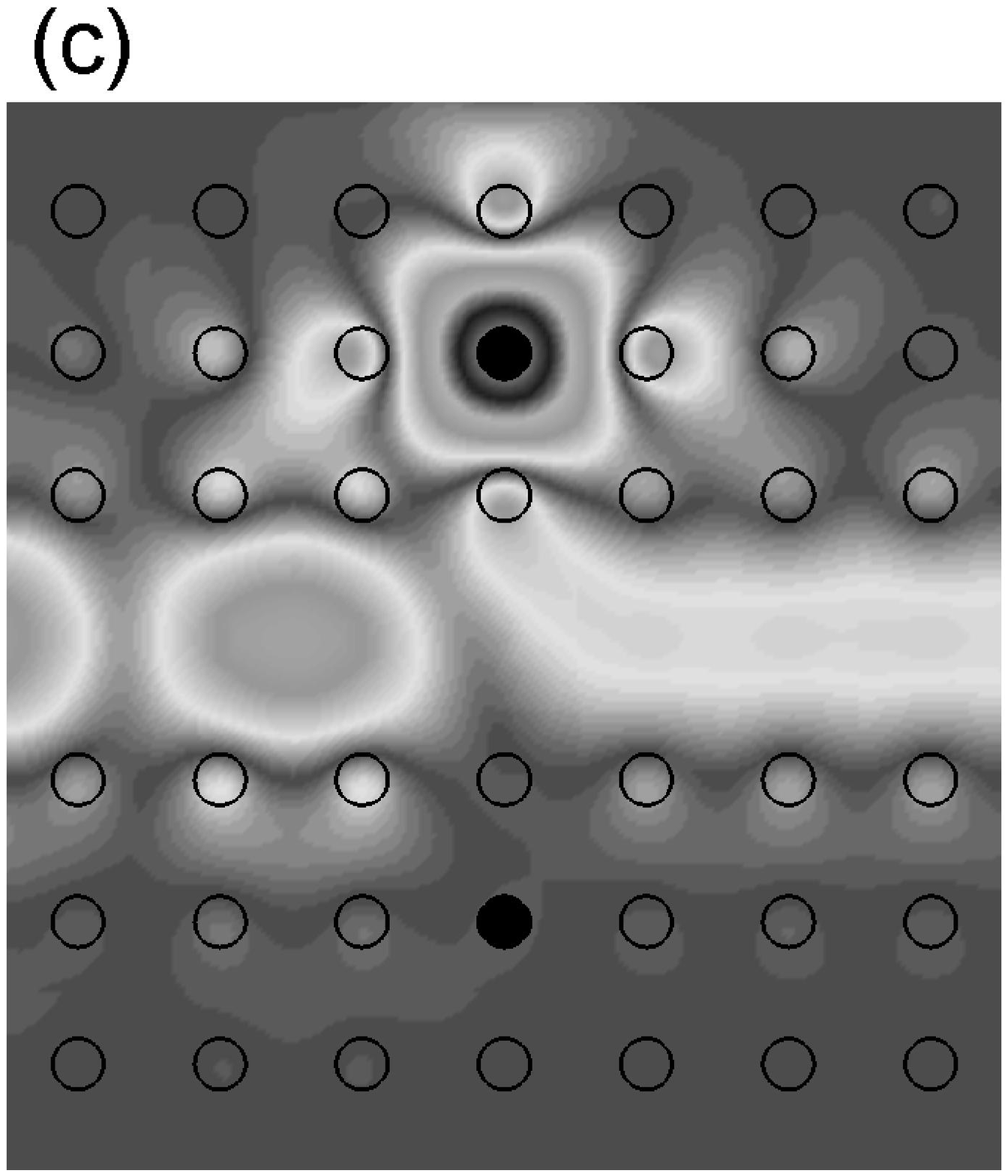}
\caption{Absolute value of the EM field solution for (a) the
symmetry preserving solution $\omega a/2\pi c=0.355$, (b) and (c)
the symmetry breaking solution  for $\omega a/2\pi c=0.355$ and
$\omega a/2\pi c=0.358$ respectively. The EM wave incidents at the
left of the waveguide.} \label{wavebroken}
\end{figure}

The self-consistent solutions are presented in the form of the
intensities in Fig. \ref{PCint}(a), which are similar to the CMT
results shown in Fig. \ref{intu0}(a). Also, one can see three
solutions in the transmission shown in Fig. \ref{PCint}(b), as was
found in the CMT model for the transmission shown in Fig.
\ref{tr}. Fig. \ref{wavebroken} shows the EM field (the absolute
value of the electric field) for the symmetry preserving solution
(a) and for the symmetry breaking solution (b) and (c). In the
latter case one can see that the field is strongly different at
bottom and top. Moreover Figs. \ref{intu0} and \ref{PCint} show
that there is a frequency at which the intensity of the EM field
might be zero at the bottom defect. Indeed, Fig.
\ref{wavebroken}(c) demonstrates this case.
\begin{figure}
\includegraphics[scale=.4]{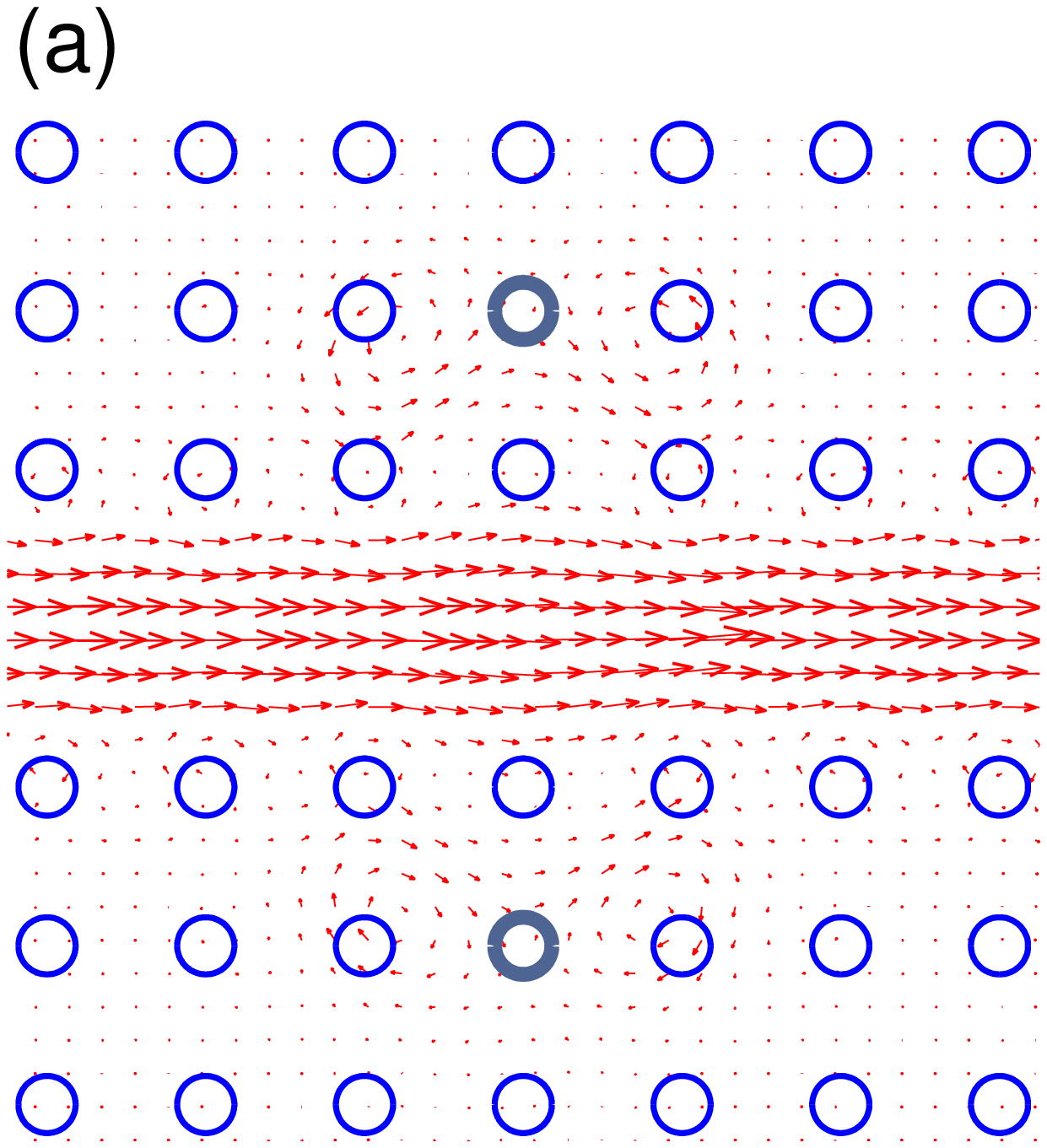}
\includegraphics[scale=.35]{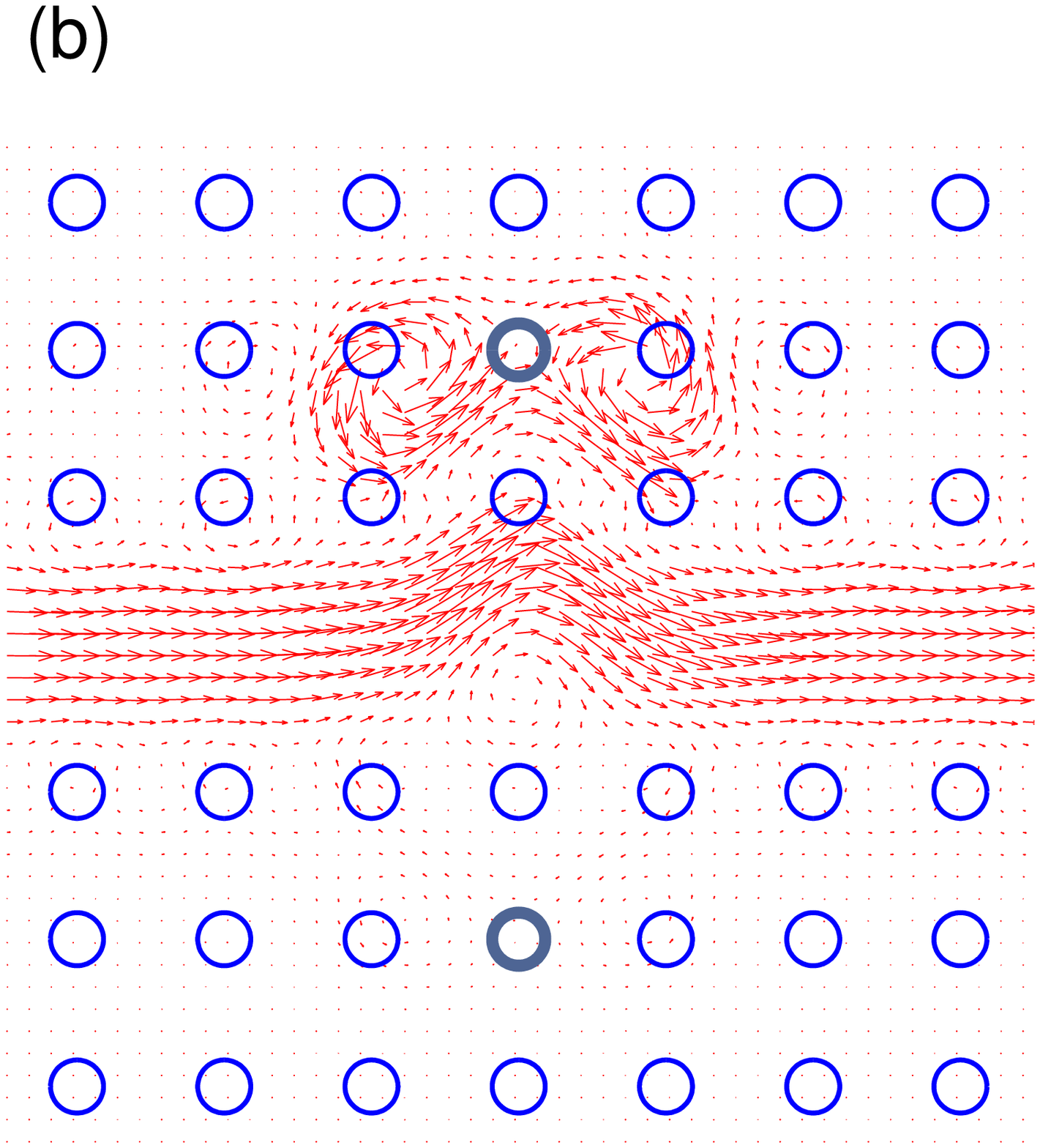}
\includegraphics[scale=.35]{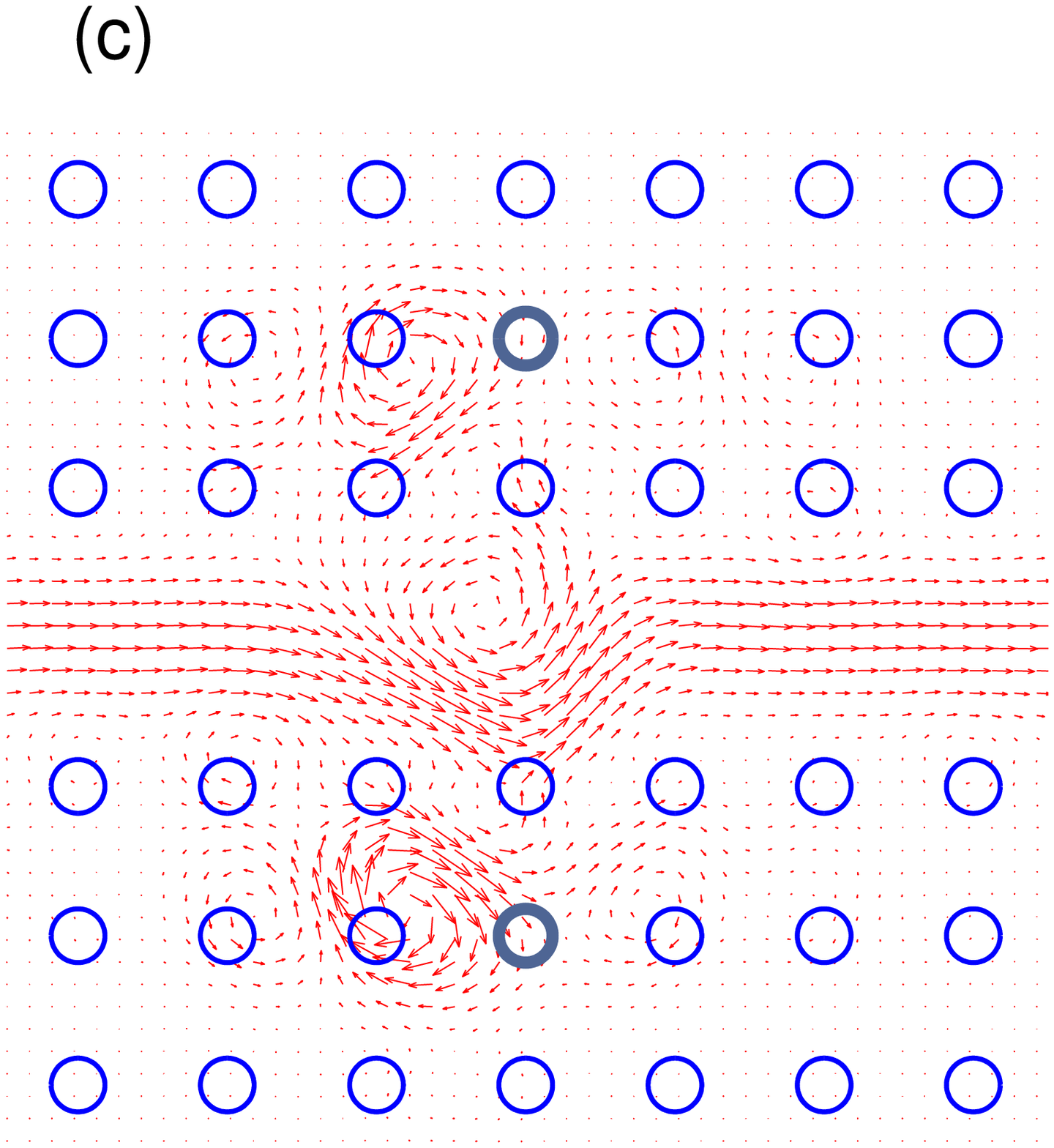}
\caption{Current flows for the symmetry preserving solution which
inherits linear case (a), the symmetry breaking solution (b), and
(c) the phase parity breaking one at $a\omega/2\pi c=0.35$. Bold
open circles mark the nonlinear defects. } \label{currents}
\end{figure}

In agreement with the model consideration Fig. \ref{currents}
shows that current flows (the Poyinting's vector patterns) are
strongly different for the different solutions. For the symmetry
preserving solution we have laminar current flow over the
waveguide with excitation of two current vortices around each
defect. The laminar flow over the waveguide and the vortical flows
around defects are well separated. The whole current pattern is
symmetrical relative to the symmetry transformation $y\rightarrow
-y$.  The picture has a similarity with ballistic electron
transport in waveguide coupled to an off-channel quantum dot
\cite{exner}. For the case of the symmetry breaking solution there
is  a current vortex inside the waveguide complemented by two
vortices near each defect, as shown in Fig. \ref{currents} (b).
The circulation in vortical flow around the upper defect is
opposite to the circulation around the bottom defect. The vortical
flow in the waveguide and the vortical flows around the defects
are well separated for both solutions. In the third case (c) for
the phase parity breaking solution one can see the current vortex
in the waveguide and single vortices around the defects are mixed.
Nevertheless because of the continuity equation in the space
beyond of the nonlinear defects $\nabla \overrightarrow{j}(x,y)=0$
the vortical flows around the defects and in the waveguide can not
cross.

\section{The T-shape waveguide coupled with two nonlinear defects}
One of the most ambitious goals in nonlinear optics is the design
of an all-optical computer that will overcome the operation speeds
in conventional (electronic) computers. Vital in this respect is
the design of basic components such as all-optical routing
switches and logic gates. It is believed that future integrated
photonic circuits for ultra fast all-optical signal processing
require different types of nonlinear functional elements such as
switches, memory and logic devices. Therefore, both physics and
designs of such all-optical devices have attracted significant
research efforts during the last two decades, and most of these
studies utilize the concepts of optical switching and bistability.
One of the simplest bistable optical devices which can be built-up
in photonic integrated circuits is a single cavity coupled with
optical waveguide or waveguides \cite{joanbook,mcdonald}.

The concept of the all-optical switching is based on a
discontinuous transition between the symmetry breaking solutions
by a small change of the input \cite{mayer}. Many of these devices
employ a configuration of two parallel coupled nonlinear
waveguides \cite{jensen,friberg,chen,boumaza,grigoriev}. Recently
Maes {\it et al} demonstrated the all-switching in the system of
two nonlinear micro-cavities aligned along the waveguide
\cite{maes1} by the use of pulses of injected light. In the
present section we use similar approach to demonstrate the
all-switching effects in the T-shaped waveguide coupled with two
nonlinear micro-cavities \cite{T,PRBT}.

\begin{figure}[ht]
\includegraphics[scale=0.25]{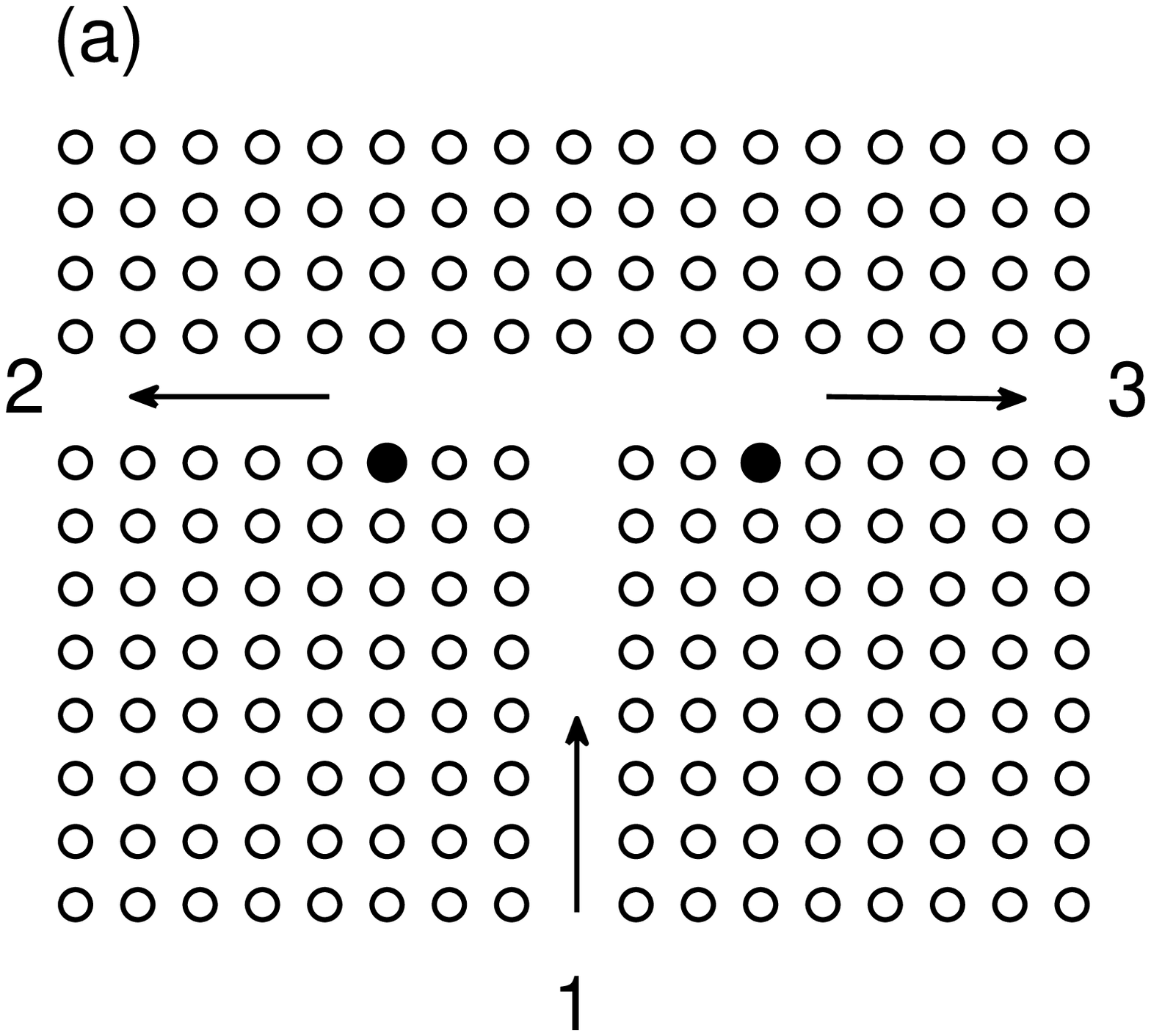}
\includegraphics[scale=0.25]{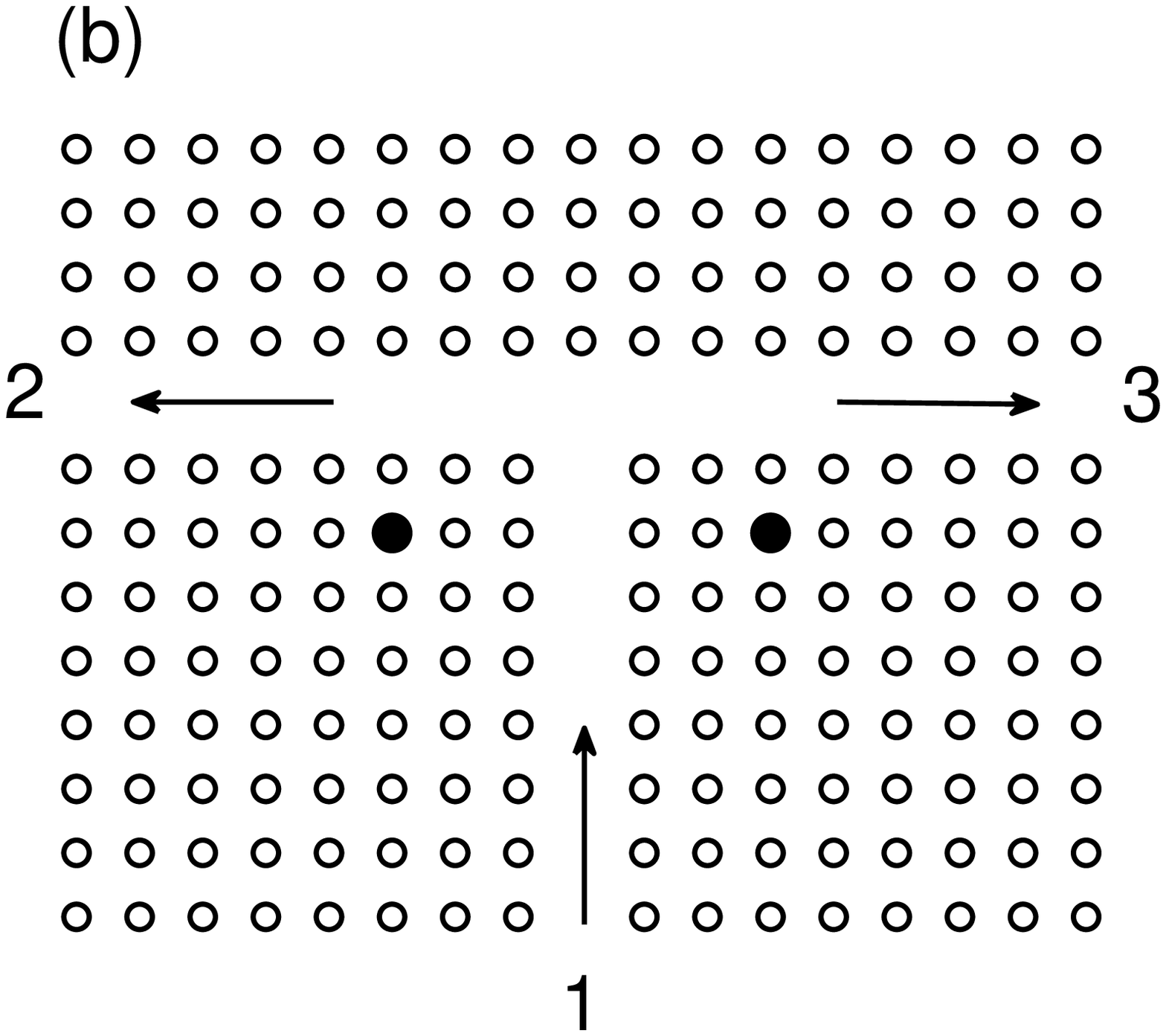}
\includegraphics[scale=0.25]{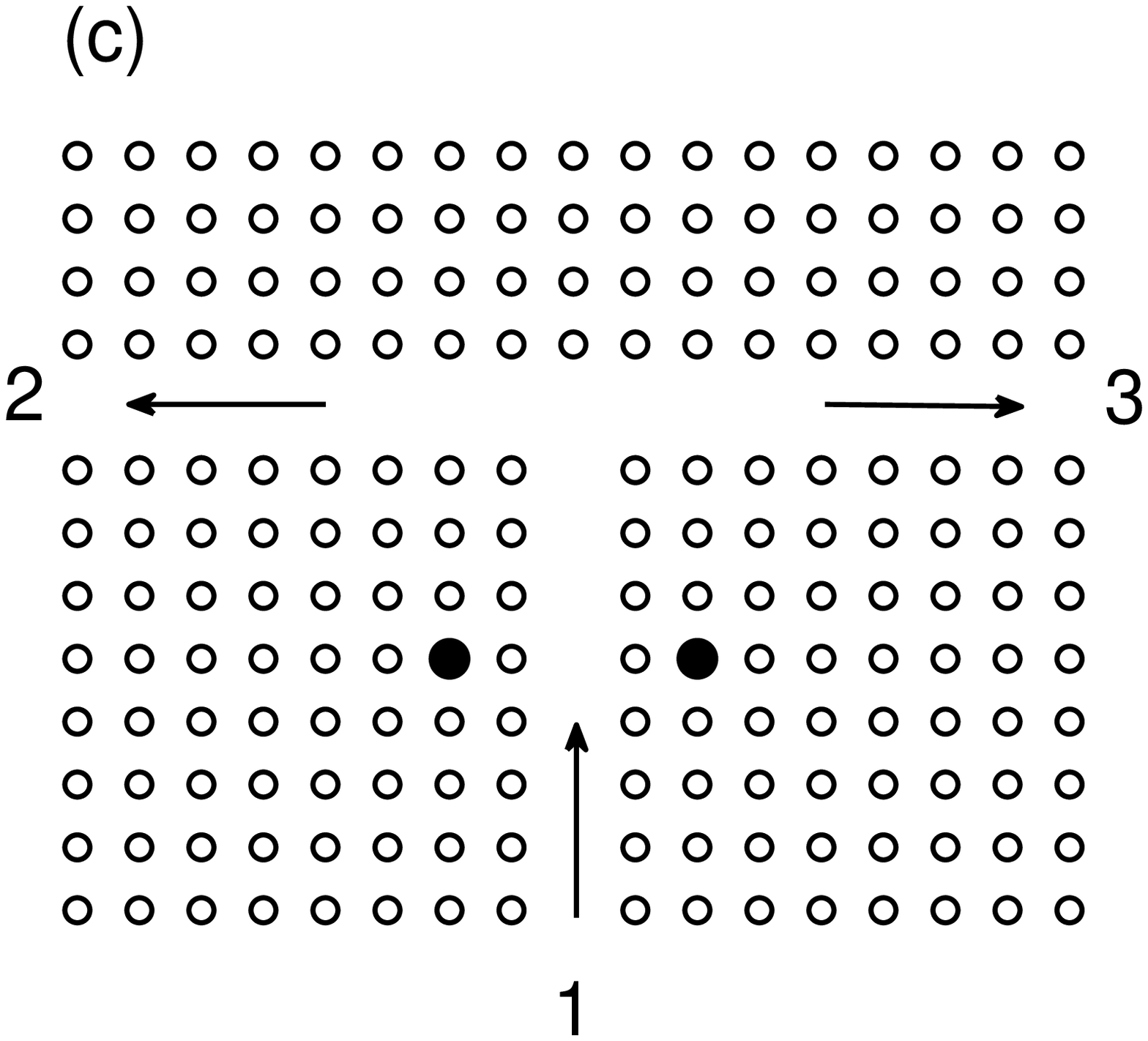}
\caption{T-shaped waveguide with two nonlinear defect rods. The
cases (a)-(c) differ by the positions of the nonlinear defects.}
\label{T1}
\end{figure}

\begin{figure}[ht]
\includegraphics[scale=0.3]{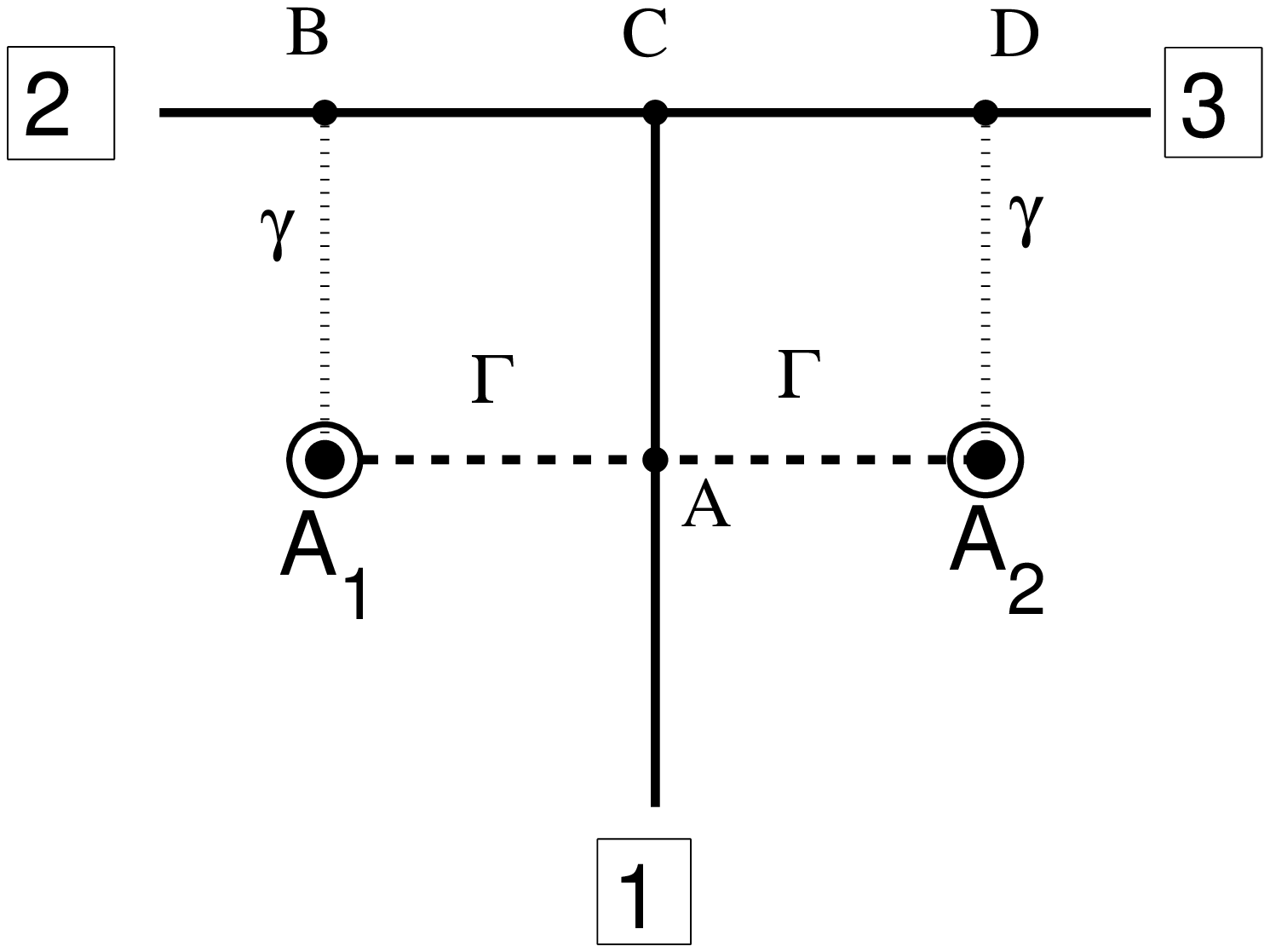}
\includegraphics[scale=0.3]{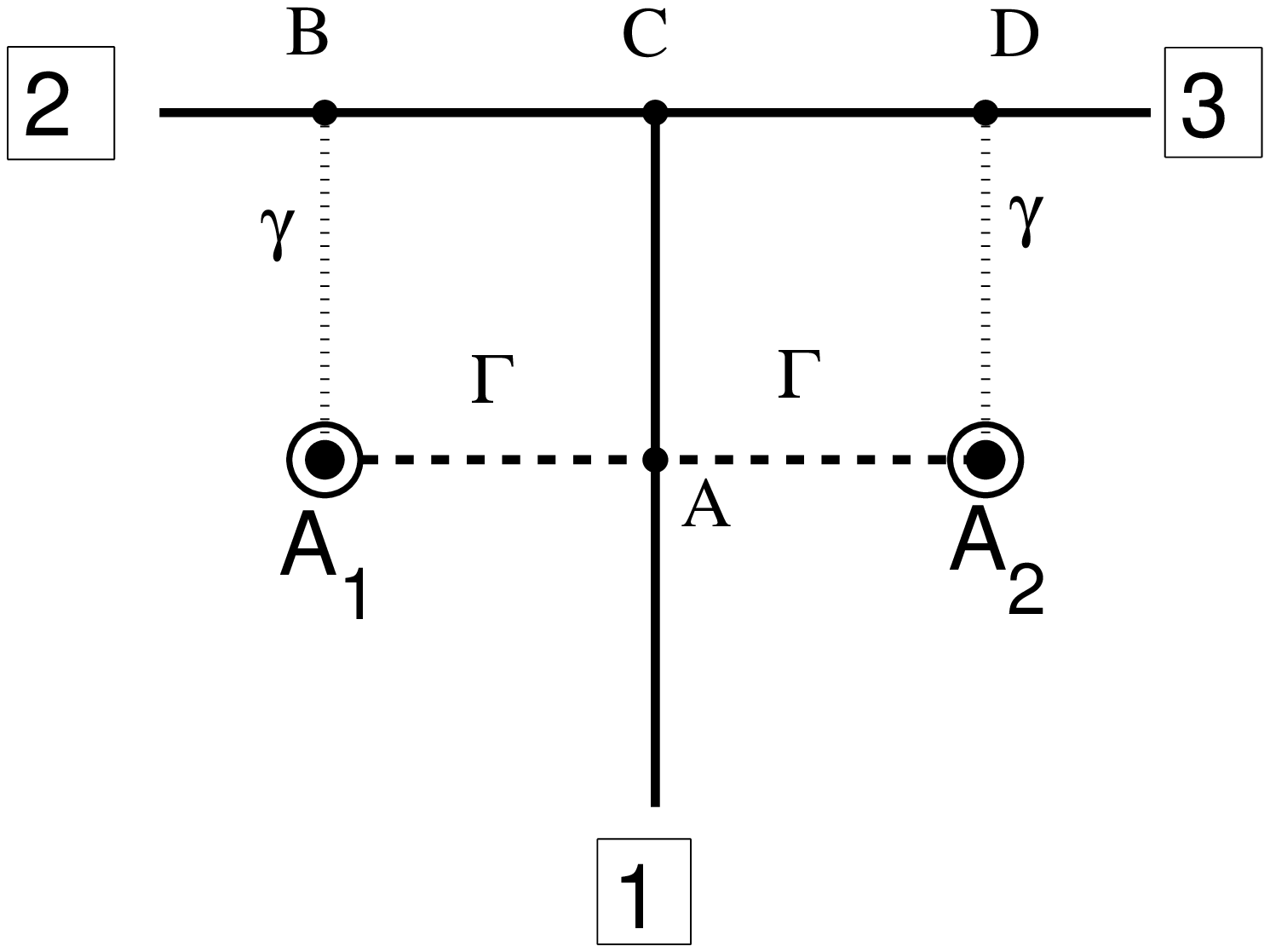}
\includegraphics[scale=0.3]{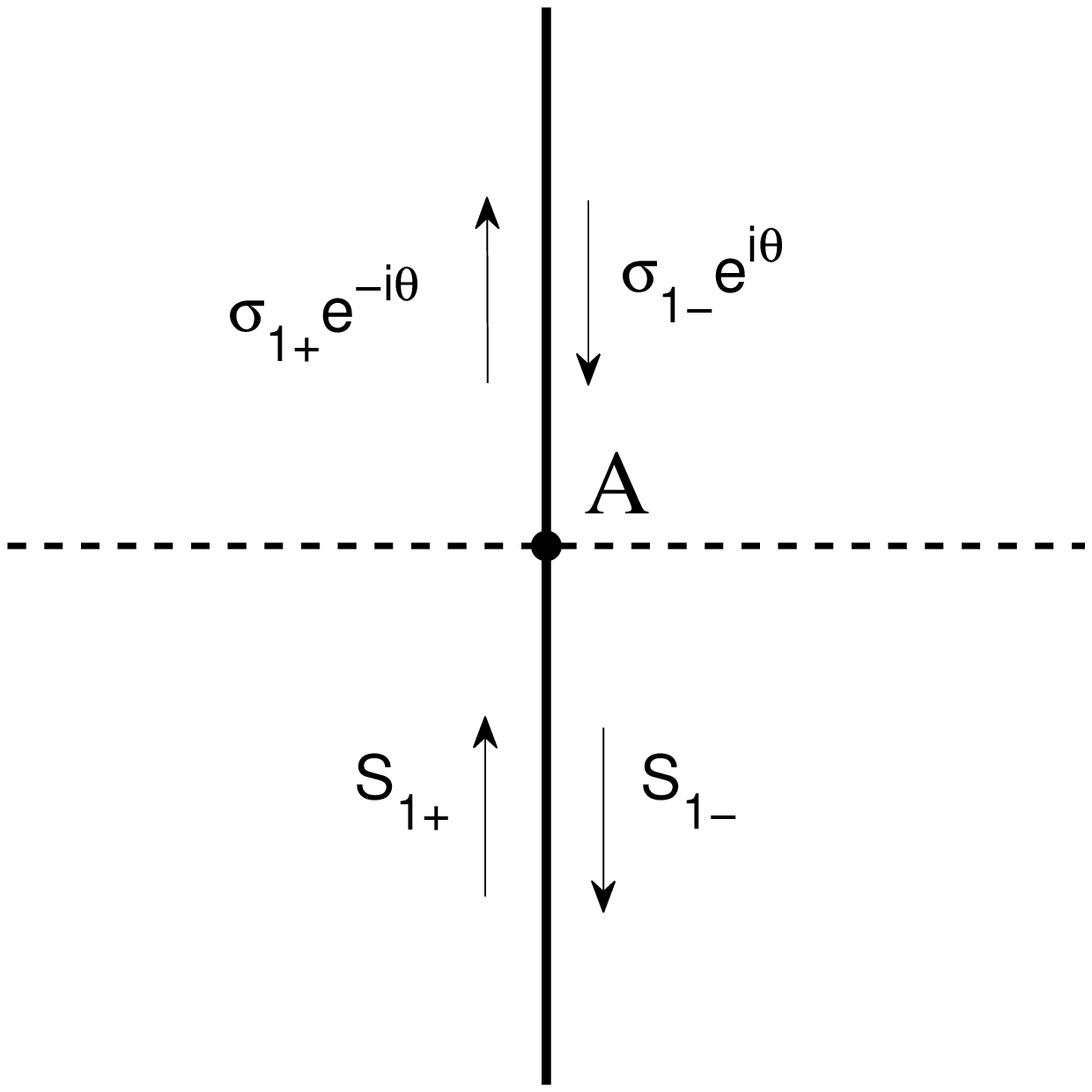}
\includegraphics[scale=0.3]{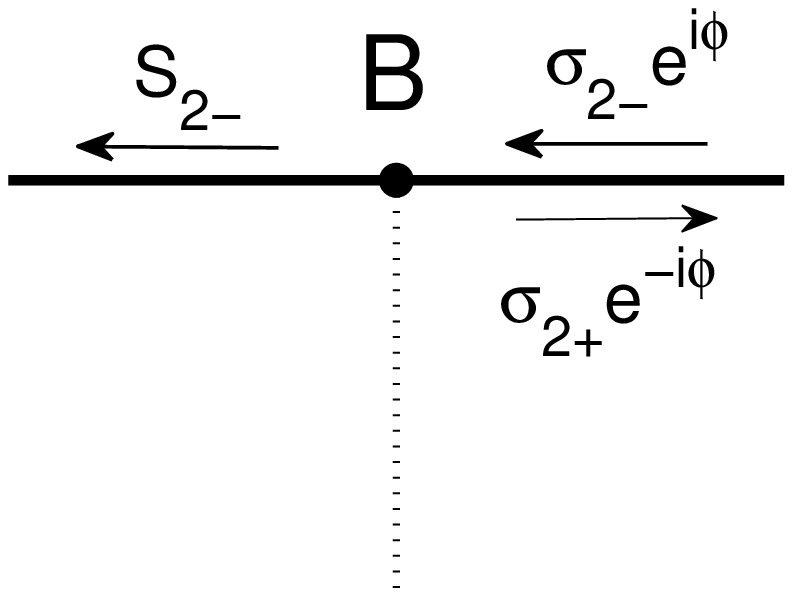}
\includegraphics[scale=0.3]{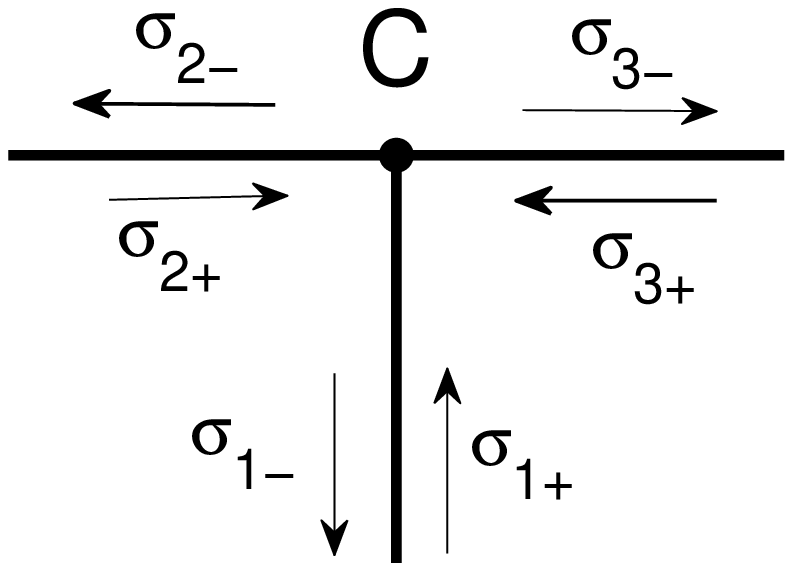}
\includegraphics[scale=0.3]{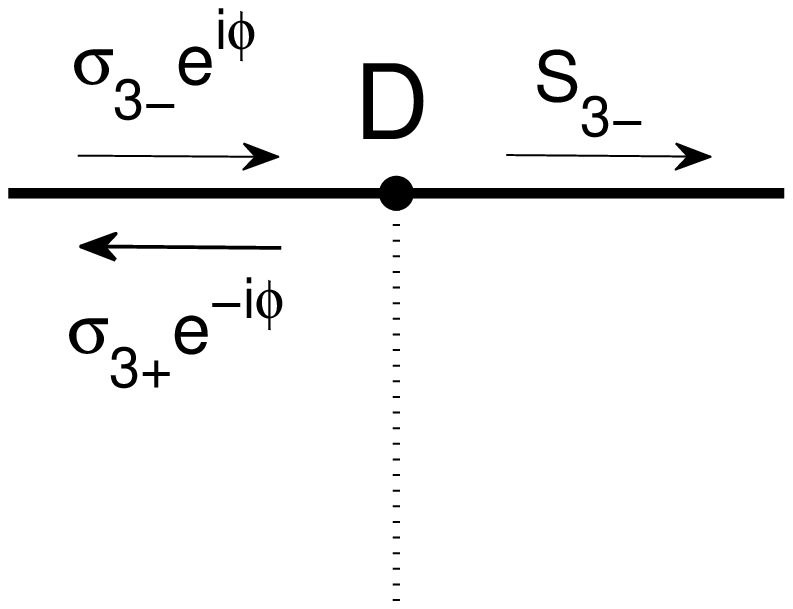}
\caption{CMT model of the T-shape photonic crystal waveguide
coupled with two nonlinear defects shown by filled bold circles.
The defects are coupled with input waveguide 1 via the constant
$\Gamma$ shown by dashed lines and with the output waveguides 2,3
via the constant $\gamma$ shown by dotted line. Separately each
connection is shown with corresponding light amplitudes.}
\label{T2}
\end{figure}

   We consider the PhC shown in Fig. \ref{T1} with the same parameters as
given in Section III:  the  lattice constant $a = 0.5\mu m$, the
cylindrical dielectric rods have radius $0.18a$ and dielectric
constant $\epsilon= 11.56$. We substitute two defect rods of the
same radius as shown in Fig. \ref{T1} made from an instantaneous
Kerr media with the nonlinear refractive index $n=n_0+n_2I_0$
where $n_0=\sqrt{3}$ and $n_2=2\times 10^{-12}cm^2/W$.
The corresponding equivalent configuration of the T-shaped
waveguide with two nonlinear defects is presented in Fig. \ref{T2}

We start with the position of the defect rods shown in Fig.
\ref{T1} (a) which have strong coupling with the output waveguides
2 and 3, and negligibly weak coupling with the input waveguide 1.
We consider a light given by the amplitude $S_{1+}$ is incident
into the waveguide 1 and outputs into all three terminals as shown
schematically in Fig. \ref{T2}. The outgoing amplitudes are
labelled as $S_{1-}, ~S_{2-}$ and $S_{3-}$. Each nonlinear optical
cavity is assumed to be given by single mode amplitudes $A_j,
~j=1,2$ and coupled with the guides 2, 3 via the coupling constant
$\gamma$ shown in Fig. \ref{T2} by dotted lines and with the guide
1 via the coupling constant $\Gamma$.

We consider that the defects are not coupled. Therefore the eigen
frequencies of the system of the defects equal the monopole eigen
frequencies of the isolated defects
$\omega_j=\omega_0+\lambda|A_j|^2, ~j=1,2$ shifted because of the
Kerr effect. Then the equations (\ref{Am}) will take the following
form
\begin{eqnarray}\label{A1A2temp}
(\omega-\omega_0-\lambda|A_j|^2+i\gamma+i\Gamma)A_1+i\Gamma
A_2=i\sqrt{\Gamma}
(S_{1+}+\sigma_{1-}e^{i\theta})+i\sqrt{\gamma}\sigma_{2-}e^{i\phi}\nonumber\\
(\omega-\omega_2+i\gamma+i\Gamma)A_2+i\Gamma A_1=i\sqrt{\Gamma}
(S_{1+}+i\sigma_{1-}e^{i\theta})+i\sqrt{\gamma}\sigma_{3-}e^{i\phi}
\end{eqnarray}
Here phases $\theta$ and $\phi$ as shown in Fig. \ref{T2} are the
optical lengths through which light goes between the connections.

These equations are to be complemented by the equations for light
amplitudes at each connection A, B, and D \cite{suh}
\begin{eqnarray}\label{A1A2}
(\omega-\omega_1+i\gamma+i\Gamma)A_1+i\Gamma A_2=i\sqrt{\Gamma}
(S_{1+}+\sigma_{1-}e^{i\theta})+i\sqrt{\gamma}\sigma_{2-}e^{i\phi}\nonumber\\
(\omega-\omega_2+i\gamma+i\Gamma)A_2+i\Gamma A_1=i\sqrt{\Gamma}
(S_{1+}+i\sigma_{1-}e^{i\theta})+i\sqrt{\gamma}\sigma_{3-}e^{i\phi}.
\end{eqnarray}
These CMT equations are to be complemented by the equations for
light amplitudes at each connection A, B, and D
\begin{eqnarray}\label{sigma}
&\sigma_{1+}e^{-i\theta}=S_{1+}-\sqrt{\Gamma}(A_1+A_2)\,&\nonumber\\
&S_{1-}=\sigma_{1-}e^{i\theta}-\sqrt{\Gamma}(A_1+A_2)\,&\nonumber\\
&S_{2-}=\sigma_{2-}e^{i\phi}-\sqrt{\gamma}A_1\,&\nonumber\\
&S_{3-}=\sigma_{3-}e^{i\phi}-\sqrt{\gamma}A_2.&\nonumber\\
&\sigma_{2+}e^{-i\phi}=-\sqrt{\gamma}A_1.&\nonumber\\
&\sigma_{3+}e^{-i\phi}=-\sqrt{\gamma}A_2.&\nonumber\\
\end{eqnarray}The T-connection at the C point connects ingoing and outgoing
amplitudes by the S-matrix as follows
\begin{equation}\label{T}
\left(\begin{array}{c} \sigma_{1-} \cr \sigma_{2-} \cr \sigma_{3-}
\end{array}\right)=\left(\begin{array}{ccc} a& b& b \cr b& c &d \cr b& d& c
\end{array}\right)\left(\begin{array}{c} \sigma_{1+} \cr \sigma_{2+} \cr \sigma_{3+}
\end{array}\right).
\end{equation}
In particular, the solution of the Maxwell equations for the
T-shaped waveguide without defects gives the matrix elements of
the S-matrix (\ref{T}) $a=-0.3547+0.308i, b=0.6+0.173i,
c=-0.4319+0.2271i, d=-0.568+0.2225i$ at $\omega a/2\pi c=0.35$.
Eqs. (\ref{A1A2temp}), (\ref{sigma}), and (\ref{T}) form a full
system of equations for 11 amplitudes $A_1, A_2, \sigma_{1+},
\sigma_{1-}, \sigma_{2+}, \sigma_{2-}, \sigma_{3+}, \sigma_{3-},
S_{1-}, S_{2-}, S_{3-}$. Substituting $S_{1+}=E_{in}e^{i\omega t},
A_{1,2}=A_{1,2}e^{-i\omega t}$ we obtain after some algebra the
following stationary CMT equations
\begin{equation}\label{eqHeff}
(\omega-H_{eff})\left(\begin{array}{c} A_1 \cr
A_2\end{array}\right)=iE_{in}F\left(\begin{array}{c} 1 \cr
1\end{array}\right),
\end{equation}
where
\begin{equation}\label{HeffT}
H_{eff}=\left(\begin{array}{cc}
\omega_1-iG&-iH\cr-iH&\omega_2-iG\end{array}\right),
\end{equation}
\begin{eqnarray}\label{AB}
&G=\gamma+\Gamma(1+ae^{2i\theta})+\gamma
de^{2i\phi}+\sqrt{\gamma\Gamma}(b+c)e^{i\theta+i\phi},&\nonumber\\
&H=\Gamma+\sqrt{\gamma\Gamma}(b+c)e^{i\theta+i\phi}+\Gamma
ae^{2i\theta}}+\gamma de^{2i\phi),&\nonumber\\
&F=\sqrt{\Gamma}(1+ae^{2i\theta})+\sqrt{\gamma}be^{i(\theta+\phi)}.&
\end{eqnarray}

We can calculate all parameters which are necessary in Eqs.
(\ref{eqHeff}), (\ref{HeffT}), and (\ref{AB}). For the light
transmission in straight waveguide coupled with the single linear
off-channel defect positioned at different positions we able to
extract the coupling constant of the cavity with PhC waveguide
$\Gamma$ and the eigen frequency of monopole mode $\omega_0$. The
results are collected in Table \ref{tab1}.
\begin{table}
\caption{Parameter sets of the CMT for PhC T-shaped waveguide
shown in Fig. \ref{T1}} \label{tab1}
\begin{tabular}{|c|c|c|c|} \hline
 Type of structure in Fig. \ref{T1}     & (a) & (b) & (c)   \cr \hline
$\Gamma$ (in terms of $a/2\pi c$)       & 0.0002  & 0.0002   & 0.00189   \\
$\gamma$ (in terms of $a/2\pi c$)     & 0.03093 & 0.00189 & 0.00002\\
$\omega_0$ (in terms of $a/2\pi c$)  & 0.3609   & 0.365 & 0.3596    \\
\hline
\end{tabular}
\end{table}
The limiting case of the T-shaped waveguide with $\Gamma=0$ is
considered in Ref. \cite{T} with results qualitatively close to
those shown in Fig. \ref{T4}. Here we find  the solutions with the
substitution of concrete parameters listed in Table \ref{tab1} and
present the output light transmission to the left and to the right
waveguides. We start with the case shown in Fig. \ref{T1} (a).

\begin{figure}
\includegraphics[scale=0.35]{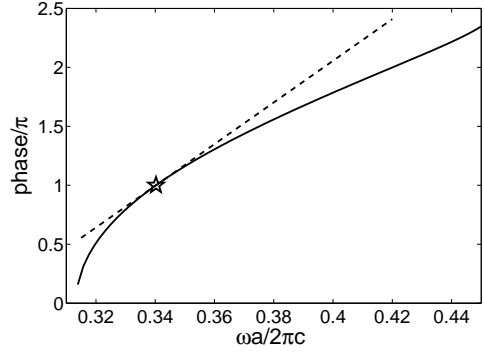}
\caption{Frequency behavior of the optical length (phase) shown by
solid line for the straight PhC waveguide which is fabricated by
removing one row of dielectric rods. The parameters of the PhC are
listed in figure caption of Fig. \ref{fig1}. The BSC point is
marked by star.} \label{T3}
\end{figure}

Moreover we present real dispersion curve $\omega(k)$ calculated
for the straight PhC waveguide to find as the optical length given
by the phase $\theta$ or $\phi$ depends on the frequency $\omega$.
This curve is shown in Fig. \ref{T3} by solid curve for the
parameters of PhC given in Fig. \ref{fig1}. In the vicinity of the
BSC frequency $\omega_ca/2\pi c=0.3402$ we approximate the
dispersion curve as linear to obtain
\begin{equation}\label{phi}
\frac{phase}{\pi}=1+\frac{17.68(\omega-\omega_c)a}{2\pi c}.
\end{equation}
parameters for the phases.

\begin{figure}
\includegraphics[scale=0.4]{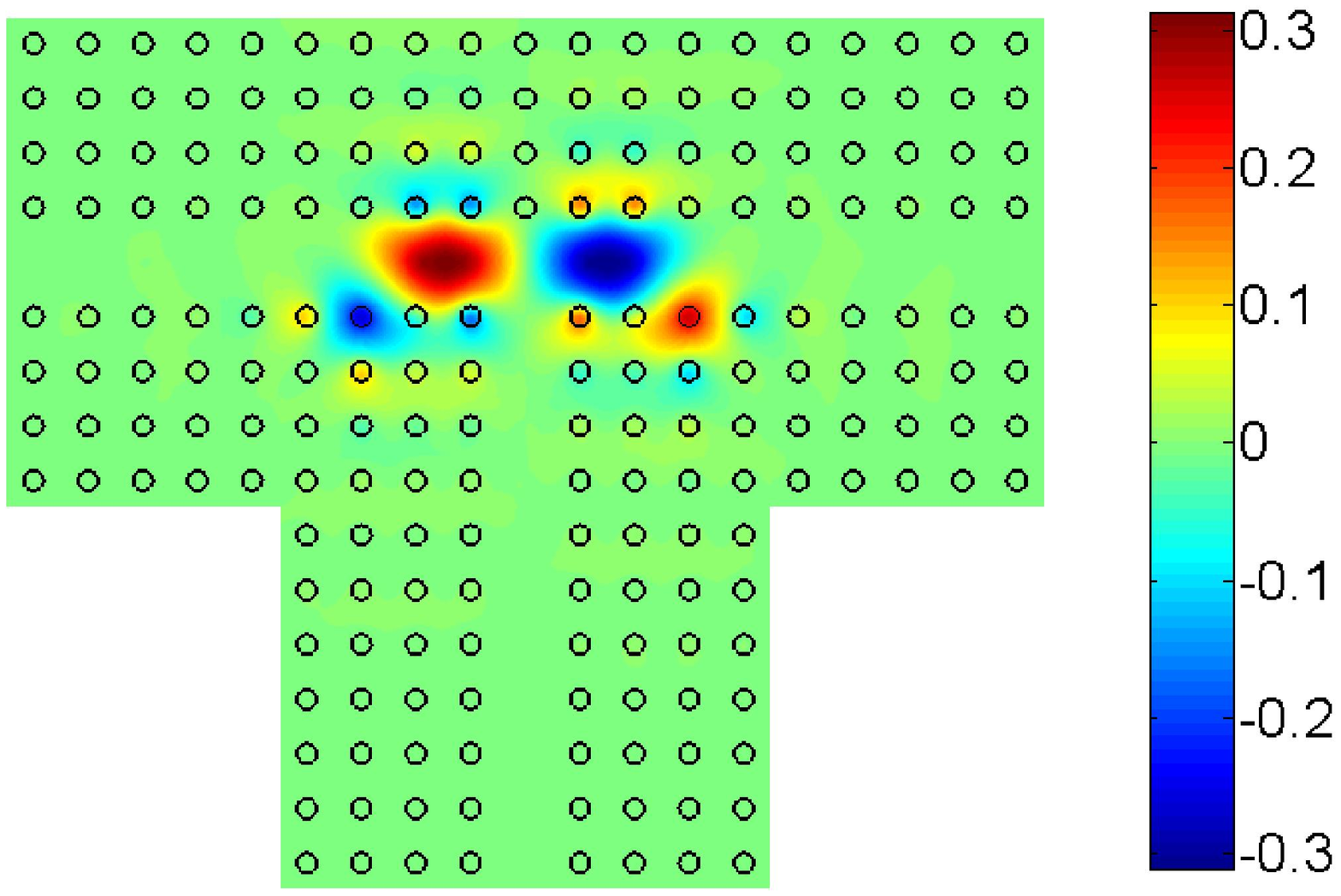}
\includegraphics[scale=0.45]{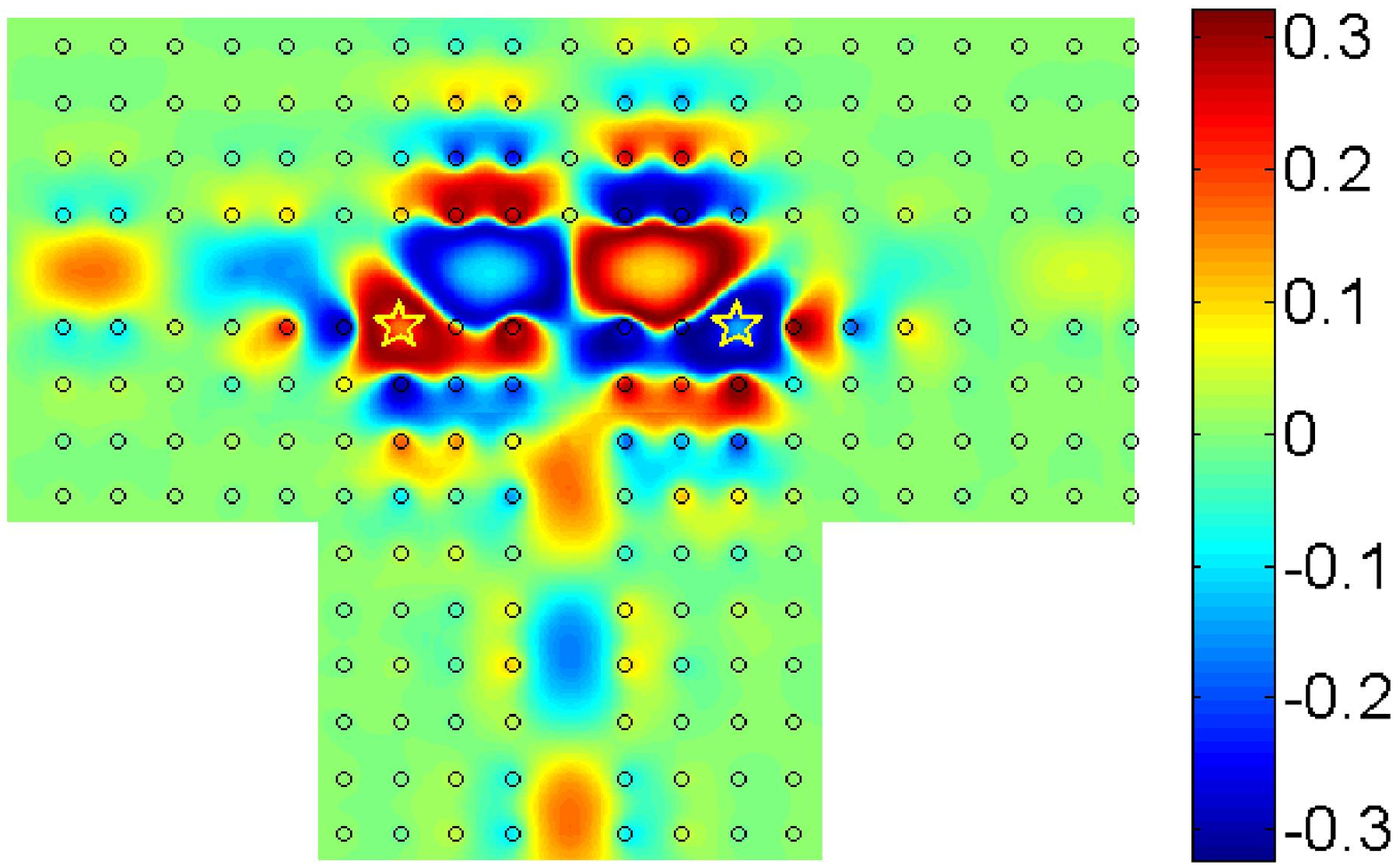}
\caption{Solutions of Maxwell equations for the case shown in Fig.
\ref{T1} which demonstrate (a)  the anti-symmetric FPI BSC
(anti-symmetric standing wave between nonlinear off-channel
defects marked by stars) with frequency $\omega_c a/2\pi c=0.3402$
for zero input power $P=0$ and (b) as this BSC mixes with incident
EM wave to give rise to the symmetry breaking because of
nonlinearity for $P=1.94W/a$. Yellow stars mark defect rods.}
\label{BSC}
\end{figure}

\begin{figure}
\includegraphics[scale=0.35]{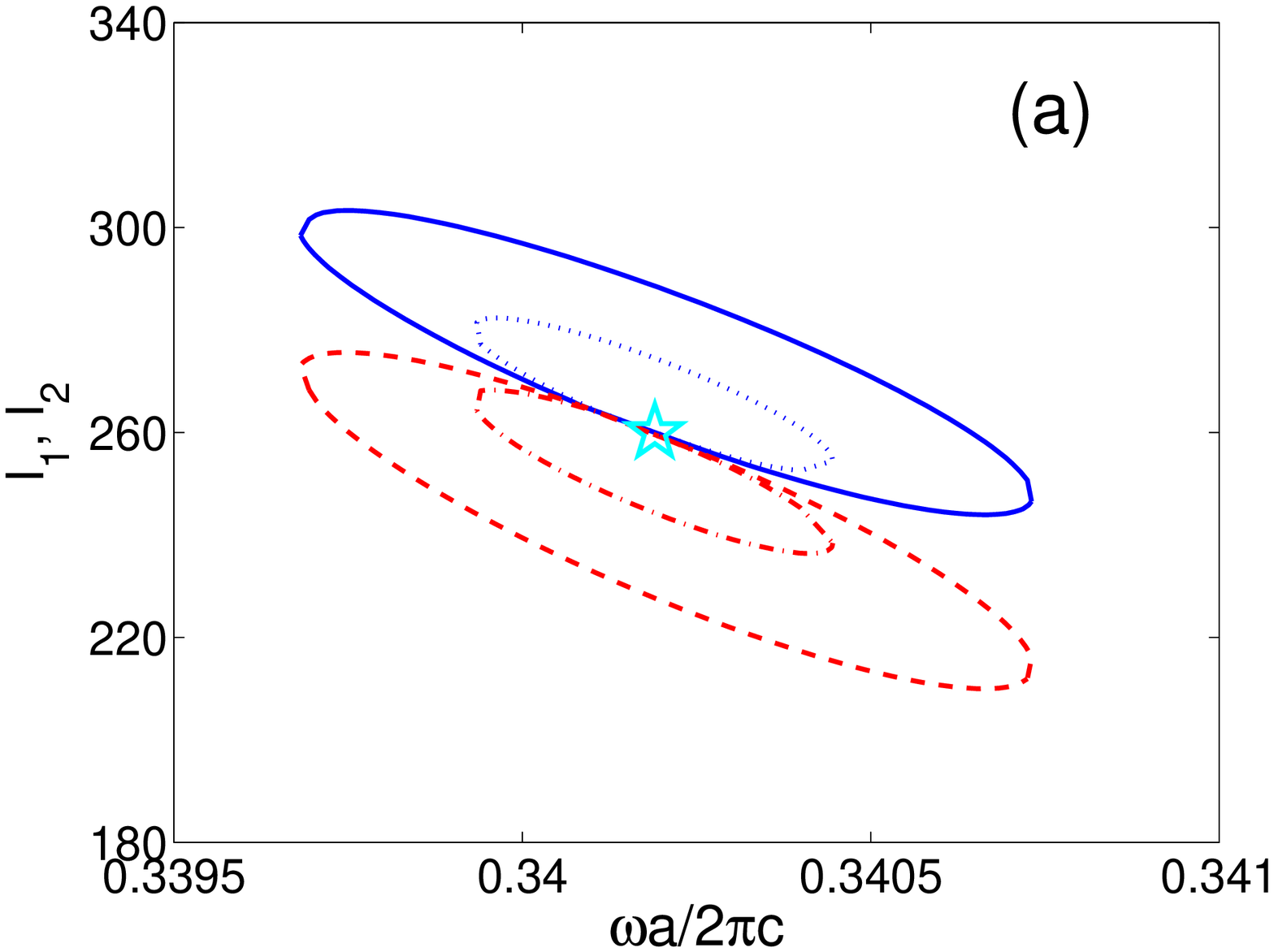}
\includegraphics[scale=0.35]{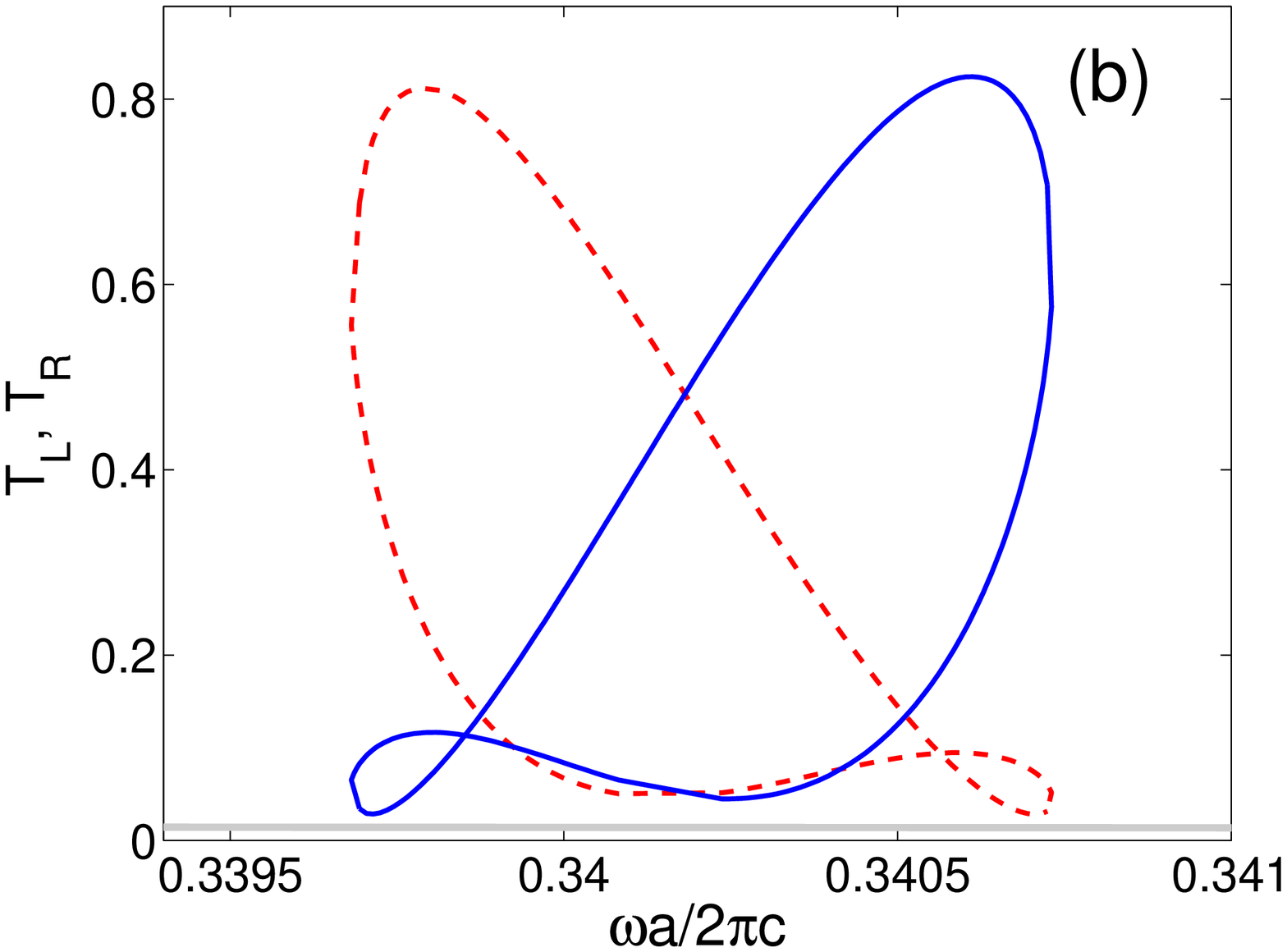}
\caption{(a) Frequency behavior of light intensities at the defect
rods given in terms of $W/a^2$. Dotted blue and dash-dotted red
lines correspond to the input power equal to $0.48W/a$, solid blue
and dashed red lines do to $1.92W/a$. Only those solution for
intensities is shown which breaks the symmetry. (b) The frequency
dependence of the transmissions to the left $T_L$ (blue solid
lines) and to the right $T_R$ (red dashed lines) for the PhC
T-shaped waveguide shown in Fig. \ref{T1} (a). The BSC point is
shown by blue star.} \label{T4}
\end{figure}

The case of the T-shaped structure shown in Fig. \ref{T1}(a) has
an analogy with the Fabry-P\'{e}rot interferometer (FPI)
comprising two off-channel nonlinear cavities. As was shown in
Ref. \cite{FPR} there is a series of the self-induced BSCs which
are the standing waves between the off-channel cavities. It was
shown that the BSCs exist for any distance between the off-channel
defect rods because of their nonlinearity \cite{FPR}. Similar BSC
solutions are expected to exist in the present case of the
T-shaped waveguide coupled with two off-channel cavities shown in
Fig. \ref{T1}(a). However these solutions must be the standing
waves with nodes at the point of connection of the waveguides.
Therefore in the present case the BSCs are the anti-symmetric
standing waves. Fig. \ref{BSC} shows one of these waves in the
T-shaped waveguide with two nonlinear defect rods with the eigen
frequency $\omega_c a/2\pi c=0.3402$ which satisfies the condition
$3k(\omega_c)a=\pi$.

For the linear case the BSC has zero coupling with the symmetric
EM wave which inputs in the waveguide 1. However as was considered
in section II  nonlinearity gives rise to the important effect of
the excitation of BSC by the transmitted wave, (i.e., an
interaction between the anti-symmetric BSC and the symmetric
transmitted wave). As a result the total solution lacks the mirror
symmetry of the PhC structure shown in Fig. \ref{T1}. That is one
of the scenarios of the symmetry breaking. Note, that Maes {\it et
al.} \cite{maes1} have already reported the symmetry breaking in
the FPI. In order for the FPI with two off-channel nonlinear
cavities to have the mirror symmetry equal input power must be
applied to both sides of the FPI \cite{maes1}. In our case of the
T-shaped waveguide this symmetry is achieved by application of the
input power via the additional waveguide positioned at the center
of the FPI.

Figure \ref{T4}(a) demonstrates the solution for the light
intensities of the cavities $I_j=c|E_j|^2/8\pi, j=1, 2$ with
broken mirror symmetry where $E_j$ are the amplitudes of the
electric field in thin defect rods for two values of input power.
The solution converges to the BSC point marked by the star if the
input power limits to zero. At this point the symmetry is
restored. As a result for the input power $P\neq 0$ the light
transmission from the input waveguide 1 to the left waveguide 2
differs from the transmission and to the right waveguide 3 as seen
from Fig. \ref{T4}(b). Moreover the difference $T_L-T_R$ crucially
depends on the frequency in the vicinity of the BSC point. Figure
\ref{T5} demonstrates the solution of the Maxwell equations for
the $z$-component of electric field (the scattering wave function)
breaks the mirror symmetry because of mixing the symmetric input
wave with the anti-symmetric BSC as shown in Fig. \ref{BSC}(b).
The intensity of the defect's modes  is centered around the BSC
intensity which is rather large as shown in Fig. \ref{T4}(a) while
the incident light intensity $P\sim 1\cdot W/a$. Thereby we have
chosen exponential scaling for the solution of the Maxwell
equations presented in Fig. \ref{BSC}(b) in order to distinguish
waves in waveguides.

\begin{figure}
\includegraphics[scale=0.35]{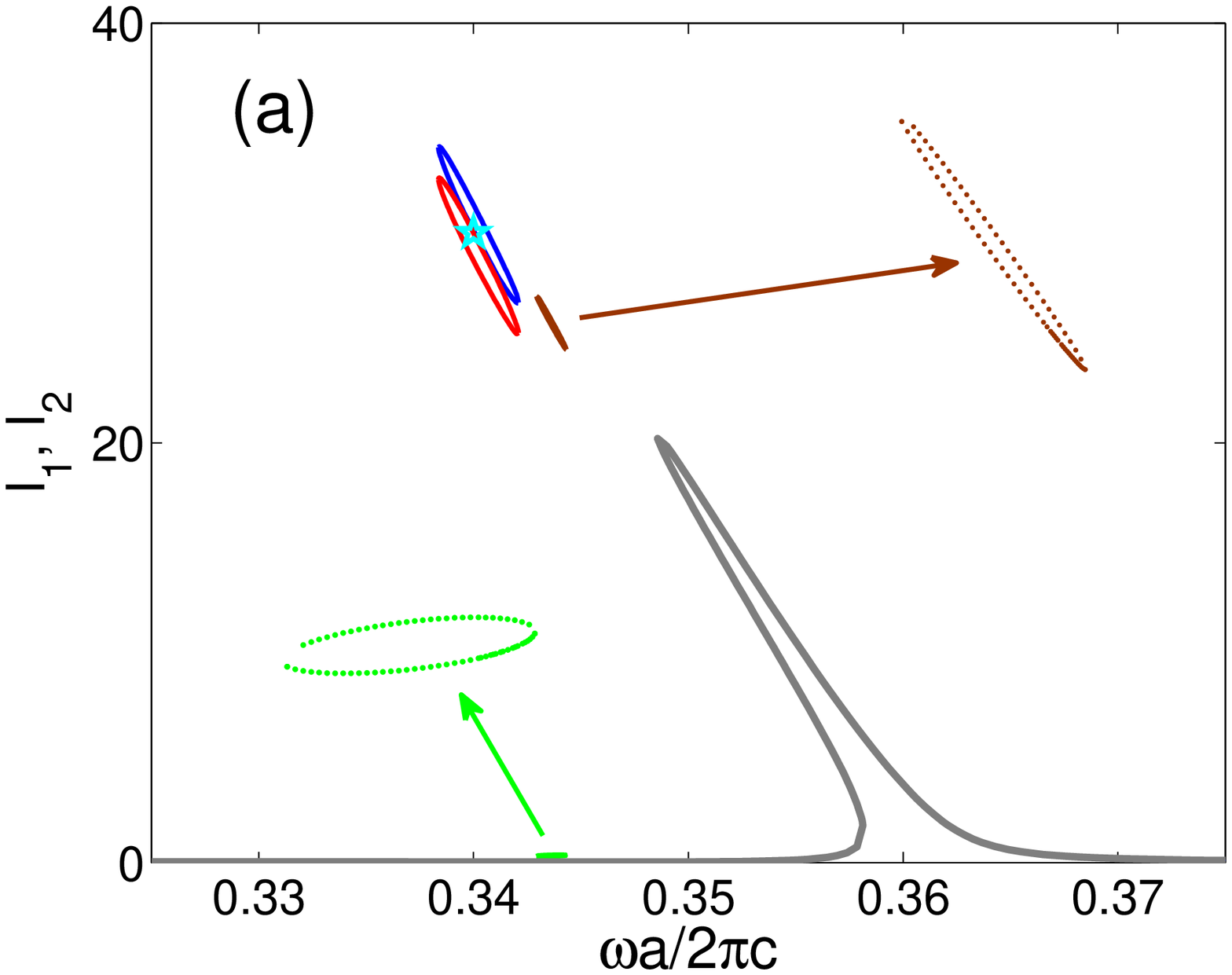}
\includegraphics[scale=0.35]{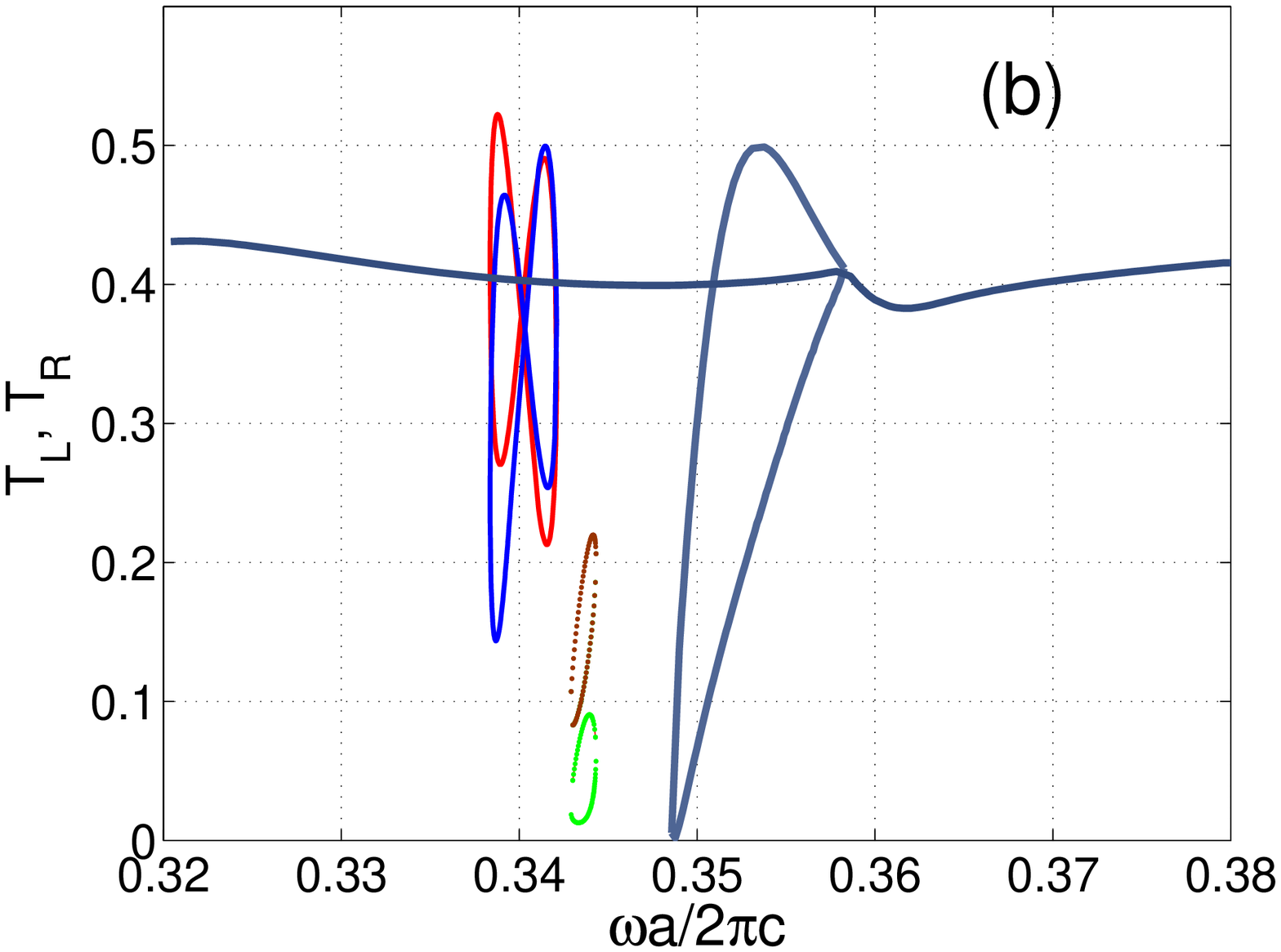}
\caption{Frequency behavior of (a) the light intensities at the
defect rods (in terms of $I_0=W/a^2$) and (b) transmissions $T_L$
and $T_R$ for the PhC T-shaped waveguide shown in Fig. \ref{T1}
(b). The input power equals $5.57W/a$. The symmetric solution is
shown by solid gray line which inherits the linear case. The
symmetry breaking solution because of the mixing the symmetric
transport solution with the anti-symmetric BSC is shown by solid
blue and red lines. The next symmetry breaking solution because of
a bistability of the transmission in each output waveguides is
shown by dotted brown and green lines. Zooming of loops are shown
by arrows in (a).} \label{T6}
\end{figure}

As was shown in the framework of the CMT there might appear an
additional branch of a loop shape for the symmetry breaking
solution with growth of the input power \cite{T}. The numerical
results for the solution of the Maxwell equations completely agree
with these model results as seen from Fig. \ref{T6}(a) by dotted
brown and green lines. The loops shown by dotted lines in Fig.
\ref{T6} are the result of individual instability that arises for
transmission in the left or right waveguide coupled with the left
or right nonlinear off-channel cavity \cite{T}.
Loops in the intensities are reflected in the loops in the
transmission for the transmission to the left and for the
transmission to the right as shown in Fig. \ref{T6}(b) by dotted
brown and green lines respectively.

\begin{figure}
\includegraphics[scale=0.3]{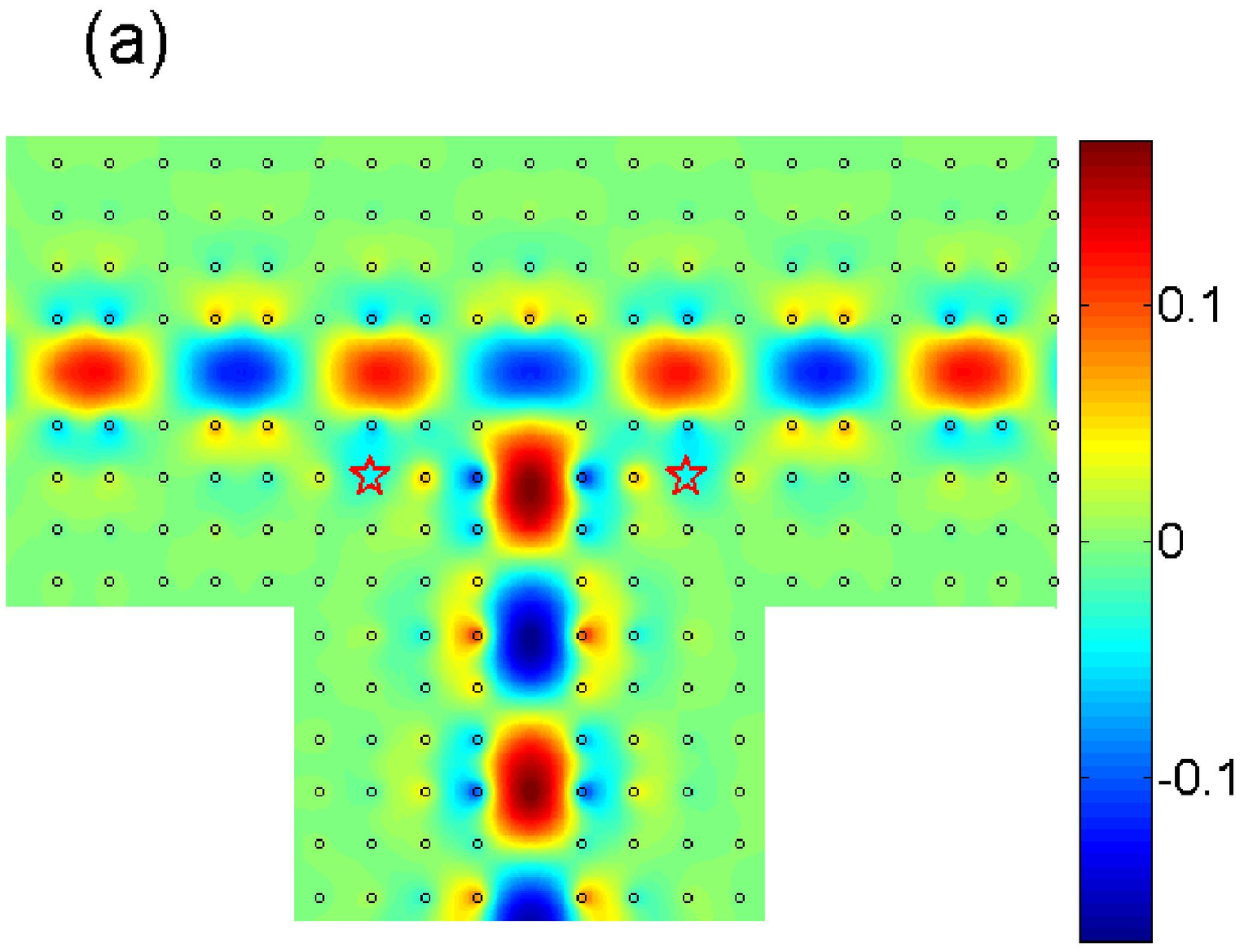}
\includegraphics[scale=0.3]{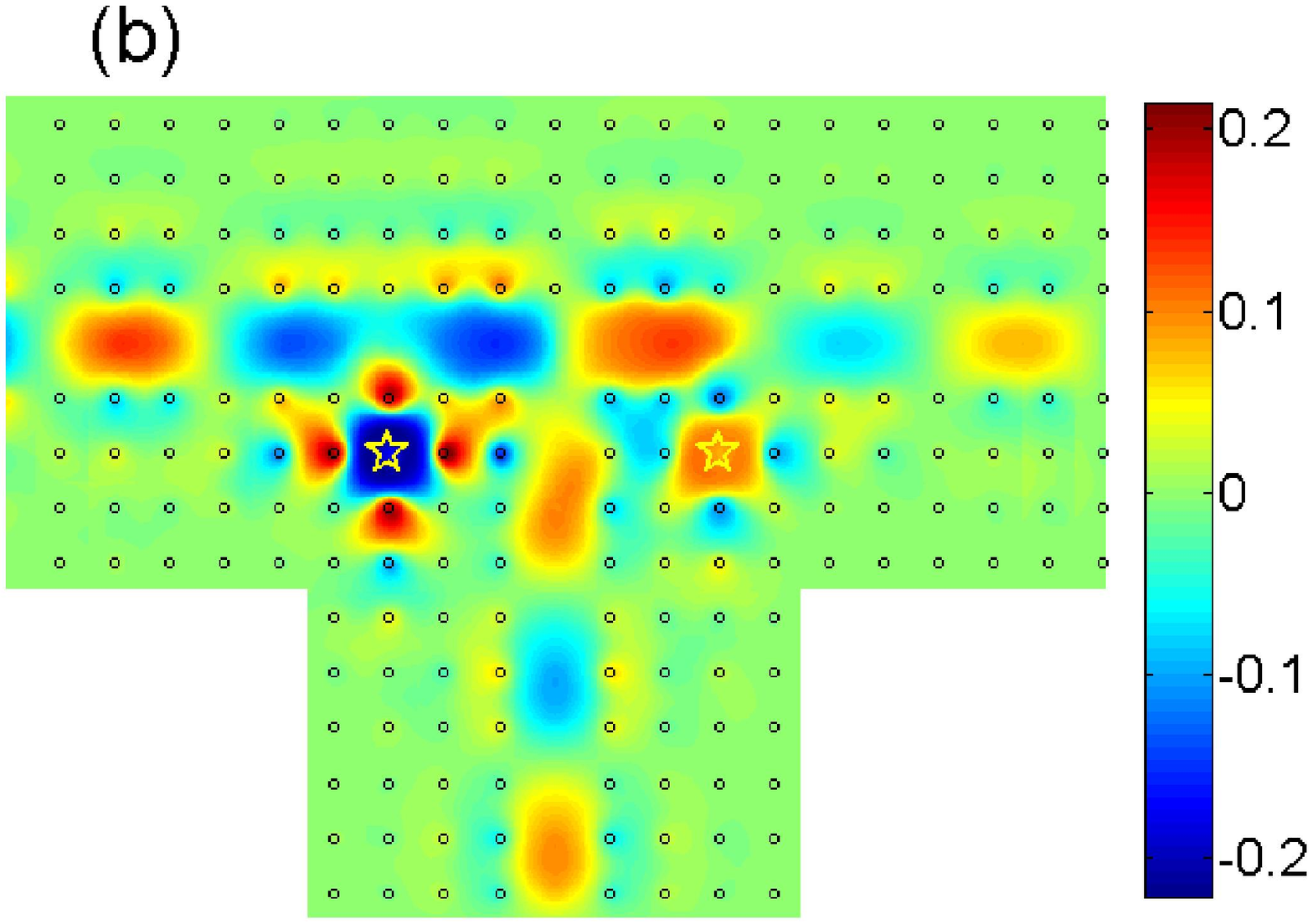}
\includegraphics[scale=0.3]{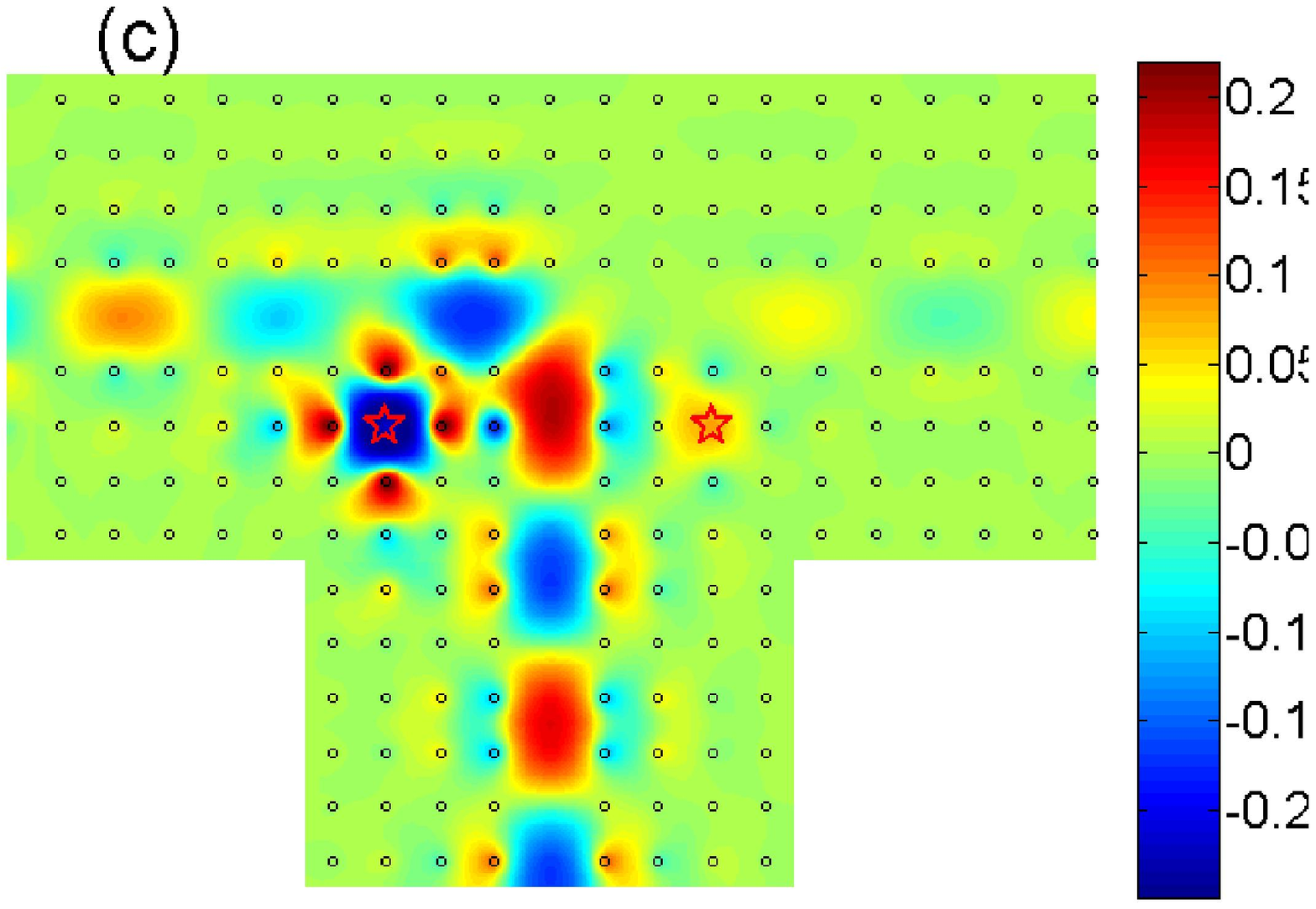}
\caption{The EM field solution in a scale $real(E_z)\exp(-|E_z|)$
for (a) the symmetry preserving solution for $\omega a/2\pi
c=0.3442$, (b)  the symmetry breaking solution caused by the BSC
for $\omega a/2\pi c=0.3388$, and (c) the for $\omega a/2\pi
c=0.3442$. Yellow stars mark defect rods.} \label{T7}
\end{figure}

Bistability of the light transmission in the PhC waveguide coupled
with nonlinear optical cavity crucially depends on the coupling:
the smaller a coupling the less input power is needed for
bistability \cite{joanbook}. For case (a) in Fig. \ref{T1} the
coupling is rather large to observe bistability in the
transmission. However case (b) has the sufficiently smaller
coupling as one can see from Table \ref{tab1}. As a result case in
Fig. \ref{T1}(b) gives rise to additional loops as shown in Fig.
\ref{T6} for larger input power $5.57W/a$. Figure \ref{T7} shows
the wave function for the symmetry preserving solution (a) and the
symmetry breaking solution (b) and (c). Cases (b) and (c) differ
by the frequency.
One can see from Fig. \ref{T7}(a) and Fig. \ref{T7}(b) that for
the symmetric solution the transmission excites the cavities
weakly, while for the symmetry breaking solution the defects are
strongly excited because of mixing the injecting symmetrical wave
with the anti-symmetric FPI BSC. Figure \ref{T7}(c) shows that for
the frequency in the loop domain $\omega a/2\pi c=0.3442$ the
first nonlinear cavity is excited much more than the second one
that is correlated with the outputs.

The T-shaped waveguide coupled with two nonlinear cavities shown
in Fig. \ref{T1} is remarkable in that it allows the limit to the
FPI [case (a)] with the FPI BSC in the form of the standing waves
between two off-channel defects \cite{FPR} as well as the limit to
case (c) in Fig. \ref{T1} with the BSC in the form of the
anti-bonding defect's state. Patterns of such anti-bonding BSC in
PhC straight waveguide coupled with two cavities positioned
perpendicular to the waveguide are shown in Fig. \ref{modes}.
\begin{figure}
\includegraphics[scale=0.35]{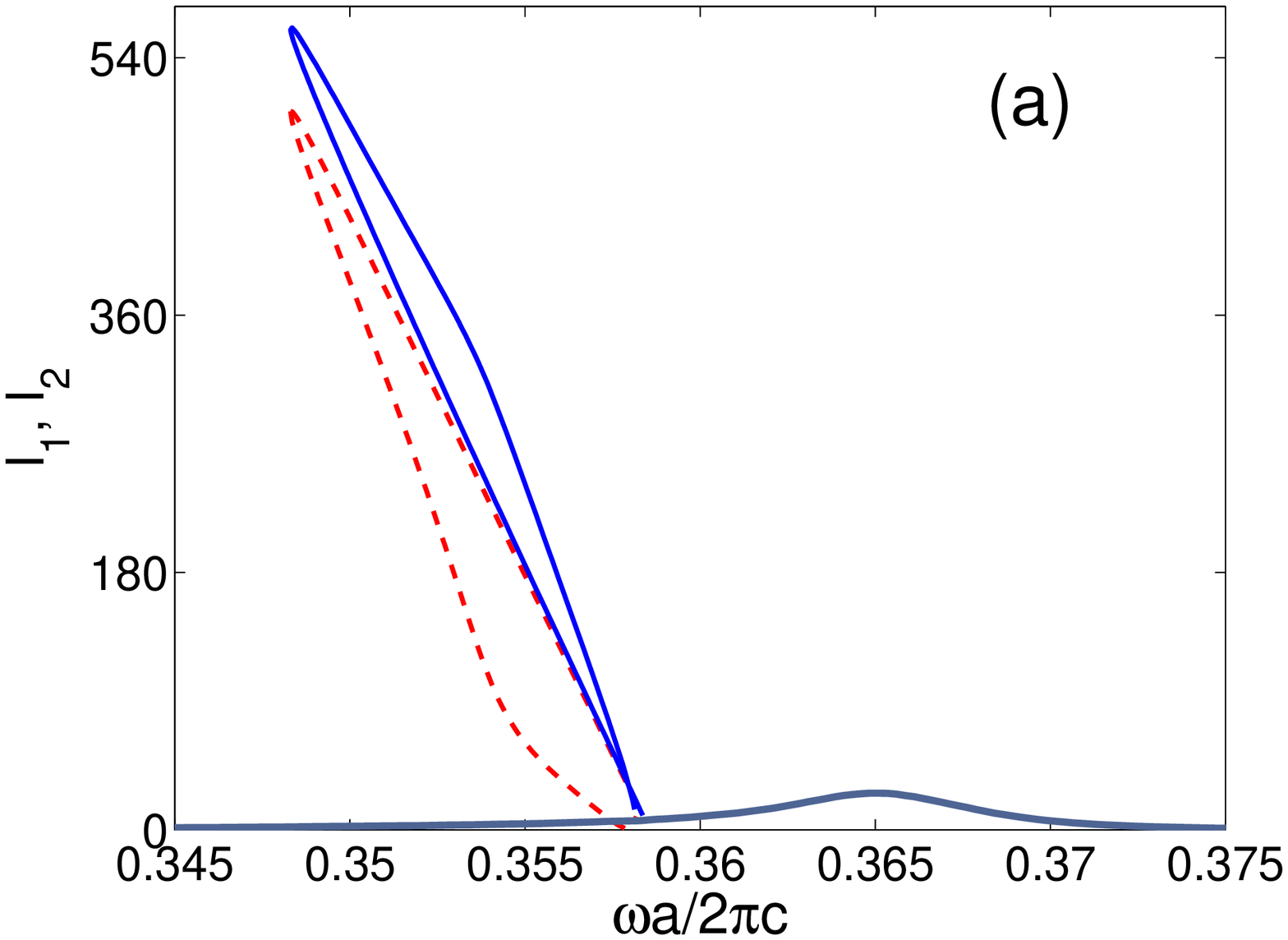}
\includegraphics[scale=0.35]{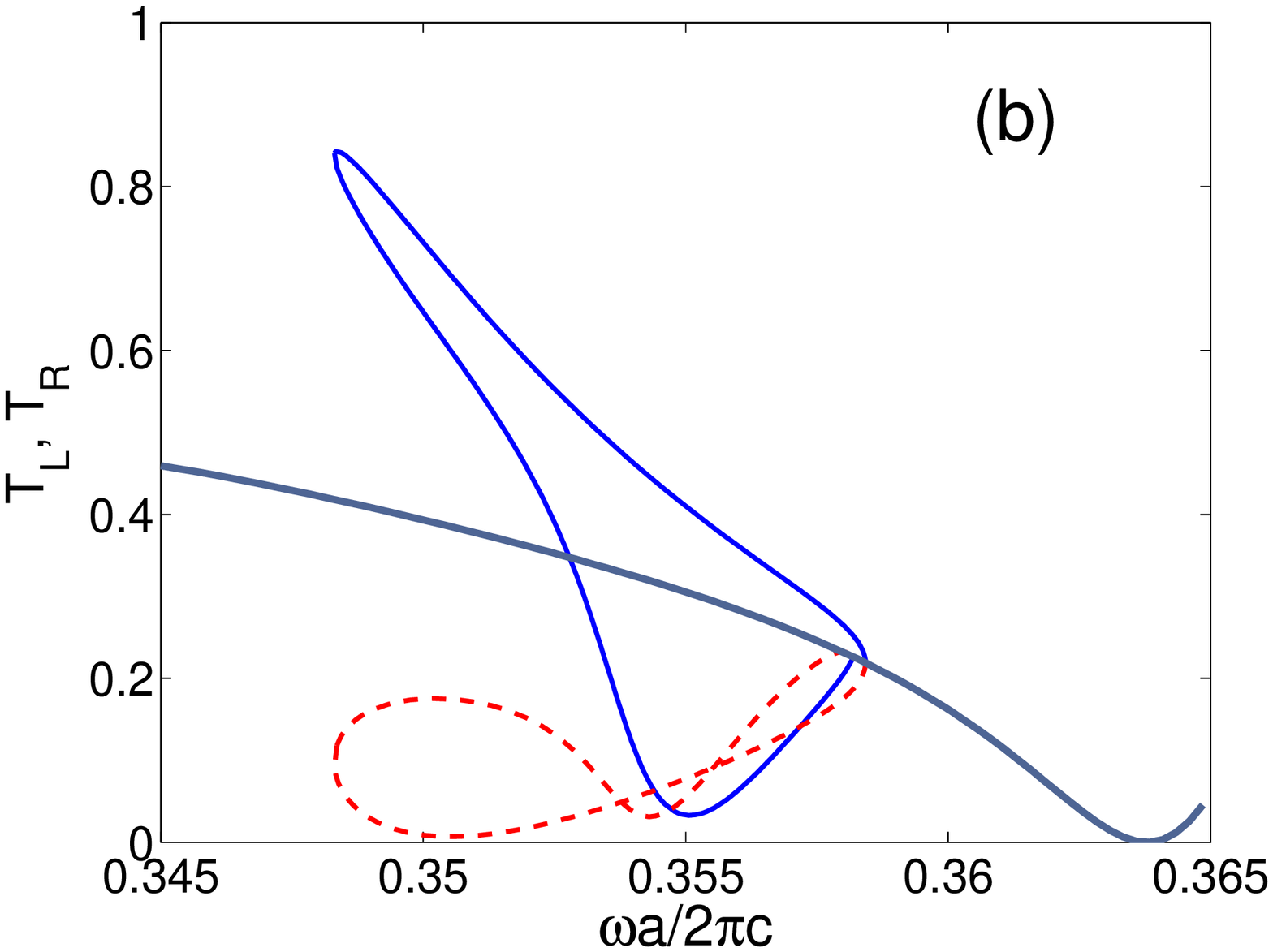}
\caption{Frequency behavior of (a) the light intensities at the
defect rods and (b) transmissions $T_L$ and $T_R$ for the PhC
T-shaped waveguide shown in Fig. \ref{T1}(c). The input power
$P=0.48W/a$. Only those solution is shown in (b) which
demonstrates different outputs in the left  (red dashed line) and
right (blue solid line) waveguides.} \label{T8}
\end{figure}
\begin{figure}
\includegraphics[scale=0.55]{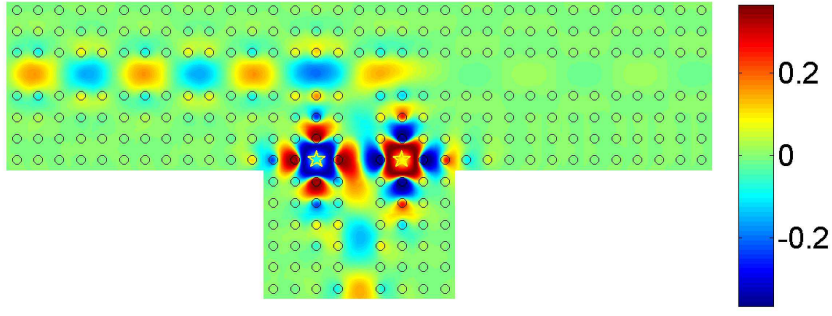}
\caption{EM field for the symmetry breaking solution $\omega
a/2\pi c=0.3505$. Light is incident into the 1 waveguide and
scatters to the left and right waveguides 2 and 3. $P=0.48W/a$.}
\label{T9}
\end{figure}

The nonlinearity gives rise to mixing the anti-bonding BSC with
the wave transmitted over waveguide 1. For the linear defect rods
this state would be the perfect BSC. For the nonlinear cavities
mixing this anti-bonding BSC with symmetric input light leads to
the breaking of the mirror symmetry to give rise to the breaking
of symmetry in the input waveguide. Then for the evolution of this
structure to the T-shaped case can expect different outputs to the
right and to the left. Indeed, in spite of the small difference of
the defect intensities presented in Fig. \ref{T8}(a) the
transmissions $T_L$ and $T_R$ demonstrate vast difference
including the case of almost perfect blocking of the transmission
to the left as shown in Fig. \ref{T8}(b) for $\omega a/2\pi
c=0.3505$. Figure \ref{T9} shows the anti-bonding BSC is mixed to
the transport over the input waveguide to give rise to the
symmetry breaking solution.

\begin{figure}
\includegraphics[scale=0.5]{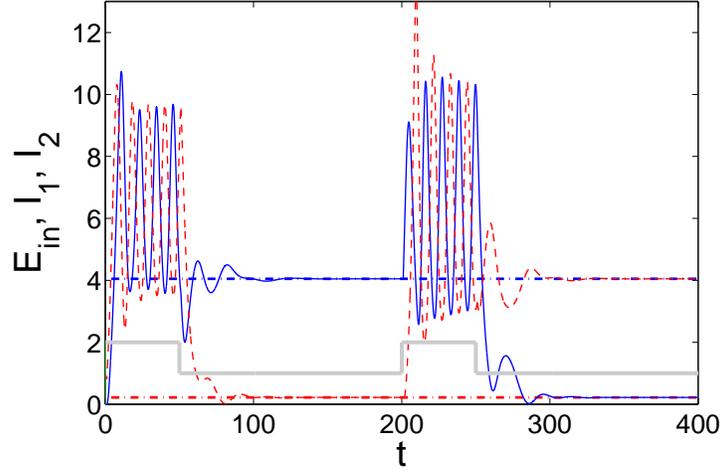}
\caption{The time dependence of the amplitudes of the light
amplitudes $|A_1|, |A_2|$ in the cavities (solid and dashed
respectively) which follow the impulses of the input amplitude
$E_{in}$ (gray). We take the cavities oscillate in non symmetric
way: $A_1=0, A_2=1$.} \label{switch}
\end{figure}

These results are extremely important for the switching of the
output power from the left waveguide to the right one. In order to
switch the system from one asymmetric state to the other we
following Refs. \cite{maes1,Otsuka} apply pulses of the input
power injected into the waveguide 1. The direct numerical solution
of the temporal CMT equation
\begin{eqnarray}\label{A1A2temp}
&i\dot{A_1}=(\omega_1-i\gamma)A_1+i\sqrt{\gamma}\sigma_{2-}e^{i\phi}\,&\nonumber\\
&i\dot{A_2}=(\omega_2-i\gamma)A_2+i\sqrt{\gamma}\sigma_{3-}e^{i\phi}\,&
\end{eqnarray}
with $S_{1+}(t)=E_{in}(t)e^{-i\omega t}$ is shown in Fig.
\ref{switch} which demonstrates the switching effect. The stepwise
time behavior of amplitude $E_{in}(t)$ is shown by gray line. One
can see that after the first impulse of the input amplitude the
oscillations of the cavity amplitude relax onto the stable
stationary solutions with broken symmetry. Moreover after each
next impulse the state of the system transmits from one asymmetric
state to the other as was observed by Maes {\it et al}
\cite{maes1}.

\section{dipole modes of the single nonlinear defect coupled with
waveguide}
If to present the defect by the single monopole mode $E_s({\bf
x})$ we obtain  from Eqs. (\ref{Am})
\begin{eqnarray}\label{Asin}
[\omega-\omega_s-\lambda
|A_s|^2-i\Gamma_s]A_s=i\sqrt{\Gamma_s}E_{in},\\t=E_{in}-\sqrt{\Gamma_s}A_s,
\end{eqnarray}
where $\lambda=-\frac{3}{4}\sigma\chi^{(3)}(\omega_s)E_s(({\bf
x}_d)^2$. That model has attracted interest over the past two
decades because of analytical treatment and its generality for
description of bistability phenomena
\cite{joanbook,mcgurn,miros,flach,longhi,ming2,miros1,miros2}.
However the monopole eigen-function presents only the trivial
identical symmetry transformation. Respectively there is no room
for the breaking of symmetry.

\begin{figure}[ht]
\includegraphics[scale=0.35]{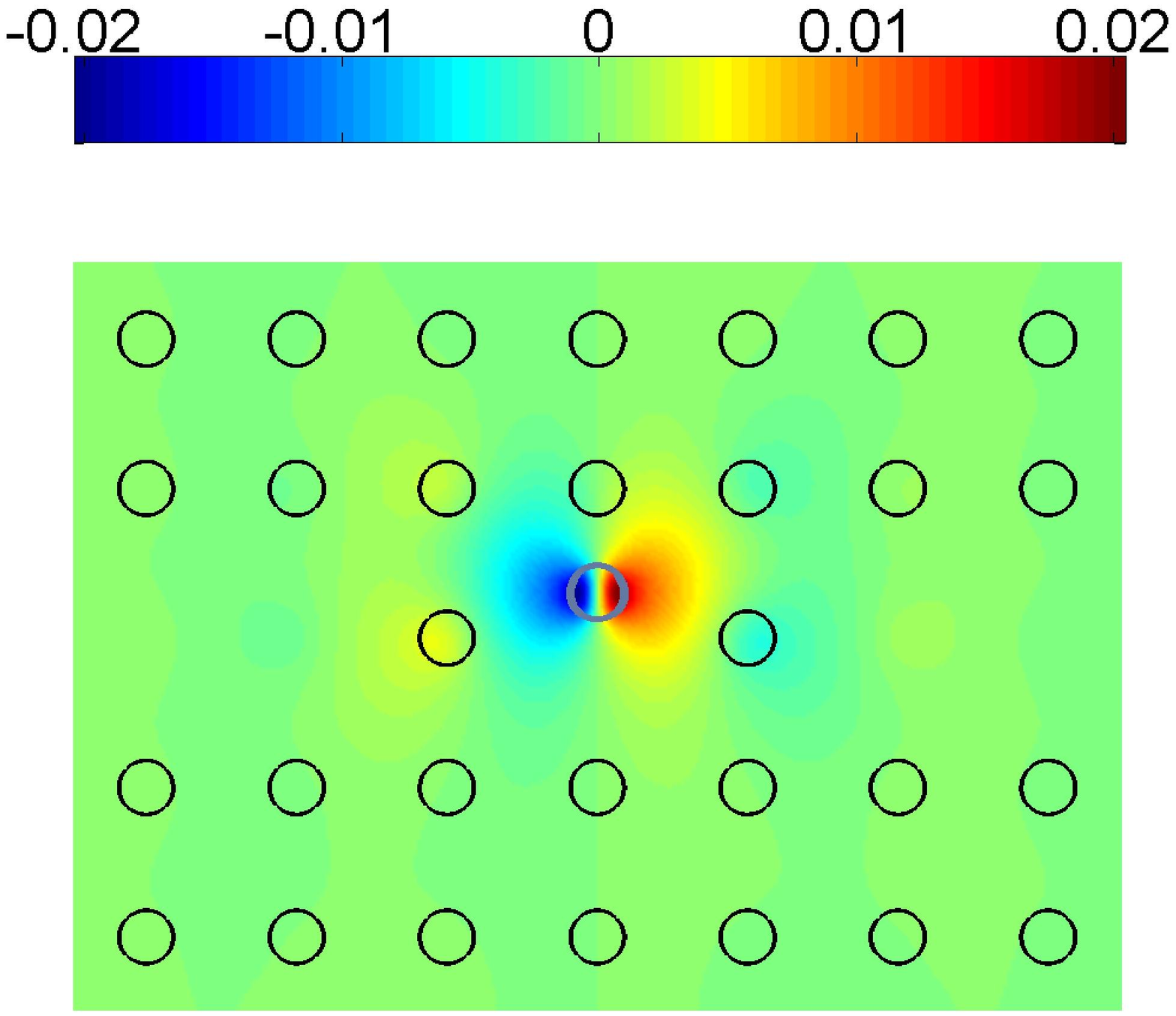}
\includegraphics[scale=0.35]{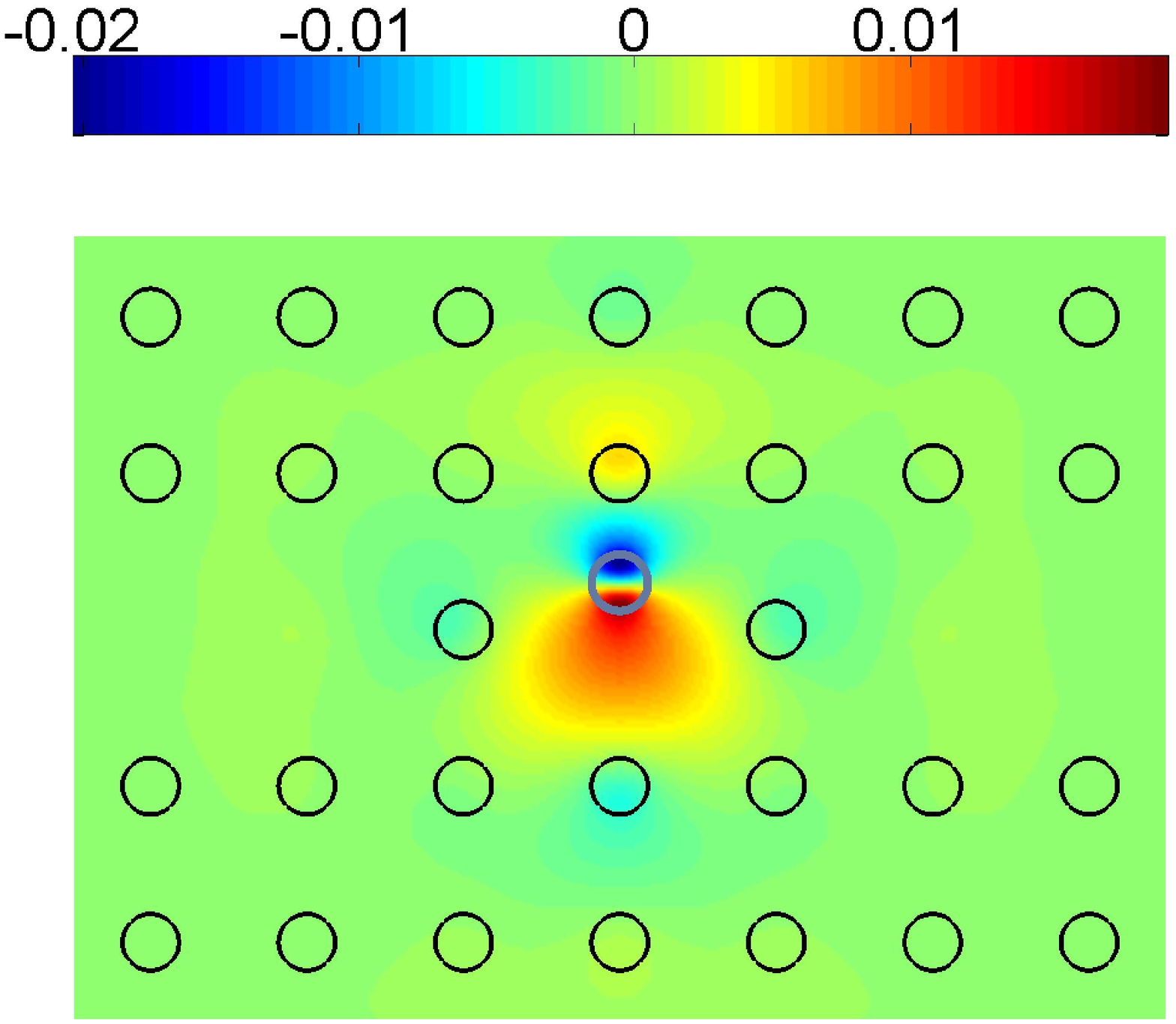}
\caption{Two defect dipole eigen-modes with the eigen frequencies
$\omega_1=0.3578\pi a/c$ and $\omega_2=0.3616\pi a/c$ in the
two-dimensional square lattice PhC consisted of the GaAs
dielectric rods with radius $0.18a$ and dielectric constant
$\epsilon=11.56$ where $a=0.5\mu m$ is the lattice unit. These
rods are shown by black open circles. The defect shown by open
gray circle has the same radius $0.18a$ and $\epsilon_0=30$. Its
center is positioned at $x_d=0, y_d=0.3a$.} \label{dip}
\end{figure}

In this respect the system becomes nontrivial if we include two
eigen dipole modes $E_1({\bf x})$ and $E_2({\bf x})$ of the defect
rod with the eigen-frequencies $\omega_1$ and $\omega_2$. The
modes are shown in Fig. \ref{dip}.
Respectively, the electric field at the thin defects can be
expanded over these dipole modes $E({\bf x}_d)=A_1E_1({\bf
x}_d)+A_2E_2({\bf x}_d)$ only. Substituting that expansion into
Eq. (\ref{MEV}) we obtain that Eq. (\ref{Am}) will take the
following form
\begin{eqnarray}\label{A1A2dip}
&[\omega-\omega_1-\lambda_{11}|A_1|^2-\lambda_{12}
|A_2|^2+i\Gamma_1]A_1-2\lambda_{12}
Re(A_1^{*}A_2)A_2=i\sqrt{\Gamma_1}(S_1^{+}-S_2^{+}),&\nonumber\\
&-2\lambda_{12}Re(A_1^{*}A_2+[\omega-\omega_1-\lambda_{22}|A_2|^2-\lambda_{12}
|A_1|^2+i\Gamma_2]A_2=i\sqrt{\Gamma_2}(S_1^{+}+S_2^{+}),&
\end{eqnarray}
where $S_{1,2}^{+}$ are the amplitudes of light injected
simultaneously into both sides of the waveguide. Here with
accordance to Eq. (\ref{Vmn1}) we have
\begin{equation}\label{lambdamn}
    \lambda_{mn}=\frac{3}{16\epsilon^{3/2}}\chi^{(3)}(\omega_m+\omega_n)Q_{mn}, ~~
    Q_{mn}=\int E_m^2({\bf x})E_n^2({\bf x})d^2{\bf x}\nonumber.
\end{equation}
Moreover we have taken into account the symmetry relations for the
coupling constants of the dipole modes with the waveguide
solutions \cite{johnson}. As seen from Fig. \ref{dip} the first
dipole mode has the coupling with the left and the right ingoing
waves opposite signs while the second dipole mode has the same
coupling with these waves. Therefore the coupling matrix in Eq.
(\ref{LS}) equals
\begin{equation}\label{Wdip}
W=\left(\begin{array}{cc}  \sqrt{\Gamma_1}  & -\sqrt{\Gamma_1}\cr
\sqrt{\Gamma_2} & \sqrt{\Gamma_2} \cr
\end{array}\right).
\end{equation}
Moreover Eq. (\ref{A1A2dip}) must be complemented by equation for
the transmission amplitude
\begin{eqnarray}\label{tdip}
    t_{LR}=S_1^{+}-i\sqrt{\Gamma_1}A_1+i\sqrt{\Gamma_2}A_2,\nonumber\\
t_{RL}=S_2^{+}-i\sqrt{\Gamma_1}A_1+i\sqrt{\Gamma_2}A_2.
\end{eqnarray}

\begin{figure}
\includegraphics[scale=0.35]{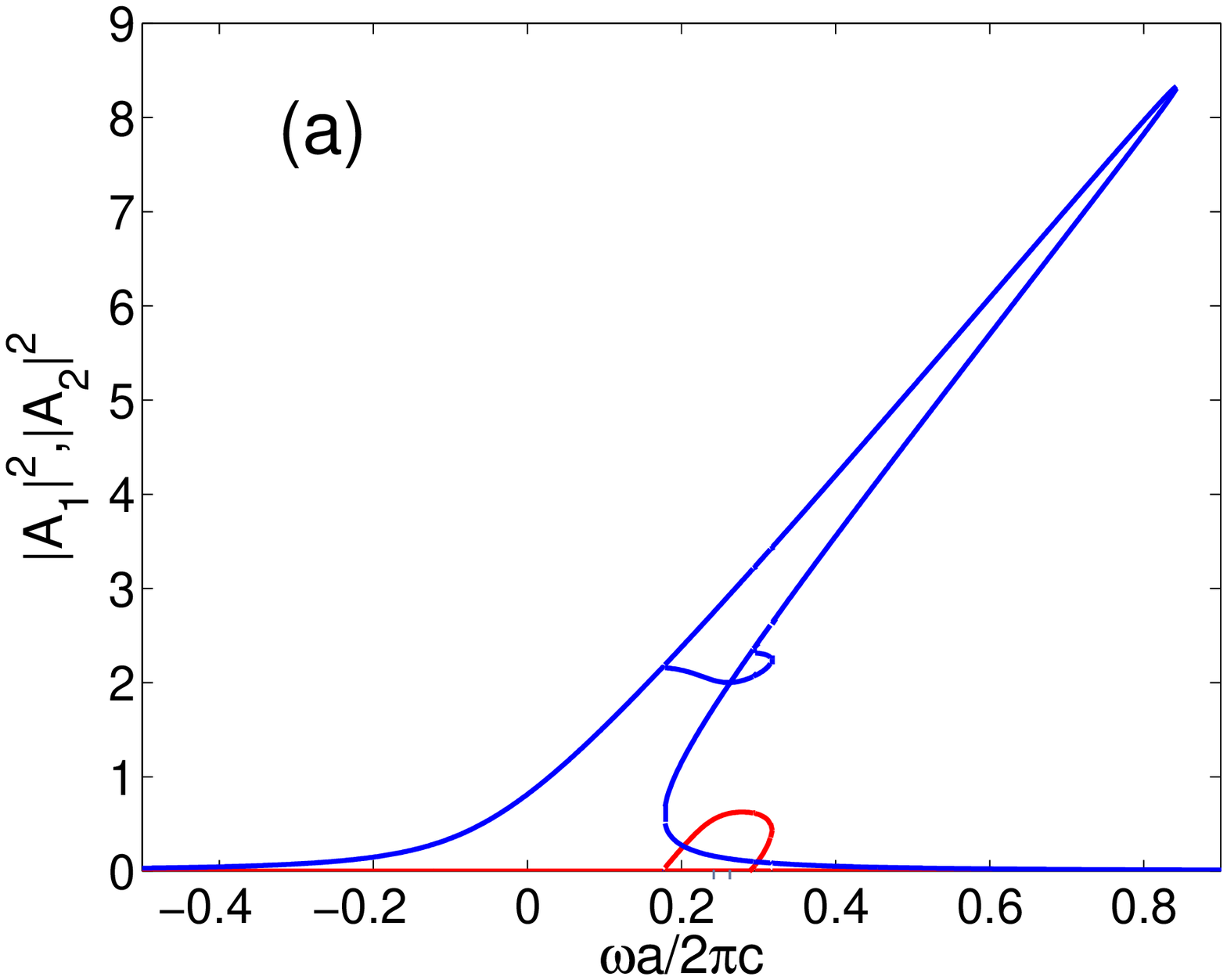}
\includegraphics[scale=0.35]{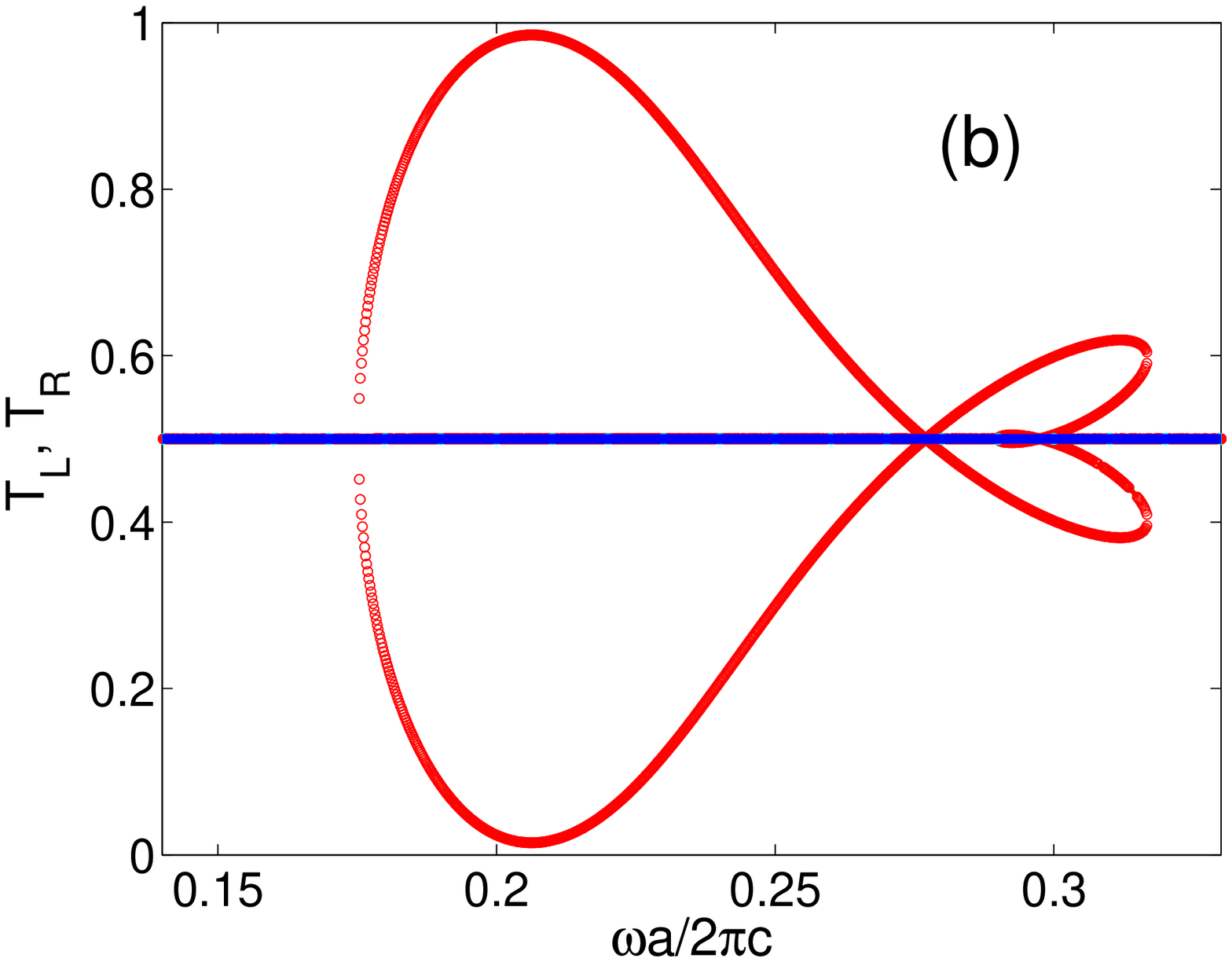}
\caption{Frequency behavior of (a) intensities of dipole modes and
(b) transmissions to the left $T_L=|t_{RL}|^2$ and to the right
$T_R=|t_{LR}|^2$ for for light injection $S_{1,2}^{+}=0.025$ onto
both sides of the straight forward waveguide. Blue lines show the
intensity of even dipole mode, while red lines show the intensity
of odd dipole mode. excitement The model parameters are listed in
the text.} \label{I1I2mod}
\end{figure}

Fig. \ref{I1I2mod} shows the self-consistent solutions of Eq.
(\ref{A1A2dip}) after substitution of the following model
parameters $\omega_1=0, \omega_2=0.01,
\lambda_{11}=\lambda_{22}=0.1, \lambda_{12}=0.05, \Gamma_1=0.1,
\Gamma_2=0.03$.
One can see that for the symmetry preserving branch the first even
dipole mode is excited with frequency behavior typical for the
single nonlinear mode described by Eq. (\ref{As}) while the second
odd dipole mode is not excited because of symmetry. Respectively,
the transmission probability to both sides of the waveguide equals
1/2 because of normalization condition for the intensity of input
light and unitarity of the S-matrix
$|S_{1}|^2+|S_{2}|^2+|t_{LR}|^2+|t_{RL}|^2=1$. However there is
also the symmetry breaking branch in some narrow frequency domain.
Respectively, that branch gives rise to an asymmetry in the light
outputs. Moreover there is a frequency at which the right (left)
output is blocked almost perfectly.
\begin{figure}
\includegraphics[scale=0.35]{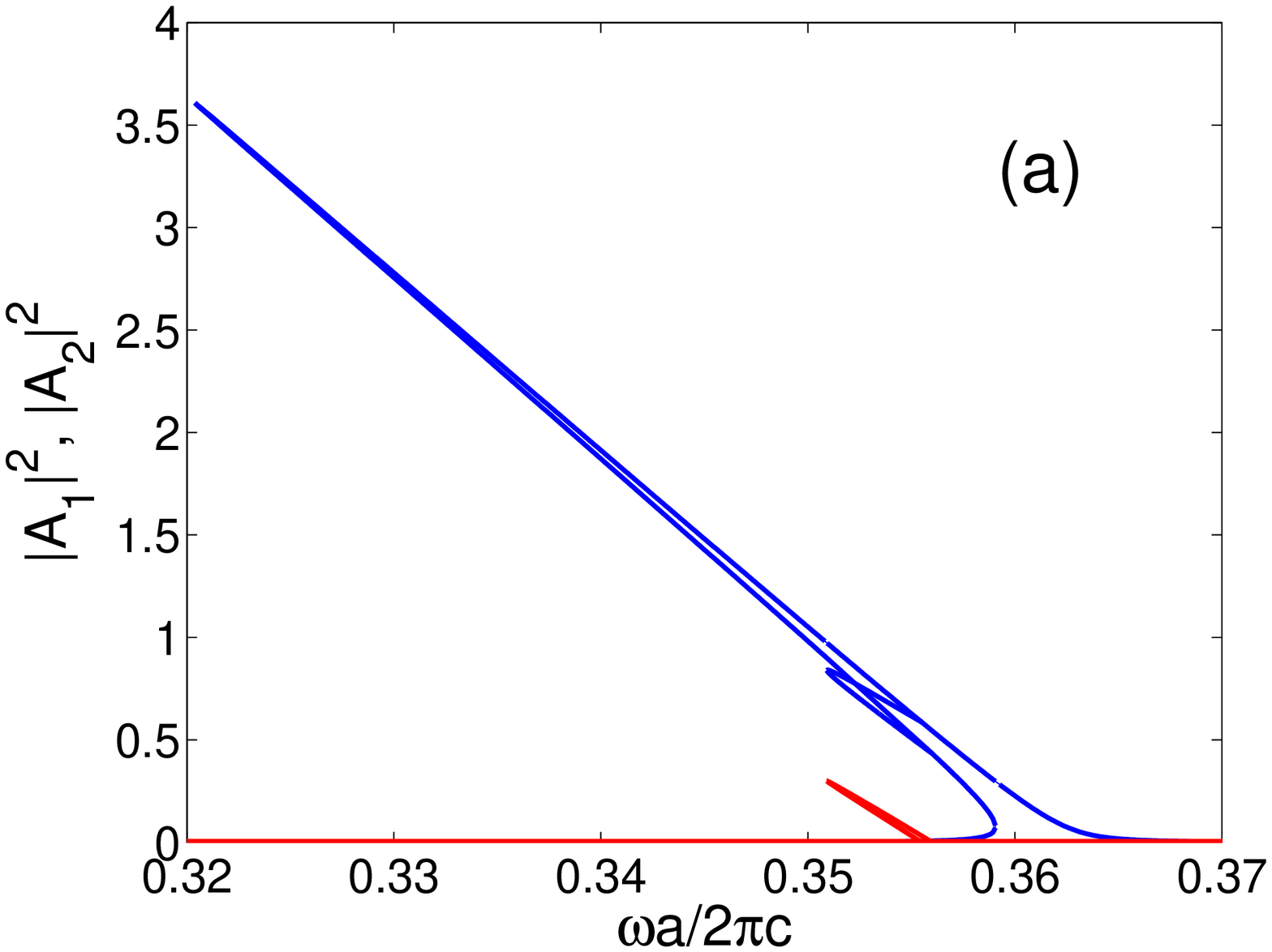}
\includegraphics[scale=0.35]{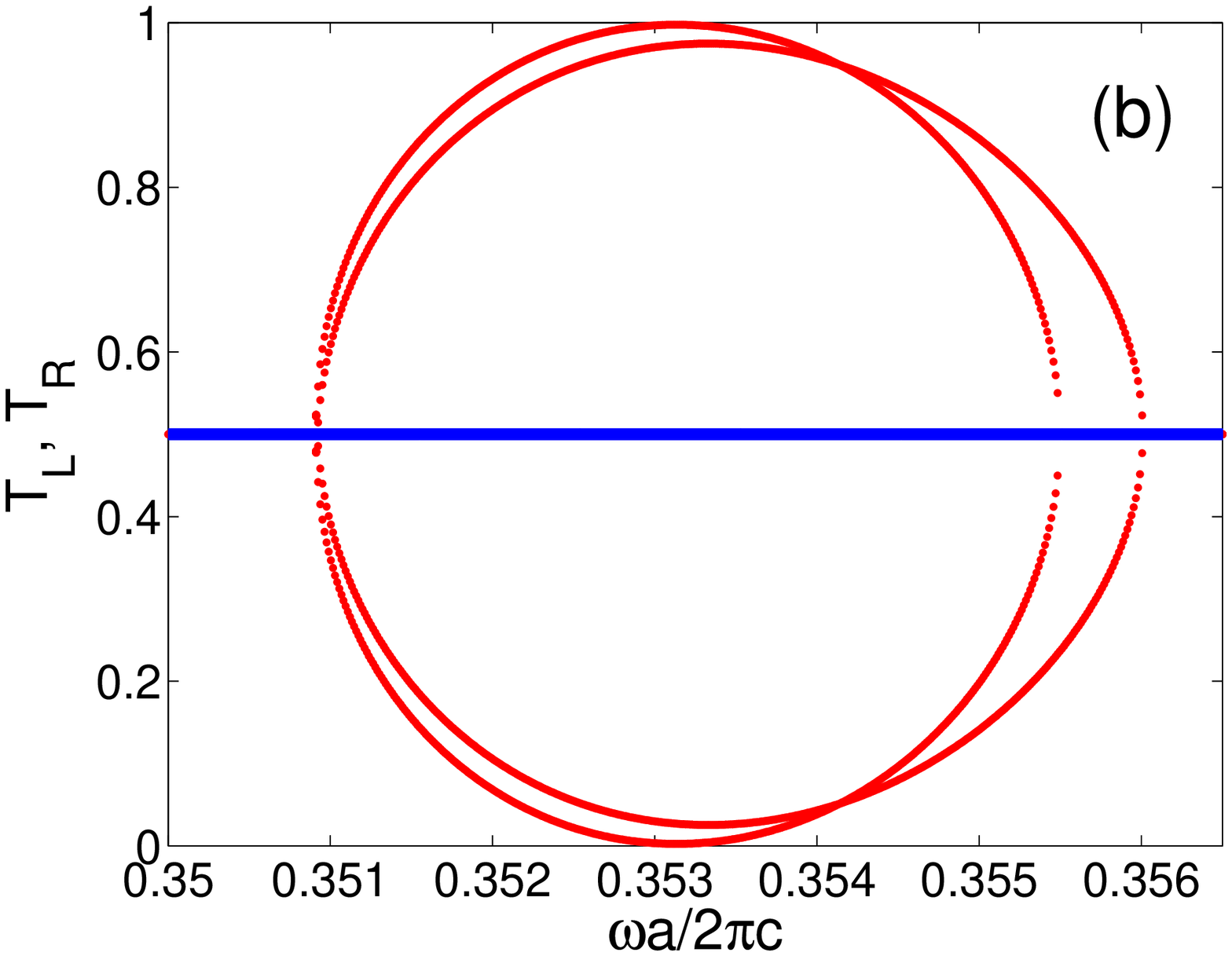}
\caption{Frequency behavior of (a) intensities of dipole modes and
(b) transmissions to the left  and to the right in real 2D PhC for
$S_{1,2}=0.015$. Dash blue line shows the symmetry preserving
solution, solid red line shows the symmetry breaking solution, and
grey thick line shows the phase parity breaking solution. }
\label{T1T2dip}
\end{figure}

Also we consider the phenomenon of the symmetry breaking based on
the parameters calculated in the 2D square lattice PhC with
parameters given earlier. The results for the eigen frequencies
are collected in Fig. \ref{dip} caption where the parameters of
the PhC are given too. The coupling constants are the following
$\Gamma_1=0.00075, \Gamma_2=0.00025$ in terms of $2\pi c/a$. In
order to enhance the coupling constants we substituted two
additional linear rods nearby the nonlinear defect rod as shown in
Fig. \ref{dip}. Let us evaluate the nonlinearity constants
$\lambda_{mn}$. With use
$\epsilon_d=\epsilon_0+2\sqrt{\epsilon_0}n_2I_0,
n_2\approx\frac{2\pi\chi^{(3)}}{n_0}$ we obtain from
(\ref{lambdamn})
\begin{equation}\label{lambdamn}
\lambda_{mn}=- \frac{3\pi}{8}\sqrt{\epsilon_0}n_2I_0\int
    E_m^2({\bf x})E_n^2({\bf x})d^2{\bf x}.
\end{equation}
We take the linear and nonlinear refractive indexes of the defect
rods are, respectively, $n_0=\sqrt{\epsilon_0}=\sqrt{3},
I_0=1.8W/a,  n_2=2\times 10^{-12}cm^2/W$ \cite{milam,boudebs}.
Moreover substituting the dipole modes into integrals in Eq.
(\ref{lambdamn}) we obtain $Q_{11}=0.0123, Q_{22}=0.0114,
Q_{12}=0.0037$. Results of solution of the self-consistency
equations for two complex amplitudes $A_1$ and $A_2$ are presented
in Fig. \ref{T1T2dip}(a). Correspondingly from Eq. (\ref{tdip}) we
obtain the transmission coefficients shown in Fig.
\ref{T1T2dip}(b).

\section{summary and conclusions}
As by direct solution of the Maxwell equations as well as by
solution of the CMT equations we have demonstrated the symmetry
breaking in the system of two nonlinear defects coupled with
waveguide through which a light is injected. The defects are
aligned symmetrically relative to the waveguide as shown in Fig.
\ref{fig1}, so there is an inversion symmetry relative to the
waveguide axis. The thin dielectric rods made from Kerr media  are
these defects which presented by the eigen monopole mode whose
eigen frequency belongs to the propagation band of the waveguide.
We assume that other eigen modes are beyond the band, and
therefore have no coupling with the injected light with accuracy
of evanescent modes.

That simplest system is remarkable in that it reveals as
nonlinearity gives rise to the breaking of symmetry. Indeed, let
us take temporarily the defects are linear. If the coupled defects
were isolated, it would have only two eigen-modes, bonding (even)
and anti-bonding (odd) with corresponding eigen frequencies
(\ref{wsa}). For light propagating over the waveguide we take that
the solution is symmetrical relative to the inversion. Then the
light can excite only the bonding mode $A_s$ to give rise typical
resonance dip at the symmetrical eigen frequency $\omega_s$ while
the anti-bonding mode $A_a$ would remain hidden for the
propagating light. Obviously, that mode is the simplest case of
the bound state in continuum \cite{BPS1,guevara,FAM}. That state
can be superposed to the scattering function with coefficient
determined by a way to excite the anti-bonding mode \cite{ring}.
However the principle of linear superposition is not correct in
the nonlinear case. Therefore the nonlinearity gives rise to
interaction of the scattering state with the anti-bonding mode,
i.e., to the interaction of the symmetric propagating light with
the anti-bonding mode. That obviously breaks the inversion
symmetry as explicitly shown in Figs. \ref{wavebroken}(b),
\ref{wavebroken}(c), \ref{T6}(b), and \ref{T6}(c). Obviously, the
mixing gives total state which is nor symmetrical neither anti
symmetrical, breaking the mirror symmetry. Further we have shown
that the symmetry can be broken not only because of different
light intensities but also because of  different phases of light
oscillations at the cavities to provoke the Josephson like current
between cavities \cite{BPS2}. For the light transmission in
two-dimensional PhC the Poyinting power current is an analog of
the Josephson currrent, which is shown in Fig. \ref{currents}(c).
It is clear that for the light transmission in PhC waveguide there
is the power current over the waveguide. However if the waveguide
is coupled with nonlinear defects the current pattern might be
rather complicated even in the symmetry preserving scenario as
shown in Fig. \ref{currents}(a). One can see that laminar current
flow over the waveguide induces two vortices around the defects
which obey the mirror symmetry. As the symmetry has broken the
vortex in the waveguide appears which is well separated from the
defect vortices. Although the mirror symmetry relative to
circulation of currents in defect vortices is broken, however, the
symmetry is remained relative to absolute value of current flows.
At last, for the phase parity breaking solution still there is
symmetry in absolute value of current, but all vortices are
exchanging by current flows as seen from Fig. \ref{currents}(c).

The T-shaped waveguide coupled with two symmetrically positioned
nonlinear defects as shown in Fig. \ref{T1} can be considered as a
combination of the previous system and the Fabry-P\'{e}rot
interferometer (FPI) consisted of two nonlinear off-channel
cavities aligned along the straightforward waveguide considered in
Refs. \cite{maes1,maes2,FPR}. As was shown in Ref. \cite{FPR}
there is a discrete set of the a self-induced bound states in
continuum (BSC) which are standing waves between off-channel
cavities which are to be anti-symmetric in order to elucidate an
escape to the input waveguide 1. As dependent on position of the
nonlinear defects shown consequently in Fig. \ref{T1} the system
goes from the FPI (case a) to the system considered in section II
(case c). In the case (a) the nonlinearity couples the FPI BSC
shown in Fig. \ref{BSC} with the incident wave which is
symmetrical relative to the inversion left to right. As the result
the inversion symmetry breaks as shown for real PhC in Figs.
\ref{T7}(b) and \ref{T7}(c) to give rise to strong asymmetry of
light outputs to the left and to the right as shown in Fig.
\ref{T8}(b). Similarly, the nonlinearity mixes the anti-bonding
hidden state with the symmetric incident wave as shown in Fig.
\ref{T9} to give rise different outputs too. At last, for the
nonlinear defects are coupled as with input waveguide as well as
with output waveguides the frequency behavior of the transmissions
is complicated as shown in Fig. \ref{T8}(b). In Ref. \cite{T} we
demonstrate as these phenomenona  can  be explored  for
all-optical switching of light transmission from the left output
waveguide to the right one by application input pulses.

In section V we show that breaking of symmetry might occur even
for the single nonlinear defect positioned in the straight forward
waveguide provided that the defect is presented by two dipole
modes. That model embraces two degrees of freedom of the defect
while in the former models each defect was presented by the only
degree of freedom.

 \acknowledgments{AS
is grateful to Boris Malomed for fruitful and helpful discussions.
The work is partially supported by RFBR grant 12-02-00483.}

\end{document}